\documentclass[12pt,preprint]{aastex}

\shorttitle{}
\shortauthors{Sembach et al.}
\slugcomment{\it To appear in the Astrophysical Journal Supplement Series}
\begin{document}

\newcommand{\kms}{\,km\,s$^{-1}$}

\title{Physical Properties and Baryonic Content of Low-Redshift 
Intergalactic Ly$\alpha$ and \ion{O}{6}
Absorption Line Systems: 
The PG\,1116+215 Sight Line\footnotemark\ $^,$\footnotemark}

\footnotetext{$^1$Based on observations obtained with the NASA/ESA Hubble Space Telescope, which
is operated by the Association of Universities for Research in Astronomy, Inc.,
under NASA contract NAS\,5-26555.}
\footnotetext{$^2$Based on observations obtained with the 
NASA-CNES-CSA Far Ultraviolet Spectroscopic Explorer.  FUSE is operated for 
NASA by the Johns Hopkins University under NASA contract NAS\,5-32985.}

\author{Kenneth~R.~Sembach\altaffilmark{3},
Todd~M.~Tripp\altaffilmark{4},
Blair~D.~Savage\altaffilmark{5},
Philipp Richter\altaffilmark{6}}
\altaffiltext{3}{Space Telescope Science Institute, 3700 San Martin Dr., 
Baltimore, MD~21218} 
\altaffiltext{4}{Department of Astronomy, University of Massachusetts, 
Amherst, MA  01003}
\altaffiltext{5}{Department of Astronomy, University of Wisconsin-Madison,
	475 N. Charter St., Madison, WI~53706}
\altaffiltext{6}{Institut fur Astrophysik und Extraterrestrische Forschung, 
Universit$\ddot{a}$t Bonn, Auf dem H$\ddot{u}$gel 71, 53121 Bonn, Germany}

\begin{abstract}
We present {\it Hubble Space Telescope} and {\it Far Ultraviolet
Spectroscopic Explorer} observations of the intergalactic absorption 
toward QSO PG\,1116+215 in the $900-3000$\,\AA\ spectral region. 
We detect 25 Ly$\alpha$ absorbers along the sight line 
at rest-frame equivalent widths $W_r > 30$ m\AA, yielding
(d$N$/d$z$)$_{{\rm Ly}\alpha} = 
154\pm18$  over an 
unblocked redshift path $\Delta X_{{\rm Ly}\alpha} = 0.162$.  
Two additional weak Ly$\alpha$ absorbers with $W_r \approx 15-20$\,m\AA\ are
also present.
Eight of the Ly$\alpha$ absorbers have large line widths (b $\gtrsim
40$ \kms).  
 The detection of narrow \ion{O}{6} absorption in the broad 
Ly$\alpha$ absorber at $z=0.06244$  supports the idea that the 
Ly$\alpha$ profile is thermally broadened in
gas with $T > 10^5$\,K. We find d$N$/d$z$ $\approx 50$ for broad 
Ly$\alpha$ absorbers with 
$W_r \gtrsim 30$\,m\AA\ and ${\rm b} \ge 40$ \kms.  This number drops to 
d$N$/d$z$ $\approx 37$ if the line widths are restricted to 
$40 \le {\rm b} \le 100$ \kms. 
If the broad Ly$\alpha$ lines are dominated by thermal broadening in
hot gas, the amount of baryonic material in these absorbers is enormous,
perhaps as much as half the baryonic mass in the low-redshift universe.
We detect \ion{O}{6} absorption in several of the Ly$\alpha$ clouds along 
the sight line.  Two detections at $z=0.13847$ and $z=0.16548$ are 
confirmed by the presence of other ions at these redshifts (e.g.,
 \ion{C}{2-III}, \ion{N}{2-III}, \ion{N}{5},
\ion{O}{1}, and \ion{Si}{2-IV}), while the 
detections at $z=0.04125$, $0.05895$, $0.05928$, and $0.06244$ are based upon 
the Ly$\alpha$ and \ion{O}{6} detections alone.  We find 
(d$N$/d$z$)$_{\rm O\,VI} \approx 17$ for \ion{O}{6} absorbers
 with $W_r > 50$\,m\AA\ toward PG\,1116+215.  
The information available for 13 low-redshift \ion{O}{6} absorbers
with $W_r \ge 50$\,m\AA\ along 5 sight lines yields 
(d$N$/d$z$)$_{\rm O\,VI} \approx 14$ and 
$\Omega_b$(\ion{O}{6}) $\gtrsim0.0027 h_{75}^{-1}$, assuming a metallicity of 
0.1 solar and an \ion{O}{6} ionization fraction $f_{\rm O\,VI} \le 0.2$.
The properties and prevalence of low-redshift
\ion{O}{6} absorbers suggest that they too may be a substantial baryon
repository, perhaps containing as much mass as stars and gas inside galaxies.
The redshifts of the \ion{O}{6} absorbers are highly correlated with the 
redshifts of galaxies along the sight line, though few of the absorbers lie
closer than $\sim600\,h_{75}^{-1}$ kpc to any single galaxy.
We analyze the kinematics and ionization of the 
metal-line systems along this sight line 
and  discuss the implications of these observations for 
understanding the physical conditions and baryonic content of 
intergalactic matter in the low-redshift universe.  

\end{abstract}


\section{Introduction}

There is strong evidence that the warm-hot 
(T $\sim 10^5-10^7$\,K) intergalactic medium (IGM) is a significant repository
of  baryons in the low-redshift universe.  The observational basis for this 
assertion is the growing sample of absorption-line systems found to contain 
highly ionized stages of oxygen (e.g., \ion{O}{6} -- 
Tripp, Savage, \& Jenkins 2000; Tripp \& Savage 2000; Savage et al. 2002)
and/or broad \ion{H}{1} Ly$\alpha$ absorption (Richter et al.\ 2004).
These measurements, which are being made  with the spectrographs on the 
{\it Hubble Space Telescope} (HST) and {\it Far Ultraviolet Spectroscopic
 Explorer} (FUSE), are available for only a small number of sight lines, 
but they already indicate that the hot gas may be prevalent
throughout the nearby universe.
In addition, X-ray detections of higher ionization 
species (e.g., \ion{O}{7}) at X-ray wavelengths with the
{\it Chandra X-ray Observatory} 
and {\it XMM-Newton} point to a large amount of coronal gas in the 
Local Group if the absorption does not arise within the interstellar
medium of the Galaxy along the lines of sight surveyed
(e.g., Nicastro et al. 2002; Fang, Sembach, \&
Canizares 2003; Rasmussen, Kahn, \& Paerels 2003).  Support for an
extended distribution of coronal gas around the Milky Way is bolstered by 
detections of \ion{O}{6} absorption at the boundaries of  
high-velocity clouds located at large distances from the Galactic plane
(Sembach et al.\ 2000; 2003).

The low-redshift \ion{O}{6} and Ly$\alpha$ 
observations favor the presence of 
warm-hot ($10^5-10^7$\,K) intergalactic gas predicted by
 theoretical and numerical
expectations for the evolution of intergalactic
gas in the presence of cold dark matter 
(e.g.,  Cen \& Ostriker 1999a; Dav\'e et al. 1999, 2001).  
Hydrodynamical simulations of the cosmos predict that Ly$\alpha$ clouds 
 evolve to form extended 
filamentary and sheet-like structures. 
These sheets and filaments gradually collapse into 
denser concentrations and form galaxies.   As this happens, shocks heat 
the gas to high temperatures.  
The gas density, temperature, and 
metallicity  of the warm-hot IGM
 depend upon the evolutionary state of the 
gas and its proximity to galaxies (Cen \& Ostriker 1999b), so it is 
important to obtain spectroscopic observations of \ion{O}{6} and other 
absorption lines to gain insight into the physical conditions
and environments of the absorbers detected.

An initial census of low-redshift \ion{O}{6}
absorbers along 5 sight lines observed with the HST and FUSE 
indicates that
the number of \ion{O}{6} absorbers per unit redshift is (d$N$/d$z$)$_{\rm O VI}
\approx 14\pm^9_6$ (Savage et al. 2002 and references therein).  
Assuming a metallicity of 0.1 solar and converting this number density to 
a mass implies a cosmic baryon density 
$\Omega_b$(\ion{O}{6}) $ \gtrsim 0.002 h^{-1}_{75}$ in these \ion{O}{6}
absorbers,
which is comparable to the baryonic mass in galaxies (Fukugita, Hogan,
\& Peebles 1998).  This estimate depends critically on knowing the 
metallicity and ionization properties of the \ion{O}{6} systems since 
the total gas column must be calculated from the observed \ion{O}{6}
column densities
(i.e., N$_{\rm H}$ = N(\ion{O}{6})\,$\times$\,N(O)/N(\ion{O}{6})\,$\times
\,{\rm (O/H)}^{-1}$).   Determining the total baryonic content of the 
\ion{O}{6} absorbers therefore requires estimates of the amount of gas 
in the different types of systems in which \ion{O}{6} is found 
(photoionized, collisionally ionized, mixed ionization, etc).  This is 
particularly important in light of recent estimates for the 
amount of gas that may be contained in broad Ly$\alpha$ 
absorbers (Richter et al.\ 2004; see also \S9.2).  Some of the broad Ly$\alpha$ 
absorbers contain detectable amounts of \ion{O}{6}, and some do not 
(see \S9.2).

Distinguishing between  different types of \ion{O}{6} 
systems and quantifying N(O)/N(\ion{O}{6}) and the gas metallicity are
 best accomplished
by comparing the \ion{O}{6} absorption to absorption by other species 
detected at ultraviolet and X-ray wavelengths.  
 High resolution ultraviolet spectroscopy 
permits detailed kinematical comparisons 
of the high ionization lines (e.g.,  \ion{O}{6}) with lower 
ionization metal-line species (e.g.,\ion{C}{2-III-IV}, \ion{Si}{3-IV},
\ion{Al}{3}) and \ion{H}{1} Lyman-series features.  Close kinematical coupling
of \ion{O}{6} with lower ionization stages would suggest that
the warm-hot and cool IGM are mixed or in close proximity. 
X-ray measurements may also hold great promise for
placing gas hotter than that traced by \ion{O}{6} 
(e.g., $T > 10^6$\,K) in context, but current X-ray instruments
provide only low-resolution ($\lambda/\Delta\lambda < 750$) 
measurements of bulk kinematics and  column densities in   
absorption toward only the very
brightest AGNs/QSOs (see, e.g., Mathur, Weinberg, \& Chen 2003). 

To date, only a handful of sight lines of been surveyed
for low-redshift \ion{O}{6} absorption and broad \ion{H}{1} Ly$\alpha$
absorption.  Even fewer have had systematic galaxy survey information
published to assess the relationship of the absorbers and galaxies.
In this paper we discuss the physical properties and baryonic content of the 
Ly$\alpha$ and \ion{O}{6} absorption line systems along the PG\,1116+215
sight line. 
Section~2 contains a description of the observations and data reduction.
In \S3 we briefly describe the sight line and the 
Galactic foreground absorption.  We also calculate the unblocked 
redshift paths available to search for Ly$\alpha$ and \ion{O}{6} absorption.
In \S4 we provide information about the 
identification of the absorption lines in the FUSE and STIS spectra.  
Section~5 contains a brief overview of each intergalactic absorption
line system identified.   Sections 6 and 7 
contain detailed information for the two metal-line systems at $z=0.13847$ and 
$z\approx0.166$.  Section~8 summarizes the information for several
additional weak \ion{O}{6} absorbers detected.
Section 9 describes the properties of the 
Ly$\alpha$ systems along the sight line and includes an estimate for 
the baryonic mass in the broad Ly$\alpha$ absorbers. 
In \S10 we examine the relationship between the absorption systems and nearby
galaxies. We conclude in \S11 with 
a discussion of the baryonic content of the warm-hot intergalactic medium.
Section 12 contains a summary of the primary results of the study.

\section{Observations and Data Reduction}

The spectral range covered by the FUSE and HST/STIS data obtained for 
this study is shown in Figure~\ref{fullspec}, where we have binned the spectra 
into 0.2\,\AA\ wide bins for clarity.  All measurements and line 
analyses in this paper were conducted on optimally sampled
data.  Unless otherwise state, all velocities in this paper are given 
in the heliocentric reference frame.  Note that in the direction of 
PG\,1116+215 the 
Local Standard of Rest (LSR) and heliocentric reference frames are nearly
identical: $v_{\rm LSR} = v_{\rm helio} +1$ \kms.

\subsection{Far Ultraviolet Spectroscopic Explorer Observations}
We observed PG\,1116+215 with FUSE in April 2000 and April 2001 as part
of the FUSE Team low-redshift \ion{O}{6} project.  For all 
observations PG\,1116+215 was aligned in the center of the LiF1 channel
 LWRS ($30\arcsec\times30\arcsec$) aperture used for guiding.  
The remaining channels were co-aligned throughout the observations.  
The initial observation in April 2000 consisted of 7 exposures totaling 11 
ksec of exposure time.  The second set of observations in April 2001 
consisted of 36 exposures totaling 66 ksec in the LiF channels and 53 
ksec in the SiC channels after screening the time-tagged photon-address
lists for valid data. 
Table~\ref{tab_fuse} contains a summary of the FUSE observations of PG\,1116+215.

We processed the FUSE data with a customized version of the standard 
FUSE pipeline 
software ({\tt CALFUSE} v2.2.2).  This processing followed the 
detailed calibration steps used in previous FUSE investigations by our group 
(e.g., Sembach et al.\ 2001, 2004; Savage et al.\ 2002).  An 
overview of FUSE and the general calibration measures employed can 
be found in Moos et al.\ (2000) and Sahnow et al.\ (2000).  For each 
channel (LiF1, LiF2, SiC1,
SiC2), we made a 
composite spectrum that incorporated all of the available data for the
channel.  The FUSE observations span the 905--1187\,\AA\ spectral
region, with at least two channels covering any given wavelength over
most of this wavelength range.

We registered the composite spectra for the four channels
to a common heliocentric reference frame by aligning 
the velocities of Galactic interstellar features with similar 
absorption lines in the 
HST/STIS band (e.g., \ion{C}{2} $\lambda1036.337$ vs.\  $\lambda11334.532$; \ion{Si}{2} $\lambda1020.699$ vs.\
 $\lambda1304.370$; 
\ion{Fe}{2} $\lambda1144.938$ vs.\ $\lambda1608.451$). 
We also made cross-element comparisons (e.g., \ion{Si}{2} $\lambda1020.699$ vs.\ \ion{S}{2} $\lambda\lambda1250.584, 1253.811$, etc.).  For all such comparisons, we considered only the low velocity
portions of the interstellar 
profiles so as not to bias the velocity comparisons 
made as part of a companion study of the high-velocity Galactic
absorption along
the sight line (Ganguly et al.\ 2004).  The fully processed and 
calibrated FUSE spectra have a nominal zero-point velocity uncertainty 
of $\sim5$ \kms\ ($1\sigma$).   The relative velocity uncertainties
across the bandpass are comparable in size to this zero-point uncertainty but can be larger near the edges of the detectors.

We binned the oversampled FUSE spectra to a spectral bin size
of 4 pixels, or $\sim0.025$\,\AA\ ($\sim7.5$ \kms).  This binning
provides approximately 3 samples per spectral resolution element of 
20--25 \kms\ (FWHM).  The data have continuum signal-to-noise ratios 
$S/N \sim 18$ and 14 per spectral resolution element 
in the LiF1 and LiF2 channels at 1050\,\AA, and $S/N \sim 8$ and 13 at 
950\,\AA\ in the SiC1 and SiC2 channels, respectively.
 
We show the FUSE data in Figure~\ref{fusespec}, where we plot the composite 
spectra in the SiC1 and SiC2 channels below $\sim1000$\,\AA, and the 
composite LiF1 and LiF2 spectra above $\sim1000$\,\AA, as a function of
heliocentric wavelength.  Between 
$\sim1075$ and $\sim1090$\,\AA, SiC data are used to cover the 
wavelength gaps in the LiF coverage caused by the physical gaps between
detector segments.  Line identifications for those features marked above
the spectra are listed to the right
of each panel. 
Redshifts are indicated in parentheses for intergalactic 
absorption features. Rest wavelengths (in~\AA) are provided for Galactic 
lines ($\lambda > 912$\,\AA: Morton 1991, 2003; $\lambda < 912$\,\AA: Verner et al.\ 1994).  
Numerous molecular 
hydrogen (H$_2$) lines in the spectrum 
are indicated by their rotational band and level (L = Lyman band, 
W = Werner band) and vibrational level according to standard transition
selection rule notations.  
The wavelengths of the molecular hydrogen lines 
are from Abgrall et al.\ (1993a, 1993b).
For some Galactic lines, a high-velocity 
feature is also present (Ganguly et al.\ 2004); 
these features are indicated with offset
tick marks connected to the corresponding 
zero-velocity tick marks above the spectra.

\begin{figure}[ht!]
\figurenum{1}
\includegraphics{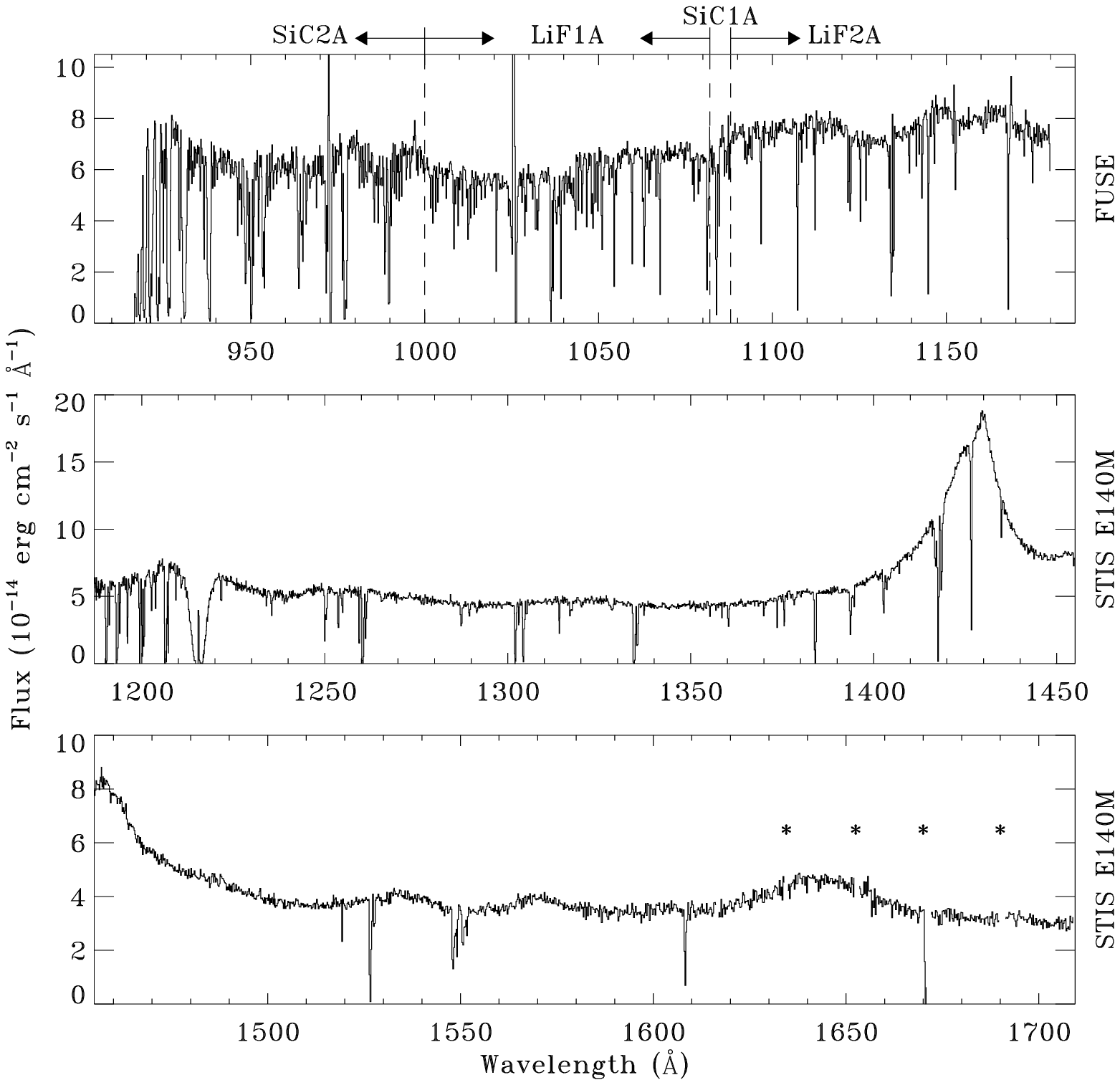}
\vspace{6.0in}
\caption{FUSE and STIS ultraviolet spectra of PG\,1116+215 obtained for this study.
The data have been binned into 0.2\,\AA\ samples for this illustration.  
The FUSE 
data segments plotted in the top panel are indicated above the spectrum.
The flux levels implied by the FUSE data are about 30\% higher than those in the 
STIS spectrum (see text); the flux discrepancy 
does not affect the line analyses in this study.  The flux level sag in
the FUSE LiF2A data between 1110 and 1150\,\AA\ is caused in part by 
an uncalibrated shadow cast by grid wires above the FUSE detector. Asterisks
mark the locations of small gaps in wavelength coverage between the STIS echelle
orders at longer wavelengths.\label{fullspec}}
\end{figure}

The fully calibrated FUSE data have a slightly higher flux than the 
STIS data in regions where the spectra overlap [$1160 \lesssim \lambda 
{\rm (\AA)} \lesssim 1190$].  The difference in the continuum levels
is approximately 30\% (see Figure~\ref{fullspec}).  The flux level differences, which
may be due to intrinsic
variability in the QSO continuum levels in the 11 months separating the 
FUSE and STIS observations, do not affect any of the analyses or results
in this paper.

\subsection{Space Telescope Imaging Spectrograph Observations}
We observed PG\,1116+215 with HST/STIS in May 2000 and June 2000 as 
part of Guest Observer programs GO-8097 and GO-8165.  We used the 
E140M grating with the $0.2\arcsec\times0.06\arcsec$ slit for our primary
observations.  The May 2000 E140M observations consist of 14 exposures 
totaling 19.9 ksec, and the June 2000 observations consist of 
14 exposures totaling 20 ksec.  
We also obtained a set of E230M 
exposures (3 exposures, 5.6 ksec total) as part of program GO-8097.
Table~\ref{tab_stis} contains a summary of the STIS E140M and E230M
observations of PG\,1116+215.

We followed the standard data reduction and calibration procedures 
used in our previous STIS investigations (see Savage et al.\ 2002;
Tripp et al.\ 2001; Sembach et al.\ 2004).  We combined the individual
exposures with an inverse variance weighting and merged the echelle
orders into
a composite spectrum after calibrating and extracting each order. We 
used the two-dimensional scattered light subtraction algorithm 
developed by the STIS Instrument Definition Team (Landsman \& Bowers 1997; 
Bowers et al.\ 1998).
The STIS data have  
a zero-point heliocentric velocity uncertainty of $\approx 1$ \kms\
and a spectral resolution of 6.5 \kms\ (FWHM). \footnotemark\  The 
spectra have $S/N\approx15$ per resolution element at 1300\,\AA.  
Details on the design and performance of
STIS can be found in articles by Woodgate et al.\ (1998) and 
Kimble et al.\ (1998). For additional
information about observations made with STIS, we refer the 
reader to the STIS Instrument Handbook (Proffitt et al.\ 2002).
\footnotetext{The STIS velocity errors may occasionally exceed this 
estimate (see Tripp et al.\ 2004), but we find no evidence of line 
shifts greater than the nominal value of $\approx1$ \kms.}

We plot the STIS E140M data in Figure~\ref{e140mspec} as a function of heliocentric 
wavelength between 1167\,\AA\  and 1649\,\AA\ 
in a manner analogous to the presentation for the FUSE data in 
Figure~\ref{fusespec}. The data shown in Figure~\ref{e140mspec}
span E140M echelle orders 90--126.  The full STIS spectrum extends from
1144\,\AA\ to 1709\,\AA.  At $\lambda < 1170$\,\AA, $S/N \lesssim 4$ per
resolution element.   

Figure~\ref{e230mspec} contains selected regions of our STIS E230M observation of 
PG\,1116+215.  
The spectrum
spans echelle orders 73--101 and covers the 2004--2818\,\AA\ wavelength range.
The spectrum is very noisy at $\lambda < 2200$\,\AA\ and has $S/N\approx10-12$
per 10 \kms\ resolution element between 2400\,\AA\ and 2800\,\AA.
All of the lines detected are Galactic in origin, but the 
data allow us to set upper limits on the presence of some metal-line species 
in the $z=0.13847$ absorber along the sight line.  

For comparison with previous 
low-resolution HST/GHRS observations of PG\,1116+215 (Tripp et al.\
1998), we show the full-resolution STIS spectrum in Figure~\ref{compfig} together
with the STIS E140M spectrum convolved to the GHRS G140L resolution of 
$\sim160$ \kms\ (FWHM).  Nearly all of the features detected in the 
earlier high-$S/N$ GHRS spectrum appear in the convolved 
STIS E140M spectrum as well.  Most of these features are composed of multiple 
absorption lines, as revealed in the full spectral resolution plots shown 
in Figure~\ref{e140mspec} and the bottom panel of Figure~\ref{compfig}.

\section{The PG\,1116+215 Sight Line}

PG\,1116+215 lies in the direction $l = 223\fdg36, b = +68\fdg21$ at a 
redshift $z_{em} = 0.1763\pm0.0008$, as measured from the Ly$\alpha$ 
emission in our STIS spectrum.  This redshift is similar to the 
value of $z_{em} = 0.1765\pm0.0004$ found by Marziani et al.\ (1996).
The sight line extends through the Galactic disk and halo,
several high-velocity clouds located in or near the Galaxy, 
and the intergalactic medium.  Absorption lines 
arising in gas in all of these regions are present in the spectra shown in
Figures~\ref{fusespec}--\ref{e230mspec}.  
Table~\ref{tab_ism} contains the wavelengths and oscillator strengths
($f$-values) for the interstellar features with observed wavelengths 
greater than 917\,\AA\ identified in Figures~\ref{fusespec} and 
\ref{e140mspec}.
Most of these lines are cleanly resolved from nearby lines in the STIS 
band and the FUSE band above 1000\,\AA.  Below 1000\,\AA, line crowding
becomes more problematic, especially at wavelengths less than 940\,\AA,
where numerous \ion{H}{1} Lyman-series and \ion{O}{1} lines are present.
The interstellar lines generally consist of two main 
absorption components centered near $v_{\rm LSR} = -44$ 
\kms\ and 0 \kms, with additional
weaker components occurring between --100 and +100 \kms.  The total \ion{H}{1}
column density along the sight line measured through \ion{H}{1} 21\,cm 
emission is N(\ion{H}{1}) = $1.2\times10^{20}$ cm$^{-2}$, with about 
60\% of the \ion{H}{1} in the stronger complex near --44 \kms\ (see 
Wakker et al.\ 2003).  Molecular 
hydrogen in the $J=0-3$ rotational levels is present  in the --44 \kms\
absorption complex; see Richter et al.\ (2003) for a detailed study of the
molecular lines in the PG\,1116+215 spectra.  The high-velocity gas, which 
is detected in a wide range of ionization stages,
is centered near +184 \kms, with some lines (e.g., 
\ion{C}{3} $\lambda977.020$, \ion{C}{4} $\lambda\lambda1548.195, 1550.770$,
\ion{O}{6} $\lambda\lambda1031.926, 1037.617$) showing extensions down 
to the lower velocities of the Galactic ISM absorption features.
The high-velocity gas and the Galactic
halo absorption along the sight line are the subjects of a separate
study (Ganguly et al.\ 2004).  
In this work, we are concerned primarily with the 
intergalactic absorption along the sight line, although some of the 
IGM features are blended with interstellar features at similar 
wavelengths.  These blends are noted in Table~\ref{tab_ism}.

For the purposes of this study, it is necessary to know the unblocked redshift
path for intervening \ion{H}{1} Ly$\alpha$ and \ion{O}{6} absorption 
(see \S\S9 and 11).
The maximum redshift path available for either species is set by the redshift
of the quasar ($z_{\rm max} = 0.1763$).  To correct this total path for the 
wavelength regions blocked by Galactic lines and other intergalactic 
absorption 
features, we calculated the redshift
intervals capable of obscuring a 30\,m\AA\ absorption feature due to either
\ion{H}{1} or \ion{O}{6}.  For Ly$\alpha$, we find a blocking interval
of $\Delta z_{\rm B} = 0.026$.  For \ion{O}{6}, we find a single-line
blocking interval $\Delta z_{\rm B} = 0.066$.  The \ion{O}{6} blocking interval
is higher than that
for Ly$\alpha$ because of the numerous Galactic H$_2$ lines present in the 
FUSE band below 1100\,\AA.  The \ion{O}{6} value is appropriate for either 
the 
$\lambda1031.926$ line or the $\lambda1037.617$ line.  Requiring both lines 
of the doublet
to be unblocked at the same redshift decreases this estimate slightly.
The blocking-corrected
 distance interval in which the absorption can be located, $\Delta X$, is 
given by the following expression

\begin{equation}
\Delta X = 0.5\{[(1+z_{\rm max}-\Delta z_{\rm B})^2-1]-[(1+z_{\rm min})^2-1]\},
\end{equation}
where we have chosen a cosmology with $q_0 = 0$ (Savage et al.\ 2002).  
In the above equation
$z_{\rm min} = 0$ since we have included the blocking for the Galactic 
Ly$\alpha$ or \ion{O}{6} lines in the estimates of $\Delta z_{\rm B}$.
We calculate $\Delta X_{{\rm Ly}\alpha} = 0.162$ and 
$\Delta X_{\rm O\,{\sc VI}} = 0.117$.

An additional factor worth considering in the calculation of $\Delta X$ is the
possibility that some of the lines observed may be associated with the 
host environment of PG\,1116+215.  The radiation 
field of the QSO may impact the ionization of the gas in the vicinity of the 
QSO and alter the ionization of the gas and therefore affect 
the detectability of Ly$\alpha$ and \ion{O}{6} (i.e., the proximity effect).  
For this reason, we also calculate the unblocked redshift path after
excluding the 5000 \kms\ velocity interval blueward of the QSO redshift.
We identify these revised values of  $\Delta X$ with a primed notation,
and find $\Delta X_{{\rm Ly}\alpha}^\prime = 0.143$ and 
$\Delta X_{\rm O\,{\sc VI}}^\prime = 0.098$

Tripp et al.\ (1998) identified numerous galaxies within $\sim1\degr$ of 
PG\,1116+215 and obtained accurate spectroscopic 
redshifts for 118 of these at redshifts $z \le z_{\rm QSO}$.  Their redshift
survey has an estimated completeness of  $\approx87$\% for $B_J\le19.0$ out to
a radius of 20\arcmin\ from the QSO.  The completeness drops to 
$\approx78$\% for  $B_J\le19.0$ out to a radius of 30\arcmin\ from the QSO.
The intergalactic absorption features considered in this study may be 
associated with some of these galaxies, a topic to which we will return
in \S10.

\section{Line Identification and Analysis}

We identified absorption lines in the PG\,1116+215 spectra 
interactively using the measured $S/N$ of the FUSE and STIS spectra 
as a guide to judging the significance of the observed features.  
All identified IGM features are labeled above the spectra in 
Figures~\ref{fusespec}--\ref{e230mspec}, but only the most prominent Galactic ISM 
features are labeled to avoid confusion.  
Some weak Galactic absorption features and 
many blends of H$_2$ lines are also present, especially at wavelengths 
$\lambda < 1000$\,\AA.  We 
refer the reader to similar plots for other sight lines (3C\,273: Sembach et al.\
2001; PG\,1259+593: Richter et al.\ 2004) and to the line 
identifications in the synthetic interstellar spectra presented by Sembach (1999)
for further examples of the absorption expected at these wavelengths.

For the FUSE data, we considered data from multiple channels to gauge the 
impact of fixed-pattern noise on the observed line strengths.  
Obvious fixed-pattern noise features are present in the 
FUSE LiF2 spectrum  at 1097.7\,\AA\ and 1152.0\,\AA\ (See Figure~\ref{fusespec}).  
Weaker detector
features are present throughout the FUSE spectra but do not 
significantly impact the intergalactic line strengths.  The uncertainties
in line strengths caused by these features are included in our
error estimates.  The $3\sigma$ 
equivalent width detection limit is $\approx 15-30$ m\AA\ over a
 20 \kms\ velocity interval, depending upon wavelength.
The $3\sigma$ 
equivalent width detection limit over a 
20 \kms\ resolution element in the 
FUSE data is  
$\approx 30$ m\AA\ at 1050\,\AA\ (LiF1A), and 
$\approx 40$ m\AA\ at 1150\,\AA\ (LiF2A).  Unless stated otherwise,
all detection limits reported in this paper are $3\sigma$ confidence 
estimates.

After identifying all candidate $z>0$ 
Ly$\alpha$ absorption lines at $\lambda > 
1216$\,\AA, we searched the spectra for additional \ion{H}{1} Lyman-series
lines and metal lines.  For all lines identified as \ion{H}{1} Ly$\alpha$, 
we measured the 
strengths of the corresponding \ion{O}{6} $\lambda\lambda1031.926, 1037.617$ 
lines and report either the measured equivalent widths or $3\sigma$ 
upper limits in the absorber descriptions presented in \S5.
We measured the equivalent widths and uncertainties 
for all detected lines using the error calculation procedures 
outlined by Sembach \& Savage (1992).  The error estimates include 
uncertainties caused by Poisson noise fluctuations, fixed-pattern noise 
structures, and continuum placement.  We set continuum levels for all
lines using low-order ($n < 5$) Legendre polynomial interpolations to line-free
regions within 1000 \kms\ of each line (see Sembach \& Savage 1992).  This 
process more accurately represents the local continuum in the vicinity of the 
individual lines than fitting a single global continuum to the entire 
FUSE and STIS spectra.  Continuum placement is particularly important for 
weak lines, and for this reason, we experimented with several continuum 
placement choices for these lines to make sure that the continuum placement
error was robust.  For lines falling in the FUSE bandpass, we measured 
the line strengths in at least two channels and report these results 
separately since these are independent measurements.  All equivalent
widths in this paper are the observed values ($W_{obs}$) measured at the 
observed
wavelengths of the lines .  To convert these observed equivalent widths 
into rest-frame equivalent widths ($W_r$), divide the observed values by 
($1+z$).

In total, we find 75 absorption lines due to the intergalactic medium 
along the sight line at observed wavelengths $\lambda > 1000$\,\AA.  Of 
these, 38 are Lyman-series lines of \ion{H}{1} and the rest are metal lines.
The most prominent intergalactic absorption system is the one 
at $z=0.13847$, which has a total of 26 lines detected at $\gtrsim3\sigma$ 
confidence.  

We searched for intergalactic lines at $\lambda < 1000$\,\AA\ as well, but 
this search was confounded by the many Galactic H$_2$ lines present at these 
shorter wavelengths.  An example of the complications caused by the Galactic
absorption is shown in Figure\,\ref{o3fig}, where we plot an expanded view
of the FUSE SiC2 spectrum between 946.7 and 950.3\,\AA. This figure
shows that the redshifted \ion{O}{2} $\lambda833.329$ and $\lambda834.466$ 
lines in the $z=0.13847$ metal-line system are 
completely overwhelmed
by Milky Way \ion{O}{1} and \ion{H}{1} absorption features, respectively.
Similarly, redshifted \ion{O}{3} $\lambda832.927$ is confused by the 
presence of interstellar H$_2$ absorption in the $J=2$ and $J=3$ rotational 
levels.
The heavy curve overplotted on the FUSE spectrum indicates the expected 
combined strength of these two lines based on comparisons with H$_2$ 
lines of comparable strength observed at other wavelengths (see comments in 
Table~3).  There may be some residual \ion{O}{3} $\lambda832.927$  absorption 
in the spectrum, but this is difficult to quantify given the strength of the 
H$_2$ lines and the slight differences in spectral resolution between these 
lines and those used as comparisons in the LiF channels.  There may also be a 
small amount of \ion{O}{2} $\lambda832.757$ absorption present.  

Similar searches for other redshifted extreme-ultraviolet lines in the 
$z\approx0.166$ and $z=0.17360$ metal-line systems did not reveal any
substantive detections, and many of these lines are blended with Galactic
features as well.  For example, \ion{O}{3} $\lambda832.927$ at $z=0.16548$ 
is not detected at 970.76\,\AA\ (see \S7).  \ion{O}{3} $\lambda832.927$ 
at $z=0.17360$ 
is severely blended with Galactic \ion{C}{3} $\lambda977.020$ absorption.
\ion{O}{4} $\lambda787.711$ at $z=0.17360$ falls close to the 
$-44$ \kms \ Galactic H$_2$ (17-0) R(1) $\lambda924.461$ line at 924.3\,\AA.
The high quality of the PG\,1116+215 datsets 
makes such searches possible, but it 
also demonstrates that it can be difficult to identify redshifted 
extreme-ultraviolet lines below observed wavelengths of $\sim1000$\,\AA\
(i.e., at $z \lesssim 0.2$).

\section{Intergalactic Absorption Overview}

Previous HST/GHRS observations of PG\,1116+215 identified at least 
13 intergalactic Ly$\alpha$ absorbers along the sight line 
(Tripp et al.\ 1998; Penton, Stocke, \& Shull 2004).  Most of these 
previously identified absorbers are confirmed in our higher quality 
STIS spectra with a few
exceptions.  Ly$\alpha$ absorbers  previously identified at 
1266.47\,\AA\ and 1269.61\,\AA\ by Penton et al.\ (2004) 
are not present in our STIS data; no absorption features occur at
these wavelengths.\footnotemark
\footnotetext{Note that the wavelengths of the absorption lines 
quoted by Penton, Stocke, \& Shull\ (2004) are systematically too red by $\sim0.1$\,\AA\
in most cases since those authors set the strong Galactic lines in their
GHRS spectra to zero heliocentric velocity.  The primary Galactic absorption 
features
in our STIS data and the extant \ion{H}{1} 21\,cm emission data 
in the literature indicate that the Galactic features have a velocity of 
$v_{\rm helio} = -45$ \kms, or  $v_{\rm LSR} = -44$ \kms\ (see \S3).}

We plot in Figure~\ref{lya_stack} 
the continuum-normalized \ion{H}{1} Ly$\alpha$ absorption for 
the stronger Ly$\alpha$ features detected in the STIS data.  
All of the profiles are plotted at the 
systemic velocity of the indicated redshift, except for several
satellite absorbers that appear near the primary absorbers.  The
redshifts of these satellite absorbers are also indicated under the spectra.
Additional features due to either the Galactic ISM or other intervening 
systems are indicated below
the spectra.    Several additional weak features identified in 
Figures~\ref{e140mspec} and \ref{lya_stack} are also excellent Ly$\alpha$ 
candidates.
 The observed equivalent widths and measured Doppler 
line widths (b-values) of all identified Ly$\alpha$ absorbers are given in 
Table~\ref{tab_summarylya}.  Many of the Ly$\alpha$ absorbers at $z < 0.063$ were  identified previously 
in GHRS intermediate resolution data.  The remaining Ly$\alpha$ lines, 
with the exception of the broad absorber at $z=0.13370$ and the weak 
absorbers at $z=0.07188$ and $z=0.10003$, were identified 
in the 
high-$S/N$ low-resolution GHRS spectrum obtained by Tripp et al.\ (1998).  
In some cases, previously identified Ly$\alpha$ lines are now 
resolved into multiple
components in the much higher resolution HST/STIS data
(e.g., the $z\approx0.05895$ system).

We calculated the  \ion{H}{1} column densities in Table~\ref{tab_summarylya} for
most of the absorbers by fitting an instrumentally convolved single-component
Voigt profile with a width of 6.5 \kms\ (FWHM)
to the observed Ly$\alpha$ absorption line.  In some 
cases (e.g., the absorbers at $z=0.13847$, $z=0.16610$), it was possible to 
construct a single-component Doppler-broadened curve of growth using the 
additional Lyman-series lines available.  These procedures followed those
outlined by Sembach et al.\ (2001) in their analysis of the \ion{H}{1}
absorption  along the 3C\,273 sight line.  

We calculated  \ion{O}{6} column densities or column density
limits for each
system using several methods.  For systems where no \ion{O}{6}
was detected, we adopted an upper limit based on a linear curve of growth
fit to the rest-frame equivalent widths of the \ion{O}{6} lines.  
For systems where \ion{O}{6} was detected, the linear curve of growth
estimate for both lines was compared to the column density estimate
based on the apparent optical depths of the \ion{O}{6} lines (see 
Savage \& Sembach 1991 for a description of the apparent optical
depth technique).  These estimates were found to be in good agreement.
The notes for Table~\ref{tab_summarylya} contain comments about the \ion{H}{1} and \ion{O}{6}
column density estimates for each sight line. 

The following subsections provide an overview of the intergalactic 
absorption features detected in the FUSE and STIS spectra.

\subsection{Weak Ly$\alpha$ Absorbers}

There are several features with $W_\lambda \lesssim 40$\,m\AA\
in the STIS E140M spectrum of PG\,1116+215 
that are likely to be weak Ly$\alpha$ absorbers. These features
occur at wavelengths other than those expected for Galactic interstellar 
absorption or metal-line absorption related to the stronger intergalactic 
systems discussed in the following sections.  
The weak Ly$\alpha$ absorbers occur at 1239.05\,\AA\ ($z=0.01923$), 
1276.31\,\AA\ ($z=0.04988$), 1303.052\,\AA\ ($z=0.07188$), and 1337.27\,\AA\ 
($z=0.10003$).  None of these have corresponding \ion{O}{6} detections.
They are listed in Table~\ref{tab_summarylya}, along with relevant notes. 

\subsection{The Intervening Absorber at $z=0.00493$}
The \ion{H}{1} Ly$\alpha$ absorption for this system falls just beyond the red 
wing of the Galactic damped Ly$\alpha$ absorption. Ly$\beta$ is not detected 
by FUSE, 
as expected, since the Ly$\alpha$ absorption has a strength 
$W_\lambda=95\pm11$\,m\AA.
\ion{O}{6} $\lambda1031.926$ at 1037.01\,\AA\ is blended with Galactic high 
velocity 
\ion{C}{2} and Galactic \ion{C}{2}$^*$.
\ion{O}{6} $\lambda1037.617$ at 1042.73\,\AA\ is blended with high-order 
Ly-series 
absorption at $z=0.13847$. 

\subsection{The Intervening Absorber at $z=0.01635$}
Moderate strength, broad \ion{H}{1} Ly$\alpha$ is clearly detected at 
1235.55\,\AA\ ($W_\lambda = 113\pm10$\,m\AA; b~$=48\pm5$ \kms).  
The only possible contaminating 
Galactic absorption nearby is weak \ion{Kr}{1} $\lambda1235.838$;  for a 
Galactic \ion{H}{1} column density of $1.2\times10^{20}$ cm$^{-2}$ and a solar 
Kr/H gas-phase abundance ratio log~(Kr/H)$_\odot$ = --8.77 (Anders \& Grevesse 
1989), the \ion{Kr}{1} line should have $W_\lambda < 0.6$ m\,\AA.
Ly$\beta$ is blended with high-order Ly-series absorption at $z=0.13847$.
\ion{O}{6} $\lambda1031.926$ at 1048.80\,\AA\ is partially blended with \ion{H}{1} Ly$\iota$
at $z=0.13847$; we set an upper limit of $W_\lambda < 30$\,m\AA.  \ion{O}{6} 
$\lambda1037.617$ at 1054.58\,\AA\ is blended with \ion{H}{1} Ly$\eta$ at $z=0.13847$.

\subsection{The Intervening Absorber at $z=0.02827$}
Strong, narrow \ion{H}{1} Ly$\alpha$ is detected near the Galactic \ion{S}{2} $\lambda1250.584$ 
line.  \ion{H}{1} Ly$\beta$ with a strength of $\gtrsim35$\,m\AA\ should occur at 1054.72\,\AA,
which is just redward of the \ion{H}{1} Ly$\eta$ line at $z=0.13847$. A feature of 
this strength 
is detected, indicating that the Ly$\alpha$ absorption is optically thin.
\ion{O}{6} $\lambda1031.926$ at 1061.10\,\AA\ is not detected at a level of 
$W_\lambda < 33$\,m\AA\ ($3\sigma$).

\subsection{The Intervening Absorber at $z=0.03223$}
\ion{H}{1} Ly$\alpha$ in this system is detected at 1254.85\,\AA\ with an observed
equivalent width $W_\lambda = 93\pm9$\,m\AA\ and a line width b$ = 32\pm3$ \kms.  
Weak absorption by
Galactic high-velocity \ion{S}{2} $\lambda1253.811$ is present at v$ < -60$ \kms\
in the rest-frame of the absorber. No Ly$\beta$ absorption is detectable at 1058.78\,\AA,
as expected.    
Neither \ion{O}{6} $\lambda1031.926$ at 1065.18\,\AA\ nor \ion{O}{6} $\lambda1037.617$ at 
1071.06\,\AA\ is detected.  For both lines $W_\lambda < 36$\,m\AA\ ($3\sigma$).

\subsection{The Intervening Absorber at $z=0.04125$}
This weak \ion{H}{1} Ly$\alpha$ line at 1265.82\,\AA\ is very broad, with b $=105\pm18$ \kms.
Its strength of $81\pm17$\,m\AA\ is considerably less than the value of 
$171\pm37$\,m\AA\
estimated by Tripp et al.\ (1988).  Because the line is so shallow, 
continuum placement is a potential source of significant
systematic uncertainty in the strength of this line.  There is no obvious 
counterpart in
Ly$\beta$, though there may be very weak \ion{O}{6} $\lambda1031.926$ near 1074.49\,\AA\
in the FUSE LiF1A segment (see Figure~\ref{weak_o6_04125}).  The line has $W_\lambda 
= 28\pm10$ m\AA\ integrated over the 
--60 to +80 \kms\ velocity range.  The line in the LiF2B segment falls at the 
edge of the detector.  In the remaining 
segments (SiC1A, SiC2B), the lower $S/N$ of the data precludes a confirmation of this 
tentative detection.  There is no obvious \ion{O}{6} $\lambda1037.617$ at 1080.42\,\AA\
corresponding to this \ion{O}{6} absorption, as expected for
the $S/N$ level of these data.
  There are no other species (e.g., \ion{C}{3}
$\lambda 977.020$) detectable at this redshift that would confirm the Ly$\alpha$ or 
\ion{O}{6} $\lambda1031.926$ detections.

\subsection{The Intervening System at $z=0.05895, 0.05928$}
\ion{H}{1} Ly$\alpha$ at $z=0.05895$ 
occurs at 1287.33\,\AA\ with an observed line strength of $172\pm11$\,m\AA.
The profile consists of at least two narrow (b $\approx$ 20, 30 \kms) components.
A weak (15\,m\AA) ``satellite'' Ly$\alpha$ absorber at +93 \kms\ (i.e., $z=0.05928$)
with b $\approx 10$ \kms\ is also present.
Ly$\beta$ would occur in noisy portions of the SiC1A and SiC2B data at 1086.19\,\AA.
It is not possible to confirm the Ly$\alpha$ identification at these $S/N$ levels
and expected line strength. \ion{O}{6} $\lambda1031.926$ at this redshift 
would occur near 1092.76\,\AA\ in the LiF2A spectrum but could be blended with 
Galactic H$_2$ absorption at 1092.73\,\AA\ (see Figure~\ref{weak_o6_05895}).  
There is indeed a feature present at the expected wavelength, but 
without data from another channel to 
confirm the possible absorption ledge next to the H$_2$ line, we can only estimate
$W_\lambda \sim 47\pm13$\,m\AA\ for the \ion{O}{6} $\lambda1031.926$ absorption.

Two lines with the
expected separation of the \ion{O}{6} doublet occur 
approximately +82 \kms\ redward of the systemic velocity of this system at 1093.06
and 1099.07\,\AA\ (i.e., at $z=0.05924$ in the \ion{O}{6} rest frame).  
The shorter wavelength line is covered by the LiF2A and SiC2B
detector segments.  In the LiF2A data, the line has  $W_\lambda =26\pm7$\,m\AA.
The SiC2B data are of insufficient $S/N$ to confirm or refute the 
LiF2A detection, and the line falls in the wavelength coverage gap of detector 1.
The longer wavelength line is detected in both the LiF2A and LiF1B
data and has an equivalent width of $20\pm6$\,m\AA\ (LiF1B) and 
$16\pm6$\,m\AA\ (LiF2A).  Figure~\ref{weak_o6_05895} shows that the 
lines detected in the different FUSE channels align well in velocity with
each other.  These lines are offset slightly from the  velocity of
a weak Ly$\alpha$ feature near +93 \kms. Weak \ion{C}{4} $\lambda1548.195$
absorption may also be present at this velocity.

\subsection{The Intervening Absorber at $z=0.06072$}
\ion{H}{1} Ly$\alpha$ absorption in this system is present at 1289.49\,\AA\ with an
observed strength
$W_\lambda = 85\pm9$\,m\AA. The line has b $= 55\pm6$ \kms.
No Ly$\beta$ absorption is present at 1088.00\,\AA\ in the LiF2A data, with a 
limit of $W_\lambda < 39$\,m\AA. 
No \ion{O}{6} $\lambda1031.926$ absorption is present at 1094.58\,\AA\ in the LiF2A data, 
with a limit of $W_\lambda < 39$\,m\AA. 

\subsection{The Intervening Absorber at $z=0.06244$}
Broad \ion{H}{1} Ly$\alpha$ absorption occurs at 1291.58\,\AA\ with an
observed equivalent width
of $77\pm9$\,m\AA\ and a line width b $= 77\pm9$ \kms.  
Weak Ly$\beta$ below the FUSE detection threshold would occur 
at 1089.77\,\AA.  \ion{O}{6} $\lambda1031.926$ is present near 1096.36\,\AA\
as a 
weak narrow feature with a negative velocity offset of --20 \kms\
 relative to the
centroid of the Ly$\alpha$ absorption. 
Given the breadth of the Ly$\alpha$ line (b = 79 \kms), 
this offset is negligible.
The \ion{O}{6} $\lambda1031.926$ 
line has equivalent widths of $19\pm8$\,m\AA\ in the LiF1B data and  
$18\pm7$\,m\AA\ in the LiF2A data.  The detection of the feature in both 
channels increases the significance of the line identification.
  The corresponding \ion{O}{6} $\lambda1037.617$ line is 
too weak to be detectable at a significant level in either FUSE
channel at the $S/N$ of the 
present data.  \ion{C}{4} $\lambda1548.195$ is not detected in the STIS 
spectrum.
A stack plot of the normalized profiles for these lines is 
shown in Figure~\ref{weak_o6_06244}.  We discuss this absorber further in 
\S8.3.

\subsection{The Intervening Absorber at $z=0.08096$}
Narrow \ion{H}{1} Ly$\alpha$ absorption occurs at 1314.09\,\AA\ with an
observed strength
of $124\pm6$\,m\AA.  An upper limit of $<30$\,m\AA\ is found for the corresponding 
Ly$\beta$ absorption at 1108.76\,\AA\ in both the LiF1B and LiF2A data.  \ion{O}{6}
is not detected in either line at a limit of $<30$\,m\AA\ ($3\sigma$).

\subsection{The Intervening Absorber at $z=0.09279$}
Of all the Ly$\alpha$ absorbers detected along this sight line, this one is the most 
difficult to quantify.  The equivalent width and velocity extent of the absorber are
highly uncertain.  Tripp et al.\ (1998) identified this Ly$\alpha$ absorber and 
quoted an observed strength of $W_\lambda = 70\pm23$
m\AA, but we find that the absorption strength could be as high as 136\,m\AA\
integrating from --150 to +125 \kms\ or as low as 70 m\AA\ if the integration
range is confined to $\pm70$ \kms\ (see Figure~\ref{lya_stack}).  
The weak, narrow feature at +34 \kms\ is Galactic 
\ion{C}{1} $\lambda1328.833$. The continuum 
placement for this system is particularly important because the Ly$\alpha$ line is so weak and 
broad (b $=121\pm15$ \kms).  
There is no detectable Ly$\beta$ or \ion{O}{6} absorption associated with this 
absorber. Over the full velocity range of --150 to +125 \kms, we find $W_\lambda$(Ly$\beta$)
$< 45$ m\AA\ and $W_\lambda$(\ion{O}{6} $\lambda1031.926$) $<60$\,m\AA.

\subsection{The Intervening Absorber at $z=0.11895$}
Narrow \ion{H}{1} Ly$\alpha$ absorption occurs at 1360.27\,\AA\ with an
observed equivalent width
of $138\pm9$\,m\AA.
Ly$\beta$ at 1147.73\,\AA\ is below the FUSE detection threshold, 
with $W_\lambda <48$\,m\AA\ (LiF1B) and 
$<39$\,m\AA\ (LiF2A).  \ion{O}{6} $\lambda1031.926$ at 1154.67\,\AA\
is also below the detection 
threshold, with $W_\lambda <42$\,m\AA\ (LiF1B) and $<39$\,m\AA\ (LiF2A).

\subsection{The Intervening Absorber at $z=0.13151$}
Narrow \ion{H}{1} Ly$\alpha$ absorption occurs at 1375.54\,\AA\ with a strength
of $132\pm8$\,m\AA.  Weak Ly$\beta$ may be present near 1160.61\,\AA\ with
$W_\lambda = 33\pm12$\,m\AA\ 
integrated over a $\pm70$ \kms\ velocity interval.  
This tentative detection needs to be confirmed with better data.  \ion{O}{6}
$\lambda1031.926$ at 1167.63\,\AA\ falls within the \ion{H}{1} Ly$\beta$ 
absorption
at $z=0.13847$.  \ion{O}{6} $\lambda1037.617$ is not detected at 1174.07\,\AA\
with $W_\lambda <33$\,m\AA\ ($3\sigma$).

\subsection{The Intervening Absorber at $z=0.13370$}
Broad \ion{H}{1} Ly$\alpha$ absorption occurs at 1378.21\,\AA\ with an
equivalent width $W_\lambda =97\pm12$\,m\AA\ and a line width b $=84\pm10$ \kms.  
Ly$\beta$ should occur near 1162.86\,\AA.  Integrating over
a velocity range of $\pm100$ \kms\ yields an equivalent width limit of 45\,m\AA\
in the LiF1 and LiF2 channels.  We find $3\sigma$ 
limits of 51\,m\AA\ (LiF1B) and 45\,m\AA\
(LiF2A) for the \ion{O}{6} $\lambda1031.926$ line at 1169.89\,\AA\ over the same
velocity interval.

\subsection{The Intervening Absorber at $z=0.13847$}
This absorber is detected in numerous \ion{H}{1} Ly-series lines as well
as lines of heavier elements (C, N, O, Si) in a variety of ionization 
stages (II-VI).  It is the strongest \ion{H}{1} absorber along the 
sight line other than the Milky Way ISM.  
\ion{H}{1} Ly$\alpha$ occurs at 1384.004\,\AA\ with an observed equivalent 
width of $535\pm12$\,m\AA.   A weak Lyman-limit roll-off is produced by
the convergence of the Lyman-series lines; higher-order lines
up through Ly$\kappa$ are resolved from neighboring lines in the series.
The weak Lyman-limit
break is visible in the top panel of Figure~\ref{fusespec}g as a small reduction in
the continuum flux of the quasar beginning at about 1043\,\AA\ 
and continuing to shorter wavelengths.

\ion{O}{6} $\lambda1031.926$ absorption in this system is detected 
at 1174.817\,\AA\ in the data from both FUSE channels and STIS.  The 
line has an equivalent width ranging from $54\pm13$\,m\AA\ to 
$84\pm21$\,m\AA\ in the various datasets.  In all cases, the detection is 
highly significant and cannot be mistaken for any lines from the 
other systems identified along the sight line.  It is also unlikely to be 
caused by an unidentified absorber.  For example,
the line cannot be \ion{H}{1} Ly$\beta$ at $z=0.14536$ since there is 
no corresponding Ly$\alpha$ absorption at 1392.37\,\AA.  
\ion{C}{3} $\lambda977.020$ at $z=0.20245$ is also ruled out by the 
lack of \ion{H}{1} Ly$\alpha$ at 1461.78\,\AA\ and \ion{O}{6} $\lambda1031.926$
at 1240.84\,\AA.  

The weaker member of the \ion{O}{6} doublet at $z=0.13847$ is at best
only marginally detected in either the FUSE or STIS data.  Assuming the 
\ion{O}{6} absorption occurs on the linear part of the curve of growth, 
the expected line strength is $\approx 25-40$\,m\AA.  The line falls
at 1181.296\,\AA, which is in a region of low $S/N$ in the STIS 
data and right at the edge of the wavelength coverage in the FUSE LiF2A
data.  In the FUSE LiF1B data, a slight depression in the continuum
at this wavelength is consistent with a line having an equivalent width
less than 50\,m\AA.  Higher $S/N$ data are needed to detect this line
at a confidence greater than 2--3$\sigma$.

The wide variety of ionization stages observed in this system indicate
that it is probably multi-phase in nature.  The lower ionization stages
arise in two closely spaced ($\Delta v \approx 7$ \kms) components.  The 
good velocity correspondence of the \ion{O}{6} with the velocities of 
these components strongly suggest that the two types of gas are in close 
proximity to each other.    
We consider the ions observed in this system and their production in \S6.

\subsection{The Intervening Absorption System at $z\approx0.166$}
This absorption system consists of a strong Ly$\alpha$ absorber flanked 
at both negative and positive velocities by ``satellite'' absorbers
within 300 \kms\ of the main absorption.  The system occurs within $\sim3000$ 
\kms\ of the QSO redshift, but we consider it to be an intervening system
rather than a system associated with the QSO host environment (see \S\S7 and
10). The primary absorber 
at $z=0.16610$ is the second strongest non-Galactic absorber along the 
sight line.  It is detected in \ion{H}{1} Ly$\alpha$, Ly$\beta$, and possibly
Ly$\epsilon$.  The Ly$\alpha$ strength is $368\pm8$\,m\AA\
over the --45 to +70 \kms\ velocity range.
An accompanying weak feature with an observed equivalent width of 
$57\pm5$\,m\AA\ at $z=0.16580$ occurs next to the Ly$\alpha$ line. 
We detect  
\ion{C}{2} $\lambda1334.532$, \ion{C}{3} $\lambda977.020$, and \ion{Si}{3}
$\lambda1206.500$ at $z=0.16610$; no \ion{O}{6} is detected ($W_\lambda \lesssim 50$\,m\AA.).  
No metal lines
are detected in the accompanying feature at $z=0.16580$.

The satellite absorber at $z=0.16686$ is seen only in Ly$\alpha$.  It too
has a weak absorption wing, but at positive velocities (+50 to +130 \kms\
in the rest frame relative to $z=0.16686$).  The absorber has an
 observed equivalent width of $209\pm10$\,m\AA\ if the wing is 
excluded, and $239\pm10$\,m\AA\ if the wing is included.  The absorber is 
not detected in any other species.

The satellite absorber at $z=0.16548$ has an observed  Ly$\alpha$ equivalent 
width of $128\pm7$\,m\AA.  It is also detected in Ly$\beta$, \ion{O}{6},
and possibly \ion{N}{5}.  Both \ion{O}{6} lines are cleanly detected
in the STIS data, with equivalents widths of $124\pm12$\,m\AA\ and 
$72\pm10$\,m\AA.

\subsection{The  System at $z=0.17340$ and $z=0.17360$}
\begin{figure}[ht!]
\figurenum{11}
\includegraphics{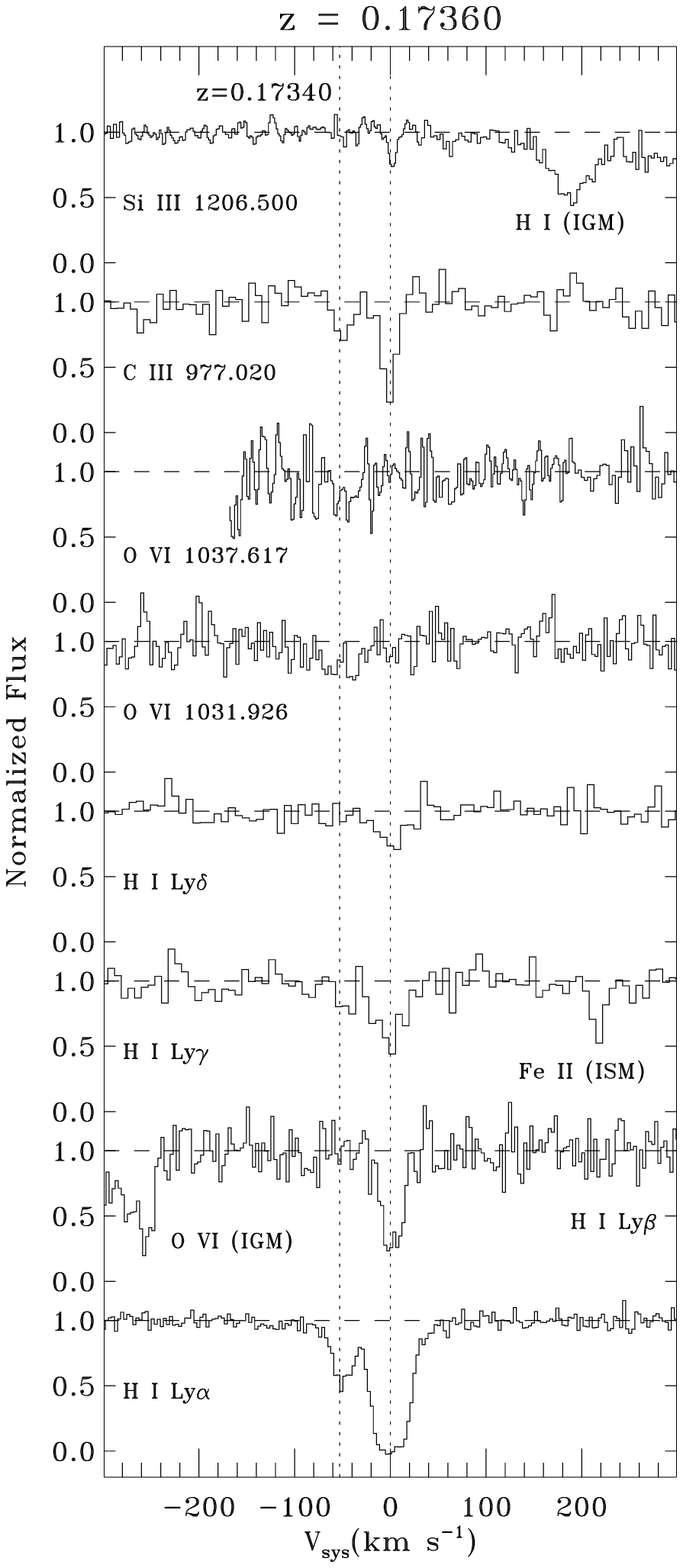}
\vspace{8.0in}
\caption{Continuum-normalized absorption profiles for lines in the 
absorption-line system at $z=0.17340-0.17360$. \label{stack_17360}}
\end{figure}

This absorption system occurs within 900 \kms\ of the Ly$\alpha$ emission
from the quasar at $z=0.177$ and therefore may be associated with the 
host galaxy of the quasar or the quasar environment.  It consists of two 
narrow (b $=13-17$ \kms) components closely spaced
components ($\Delta v = 49$ \kms) observed in \ion{H}{1} Ly$\alpha$
and \ion{C}{3} $\lambda977.020$.  The main component at $z=0.17360$ 
is also detected in \ion{H}{1} Ly$\beta$--Ly$\delta$ and \ion{Si}{3}
$\lambda1206.500$ (see Figure~\ref{stack_17360}).  
The weaker absorber at $z=0.17340$ may exhibit some 
\ion{O}{6} $\lambda1031.926$ absorption, but this detection is tentative
since the spectrum has a relatively low $S/N$ ratio at these wavelengths and
the continuum placement is somewhat uncertain.  The \ion{O}{6} lines
for both components fall in or near the broad damping wings of the 
Galactic \ion{H}{1} Ly$\alpha$ line at 1216\,\AA.  The $\lambda1031.926$ 
line is the better detected member of the doublet.  It occurs at 1211.07\,\AA,
which is a wavelength region with no known Galactic features.
Because this system occurs so close to the redshift of PG\,1116+215, we
consider the possibility that it is an associated system in our 
discussions of the Ly$\alpha$ and \ion{O}{6} absorption systems along the 
sight line.

\subsection{Unidentified Features}

Two additional weak features worth noting are also present in the 
data.  A weak feature ($W_\lambda =19\pm6$\,m\AA) at 1312.22\,\AA\
has a width narrower than the instrumental
resolution.  No ISM or IGM features are expected at this wavelength.  The 
closest match is \ion{P}{2} $\lambda1152.818$ in the $z=0.13847$ IGM absorber,
but the velocity of the feature is off by $-52$ \kms\ from the expected
position of the \ion{P}{2} line. This feature is 
most likely caused by noise in the data.  The second feature 
occurs near 1582.5\,\AA\ with $W_\lambda =28\pm11$\,m\AA.  It is not
Ly$\alpha$ since the implied redshift of $z\approx0.302$ is much
greater than the redshift of the quasar.  The feature occurs near
the edge of STIS echelle order 93, but is not caused by combining data
in the order overlap region.  The wavelength of the feature does not correspond
to that of any metal lines in the $z=0.13847$, $z\approx0.166$, or $z=0.17360$
absorption-line systems.

\section{Properties of the \ion{O}{6} System at $z = 0.13847$}

The metal-line system at $z=0.13847$ has the richest set of absorption
lines of any absorber along the PG\,1116+215 sight line.  As noted 
previously, numerous \ion{H}{1} and metal lines can be seen in the 
FUSE and STIS spectra.  

\subsection{Column Densities}

\subsubsection{Neutral Hydrogen}

The large number of \ion{H}{1} lines at this redshift permits the derivation
of an accurate \ion{H}{1} column density for this system.  Figure~\ref{stack_13847} 
contains a set of normalized \ion{H}{1} profiles over the --300 to +300
\kms\ velocity range centered on $z=0.13847$.  The observed equivalent
widths of the lines are presented in Table~\ref{tab_ew13847}.  We calculated an \ion{H}{1} 
column density for this system from two different methods, which yield
consistent results.

\begin{figure}[ht!]
\figurenum{13}
\includegraphics{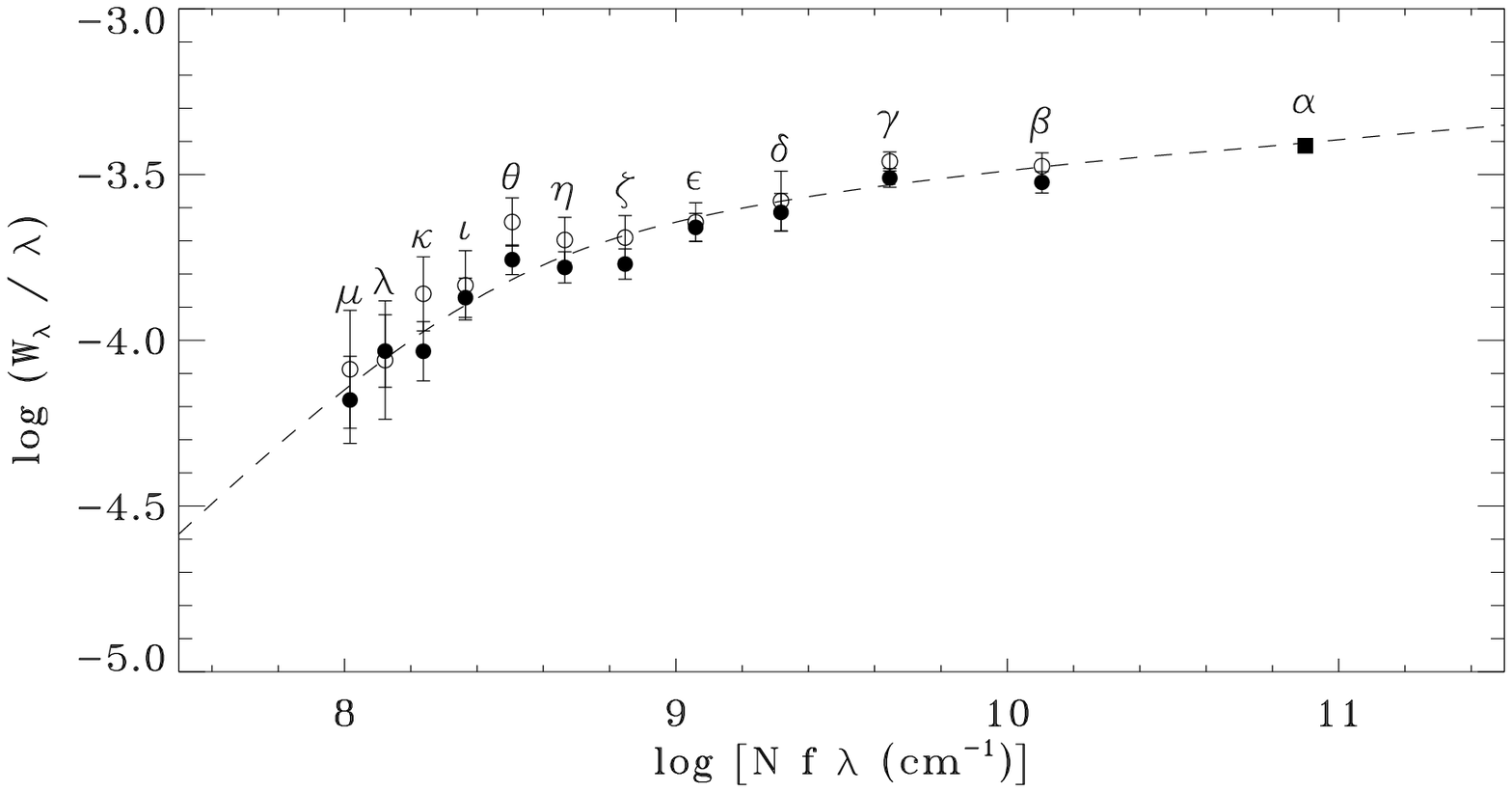}
\vspace{4.0in}
\caption{Single-component curve of growth for the \ion{H}{1} Lyman-series
lines in the $z=0.13847$ metal-line system.  
 The STIS  Ly$\alpha$ measurement is shown as a filled 
square data point.  The FUSE measurements for the other lines are represented
by either filled circles (LiF1, SiC1) or open circles (LiF2, SiC2).
Error bars are $1\sigma$ estimates.
The Ly$\theta$ line 
was not used in the fit because it is partially blended with a Galactic 
H$_2$ line (see Table~\ref{tab_ew13847}). 
The best fit COG has N(\ion{H}{1}) = $(1.57\pm^{0.18}_{0.14})\times10^{16}$
cm$^{-2}$ (log N = $16.20\pm^{0.05}_{0.04}$) and b = $22.4\pm0.3$ \kms.\label{cog13847}}
\end{figure}

First, we fit a single-component Doppler-broadened curve of growth  
to the rest-frame equivalent widths.  This curve of growth is shown in
Figure~\ref{cog13847}.  The STIS Ly$\alpha$ measurement and the two FUSE measurements 
available for the Ly$\beta$--Ly$\kappa$ lines were fit simultaneously, with 
the exception of the Ly$\theta$ line, which is partially blended with a
Galactic H$_2$ absorption line.  The fit yields 
N(\ion{H}{1}) = $(1.57\pm^{0.18}_{0.14})\times10^{16}$
cm$^{-2}$ and b = $22.4\pm0.3$ \kms.  We expect this result to be an
excellent approximation to the actual column density even though the 
absorber consists of at least two components. The two most prominent 
components are separated by about 7 \kms\ as evidenced by the 
metal-line absorption. Most of the column
density is contained in the stronger of the two components (see below).  
Nonetheless, the \ion{H}{1} b-value should be considered an ``effective''
b-value that includes a contribution from the presence of the other
component(s).  The true b-values of the individual components will be 
smaller than this estimate.

Second, we considered the absorption caused by the convergence of the 
\ion{H}{1} Lyman series at this redshift.  The continuum of the quasar
at wavelengths shortward of the expected Lyman limit is depressed by a small 
amount relative to longer wavelengths.  This depression is visible 
in the top panel of Figure~\ref{fusespec}g.  We estimate a depression depth of
$d_c = 10.0\pm1.5$\%.  Converting this into an optical depth, 
$\tau = \ln(1/d_c)$, yields an \ion{H}{1} column density

\begin{equation}
{\rm N} = \tau / \sigma_{_{912}} = (1.67\pm{0.27})\times10^{16}~{\rm cm^{-2}},
\end{equation}

\noindent 
where we have set the absorption cross section at 912\,\AA\ equal to
$\sigma_{_{912}} = 6.304\times10^{-18}$ cm$^2$ (Spitzer 1978).
This value of N(\ion{H}{1}) is well within the $1\sigma$
error estimate of the column density derived from the curve
of growth.

Finally, we also calculated a lower limit on the column density using
the apparent optical depth of the Ly$\kappa$ ($\lambda919.351$) line.  
An apparent column density profile is defined as 

\begin{equation}
{\rm N_a} = \int {\rm N_a}(v) dv
= \frac{3.768\times10^{14}}{f\lambda} \int \tau_a(v) dv~{\rm (cm^{-2})}
\end{equation}

\noindent
where $\tau_a(v)$ is the apparent optical depth of the line (equal to the 
natural logarithm of the estimated continuum divided by the observed intensity)
at velocity $v$ (in \kms), $f$ is the oscillator strength of the 
line, and $\lambda$ is the wavelength of the line (in \AA).  
Direct integration of {\rm N$_a$} over velocity yields an estimate of the 
column density of the line.  In the event that unresolved saturated 
structure exists, the value of {\rm N$_a$} should be considered a lower limit
to N (see Savage \& Sembach 1991). For the \ion{H}{1} Ly$\kappa$ line 
we find N(\ion{H}{1}) $ \ge$ {\rm N$_a$(Ly$\kappa$)} = $1.4\times10^{16}$ cm$^{-2}$, which is 
consistent with the COG value derived above.

The \ion{H}{1} column density in this absorber is 18--20 times higher than 
the combined \ion{H}{1} column in all the other systems along the sight 
line other than the Milky Way ISM.
The \ion{H}{1} column density is similar to that of 
the $z=0.00530$ absorber in the Virgo Cluster toward 3C\,273
(Tripp et al. 2002;  Sembach et al. 2001).

\subsubsection{Metal-line Species}

Continuum normalized versions of the metal lines at $z=0.13847$ 
detected in our STIS and FUSE spectra  are shown in 
Figure~\ref{stack_ion_13847}.  
For some species, absorption in only a single transition is detected or
observable (e.g., \ion{C}{3},
\ion{N}{2}, \ion{N}{3}, \ion{N}{5},
\ion{O}{1}, \ion{O}{6}, \ion{Si}{3}, \ion{Si}{4}).  
For others, absorptions from multiple transitions are observed 
(e.g., \ion{C}{2}, \ion{Si}{2}).
In still other cases, it was possible to set upper limits on line strengths
based on non-detections (e.g., \ion{S}{2}, \ion{S}{3}, \ion{Fe}{2}). 
Unfortunately, blending with Galactic lines obscures some interesting 
extreme-ultraviolet transitions that would otherwise be observable
at this redshift (e.g., \ion{O}{3}); we are able to estimate the 
\ion{O}{2} column density from the $\lambda832.757$ line 
(see Figure~\ref{o3fig}).
We calculated column densities for the metal-line species using 
curve-of-growth, apparent optical depth, and profile fitting techniques.  
A summary of the 
column densities is provided in Table~\ref{tab_col13847}, together with the 
analysis methods used
to calculate the column densities.  

The \ion{Si}{2} lines shown in Figure~\ref{stack_ion_13847} indicate that there are two
components near $z=0.13847$ separated by $\approx10$ \kms.  The dominant
component contains roughly 90\% of the total column density of the 
system, with an effective b-value of $\approx5$ \kms.  
The width of the 
weaker component is more difficult to ascertain, but it is probably comparable
to that of the dominant component. 
The dominant component
may consist of unresolved sub-components, but we are unable to place strong 
constraints on the properties of the sub-components with the existing data.
We estimated the \ion{Si}{2} column density using a single-component curve of growth,
profile fitting, and the apparent optical depth method.  For the first 
two methods, we used the information available for the $\lambda\lambda1260.422, 1193.290,
1190.416, 1304.370$ lines.  For the apparent optical depth approach, we considered only
the $\lambda1193.290, 1190.416$ lines since the $\lambda1260.422$ line is strong enough
to contain unresolved saturated structures.  
A comparison of the apparent optical
depth results to those for the curve of growth or profile fit shows that 
N(\ion{Si}{2}) $\approx 10^{13}$ cm$^{-2}$ with a large error (see Table~\ref{tab_col13847}).  
The main uncertainty in these estimates results from the unknown velocity 
structure of the main absorption component in the stronger lines.  For this 
reason, we quote column density limits from the apparent optical depth method 
for most of the low ionization species observed in this system.

\subsection{Kinematics}

\begin{figure}[ht!]
\figurenum{12}
\includegraphics{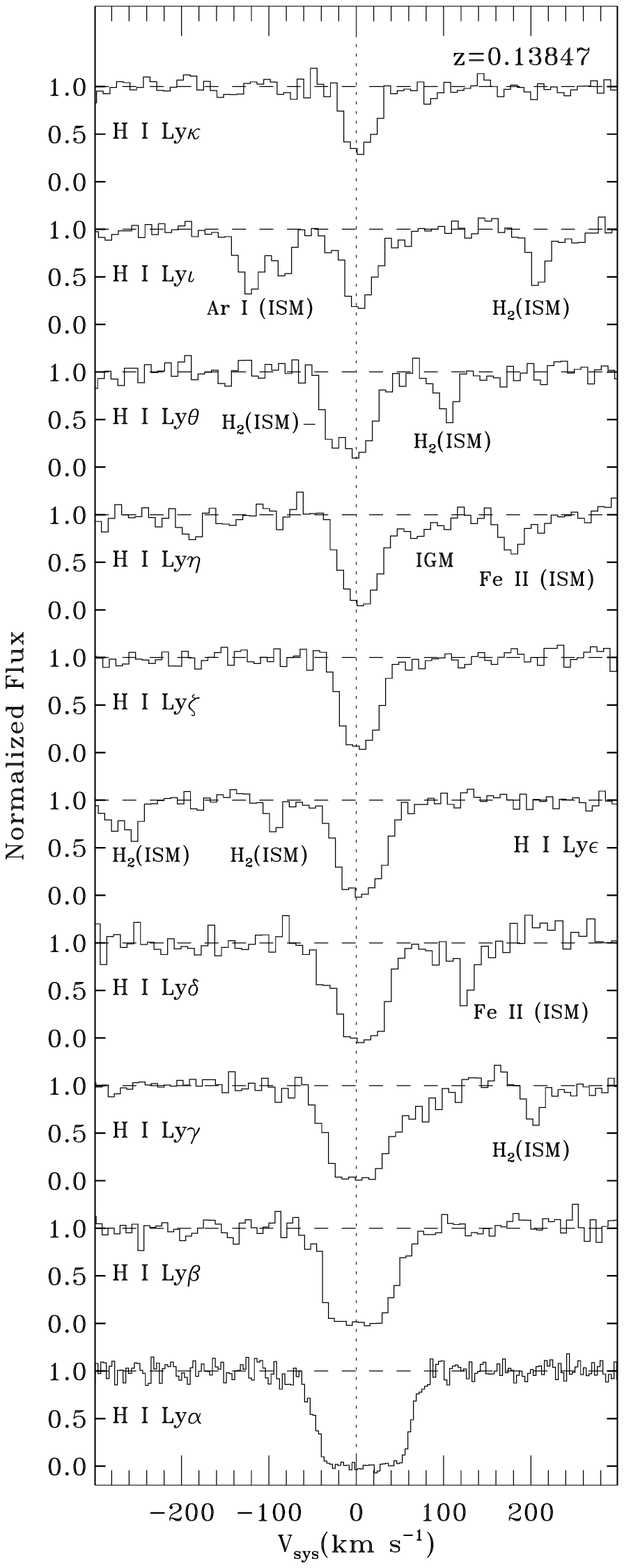}
\vspace{8.5in}
\caption{Continuum-normalized \ion{H}{1} lines in the $z=0.13847$ 
absorber.  The Ly$\alpha$ profile is from HST/STIS.  All others are from 
the FUSE LiF1 channel.\label{stack_13847}}
\end{figure}

As noted above, the low ionization metal lines have at least two components that are 
closely spaced in velocity in this system.  Examination of the \ion{H}{1} profiles in
Figure~\ref{stack_13847} shows that the same appears to be true for \ion{H}{1}.  The higher-order lines
in the Lyman series have a shape resembling that of the metal lines.  The stronger \ion{H}{1}
lines (Ly$\beta$, Ly$\alpha$) show clear extensions toward more positive velocities,
with steep absorption walls between +50 and +70 \kms.  The intermediate
ionization stages (\ion{C}{3}, \ion{N}{2-III}, \ion{Si}{3-IV}) have profile shapes 
and velocity extents similar to those of the lower ionization stages (see Figure~\ref{stack_ion_13847}).
The absorption is confined between about $-25$ and $+30$ \kms.
They too may contain at least two components as evidenced by the inflection on the 
positive velocity side of the \ion{Si}{3} $\lambda1206.500$ profile observed by STIS.
Like the low ionization gas, the intermediate ionization gas is probably dominated 
by the single primary component near 0~\kms.  The weak \ion{Si}{4} $\lambda1393.755$ line
is visible in only the dominant component.
The \ion{O}{6} $\lambda1031.926$ line is centered at the same 
velocity as the peak absorption in the lower ionization stages, but it is considerably
broader.  This line has an effective b-value 3-5 times
that of the lower ionization metal lines (b~$\approx 25$ \kms\ vs. b~$\approx 5-10$ \kms).

Several factors could contribute to the  greater \ion{O}{6} line breadth, including 
thermal broadening, turbulent motions in the highly ionized gas, and greater spatial 
extent of the absorbing region.  All of these possibilities are consistent with the 
kinematics of the \ion{H}{1} lines.  The total velocity extent of the \ion{O}{6}
(roughly $\pm40-50$ \kms, depending upon which 
FUSE or STIS data are used) falls within the velocity range covered by the strong 
\ion{H}{1} lines.

\subsection{Ionization}

The wide range of ionization stages observed in the $z=0.13847$ system is 
strong evidence 
that the absorber is multi-phase in nature.  In particular, the presence
of \ion{O}{6} with a large amount of neutral and low ionization gas indicates
that the medium probed probably does not have a uniform ionization throughout.
We find  N(\ion{H}{1})/N(\ion{O}{6}) = $328\pm^{62}_{92}$, which is the highest
ratio found for any of the systems along this sight line (Table~\ref{tab_summarylya}) or other
sight lines.  For example, the \ion{O}{6} absorbers toward H\,1821+643 
have  N(\ion{H}{1})/N(\ion{O}{6}) $\lesssim 10$ (Tripp et al.\ 2000; Oegerle 
et al.\ 2000), while those
toward PG\,0953+415 have values of 0.2 and 1.5 (Savage et al.\ 2002).
The \ion{O}{6} absorbers toward PG\,1259+593 have N(\ion{H}{1})/N(\ion{O}{6}) 
$\sim1-10$, with the exception of the system at $z=0.04606$, which has 
N(\ion{H}{1})/N(\ion{O}{6}) $\sim125$ (Richter et al.\ 2004).

\begin{figure}[ht!]
\figurenum{15}
\includegraphics{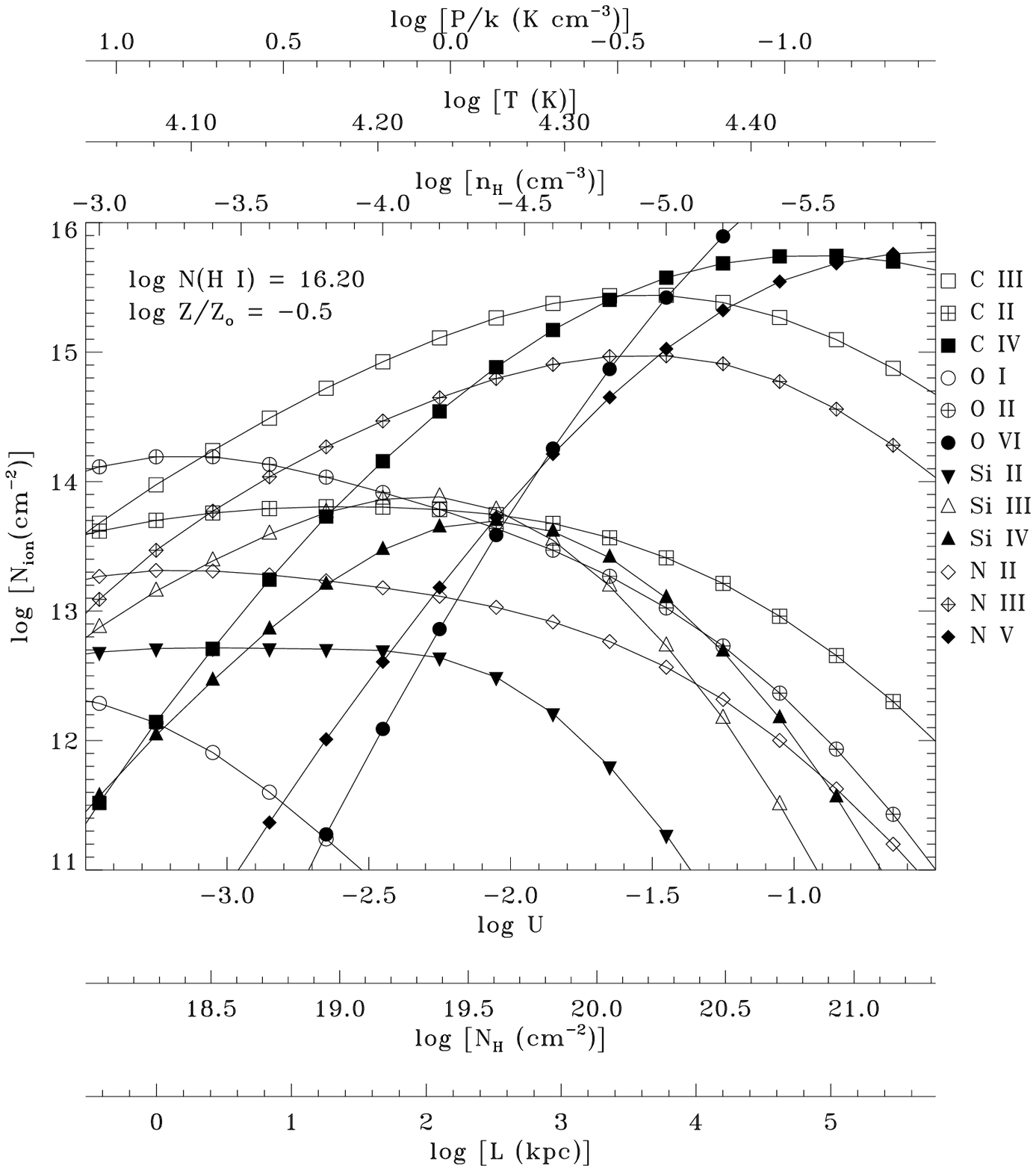}
\vspace{7.1in}
\caption{Predicted column densities for the photoionization model of the 
$z=0.13847$ system.  Relevant quantities for the physical conditions in
this plane-parallel, uniform density slab of gas (pressure, temperature,
density, ionization parameter, total hydrogen column density, and cloud thickness)
are plotted along the x~axes.  See text for details.\label{photo13847}}
\end{figure}

We considered whether the absorption in this system could be produced in a
uniform, low-density photoionized medium of the type that has been suggested 
as a possible explanation for other IGM \ion{O}{6} absorbers (see, e.g., 
Savage et al.\ 2002).  Using the \ion{H}{1} 
column density, log~N(\ion{H}{1}) = 16.20, as a boundary condition, we 
calculated photoionization models for a plane-parallel distribution of 
low density gas with the {\tt CLOUDY} ionization 
code (v94.00; Ferland 1996).  We adopted a redshifted
 ionizing spectrum produced by the 
integrated light of QSOs and AGNs normalized to a mean 
intensity at the Lyman limit 
J$_{\nu_0}$ = 1$\times$10$^{-23}$ erg~cm$^{-2}$~s$^{-1}$~Hz$^{-1}$~sr$^{-1}$
(Donahue, Aldering, \& Stocke 1995; Haardt \& Madau 1996; Shull et al.\ 1999;
Weymann et al.\ 2001)
 and the solar 
abundance pattern given by Anders \& Grevesse (1989), with updates for 
abundant elements (C, N, O, Mg, Si, Fe) from Holweger (2001) and 
Allende~Prieto, Lambert, \& Asplund (2002).  
The results from one such model with a 
metallicity of 1/3 solar [$\log (Z/Z_\odot) = -0.5$] and no dust 
are shown in Figure~\ref{photo13847}.  In this figure, the predicted ionic column densities
are plotted as a function of ionization parameter, $U = (n_\gamma/n_{\rm H})
\propto (J_\nu/n_{\rm H})$.
Similar types of models have been discussed
for other IGM absorption systems (e.g., Savage et al.\ 2002;
Tripp et al.\ 2002) and the 
ionized clouds in the vicinity of the Milky Way (e.g., Sembach et al.\ 2003;
Tripp et al.\ 2003).

In Table~\ref{tab_mod13847} we list the photoionization constraints set by the observed
column densities in the $z=0.13847$ absorber.  We consider three 
gas metallicities -- solar, 1/3 solar, and 1/10 solar.  For each ion
we list the range of ionization parameters satisfying the observed
column density value or limit.  Our upper limits on the
\ion{O}{1}, \ion{S}{2}, and \ion{Fe}{2} column densities
provide no suitable constraints for these models.
 
Photoionization in a uniform medium cannot explain all of the observed 
column densities in the $z=0.13847$ absorber simultaneously at a single
ionization parameter.  We list the column density
constraints and the allowed ranges of $\log U$ for models with three
different metallicities (1/10 solar, 1/3 solar, and solar).  Despite
being able to satisfy many of the observed constraints near $\log U 
\approx -2.5$ ($n_{\rm H} \approx 10^{-4}$ cm$^{-3}$, $L \sim 38$ kpc), 
the model shown in Figure~\ref{photo13847} has several shortcomings.  It 
under-produces the amount of \ion{N}{2} and \ion{Si}{2} at all ionization
parameters. \ion{Si}{2} and \ion{O}{2} are under-produced by a factor of 2--4. 
The model also over-produces the amount of \ion{Si}{4} expected by at 
least a factor of 2.5.  Incorporating dust into the model to alleviate the 
\ion{Si}{4} problem only exacerbates the \ion{Si}{2} discrepancy.  Finally,
the predicted value of N(\ion{O}{6}) at $\log U = -2.5$ 
is a factor of $\sim60$ less than observed; an ionization parameter of
$\log U \approx -2.0$  ($n_{\rm H} \approx 3.5\times10^{-5}$ cm$^{-3}$, 
$L \sim 450$ kpc) is required to produce the observed \ion{O}{6} column.
Increasing the gas metallicity to the solar value 
does not reduce these discrepancies significantly.  We conclude that 
a single-phase photoionized plasma is not an adequate description of the 
absorber.

Similarly, a single temperature collisionally-ionized plasma is also
ruled out.  \ion{O}{6} peaks in abundance at temperatures near 
$3\times10^5$\,K, which is much too hot to produce significant quantities 
of lower ionization stages (Sutherland \& Dopita 1993).   If we assume
that all of the highly ionized gas traced by \ion{O}{6} and \ion{N}{5}
occurs in a plasma under conditions of collisional ionization 
equilibrium, then the expected temperature of the gas is 
$T\sim(2.0-2.5)\times10^{5}$\,K.  This temperature would produce an
\ion{O}{6} line with an observed b-value of $\approx20$ \kms\ after 
convolution with the FUSE line spread function.  The larger observed 
breadth of the line in both the FUSE and STIS data indicate that 
either the \ion{O}{6}-bearing gas is hotter than this estimate, or that 
non-thermal broadening mechanisms contribute to the line width.  
We conclude
that a single-phase collisionally ionized plasma is not an adequate description
of the absorber.

A multi-temperature absorption structure is necessary to explain the 
absorption properties of the $z=0.13847$ absorber.  A cooling flow of the 
type described by Heckman et al.\ (2002) may be able to explain the column
densities of the higher ionization species (\ion{Si}{4},
\ion{N}{5}, \ion{O}{6}) but 
would probably require additional ionization mechanisms to establish the 
ionization pattern seen in the lower ionization stages.  One possible 
solution would be to combine a radiatively cooling flow with a photoionizing
spectrum.  It is interesting that the
ionization pattern of this cloud is in some ways similar to that of 
some of the highly ionized high-velocity clouds (HVCs) in the 
vicinity of the Milky Way.  A hybrid ionization solution is a 
strong possibility for the HVC at +184 \kms\ along 
the PG\,1116+215 sight line (Ganguly et al.\ 2004) and the high
velocity clouds along the Mrk\,509 and PKS\,2155-304 sight lines
(Sembach et al.\ 2000; Collins et al.\ 2004).  These HVCs have 
\ion{H}{1} column densities that are comparable to 
the \ion{H}{1} column of the $z=0.13847$ 
absorber. 

Regardless of the ionization mechanism, the $z=0.13847$ absorber contains a 
large amount of ionized gas relative to its neutral gas content.  If 
we consider only the amount of ionized hydrogen associated with the 
\ion{O}{6}, we can write the following expression to estimate N(H$^+$):

\begin{equation}
{\rm N(H^+)} = \frac{{\rm N(O\,VI)}}{f_{\rm O\,VI}}~\frac{{\rm (O/H)_{\,}}}{{\rm (O/H)}_\odot}
\end{equation}

\noindent 
where $f_{\rm O\,VI}$ is the fraction of oxygen in the form of \ion{O}{6},
and (O/H) is the abundance of oxygen relative to hydrogen.  Using a 
value of (O/H)$_\odot = 4.9\times10^{-4}$ (Allende Prieto et al.\ 2002) and 
$f_{\rm O\,VI} \lesssim 0.2$ (Tripp \& Savage 2000; Sembach et al.\ 2003),
we find N(H$^+$) $\gtrsim 5\times10^{17} (Z/Z_\odot)^{-1}$ cm$^{-2}$.  For 
metallicities between 1/10 and 1/3 solar, the system contains at least 
100-300 times as much H$^{+}$ as \ion{H}{1}.  This estimate accounts 
only for the H$^+$ directly associated with the \ion{O}{6}, which means 
that the total H$^+$ content must be even greater.

A useful piece of missing information for this system is the \ion{C}{4}
column density.  At this redshift, \ion{C}{4} $\lambda\lambda1548.195, 
1550.770$ absorption would be observed
at 1762.57 and 1765.51\,\AA.  These wavelengths are not covered by 
our existing STIS E140M data ($\lambda < 1709$\,\AA) or our E230M data
($\lambda > 2004$\,\AA). The column density ratio 
N(\ion{O}{6})/N(\ion{C}{4}) is an excellent discriminant between 
collisional ionization and photoionization when combined with 
N(\ion{O}{6})/N(\ion{N}{5}) and the column densities of 
moderately ionized species such as \ion{C}{3}, \ion{Si}{3}, and \ion{Si}{4}.

\section{Properties of the \ion{O}{6} System at $z\approx0.166$}

The redshifts of the absorbers at $z\approx0.166$
are  $\sim3000$ \kms\ blueward of the 
redshift of PG\,1116+215 ($z_{em} = 0.1763$).  This is close to the 
somewhat arbitrary velocity cutoff often adopted for separating intervening
intergalactic absorbers from those associated with quasars in general.
The Ly$\alpha$ lines at $z \approx$ 0.166 are well-aligned with
one of the most prominent peaks in the galaxy redshift distribution
within $\sim 1^{\circ}$ of the sight line (see Tripp et al.\ 1998 and
Figure~\ref{galaxyhist}). Ejected associated absorption would have 
a random velocity
with respect to the galaxy distribution and would be unlikely to be so
well-aligned with nearby galaxies.
Furthermore, the three primary absorbers have
considerably different ionization properties, suggesting that they trace
different environments.  For these reasons, we
treat the absorbers at this redshift as intervening systems.

A finding list for the \ion{H}{1} and metal-lines detected in the three
strongest
$z\approx0.166$ absorbers can be found in Table~\ref{tab_ew166}.
Continuum normalized line profiles are shown in Figure~\ref{stack_166} 
plotted against the rest-frame velocity of the $z=0.16610$ absorber.
We estimated the \ion{H}{1} content of the three absorbers 
using both profile fitting of the Ly$\alpha$ lines 
and curve of growth techniques.  Two weak flanking absorbers are detected 
only in the Ly$\alpha$ transition.  Single-component Voigt profiles provide 
excellent approximations to these lines and yield the column densities 
listed in Table~\ref{tab_col166}.  The primary absorber at 
$z=0.16610$ is detected in Ly$\alpha$, Ly$\beta$, and Ly$\epsilon$ 
(see Table~\ref{tab_ew166}).
A single-component 
curve of growth with b = $22.0\pm0.9$ \kms\ provides an excellent 
approximation to the rest-frame equivalent widths of these \ion{H}{1}
lines (Figure~\ref{cog16610}).  The fits to the 
Ly$\alpha$ lines for all three absorbers are shown in Figure~\ref{lya_stack}.

\subsection{The $z=0.16548$ Absorber}
\subsubsection{Column Densities and Kinematics}

Of the three absorbers near $z\approx 0.166$, this is the only one
that exhibits \ion{O}{6} absorption.  We estimated an \ion{O}{6} 
column density for this absorber from the apparent optical depth profiles 
constructed for the two lines, as shown in the top panel of 
Figure~\ref{acd16548}.  The run of N$_a$(v)
for both profiles is very similar over the $-50 \le v_{\rm sys} \le +50$ 
\kms\ velocity range, indicating that there are no significant unresolved 
saturated structures within the lines.  This is as expected because the 
STIS data have sufficient spectral resolution to resolve lines of oxygen
at temperatures $T \gtrsim 1.5\times10^4$\,K, which is well below the
temperatures expected for all reasonable ionization scenarios involving 
\ion{O}{6}.  The integrated values of N$_a$ for the two lines are nearly
identical.  We averaged the two values to produce the adopted value of 
N(\ion{O}{6}) = $(1.21\pm0.15)\times10^{14}$ cm$^{-2}$ listed in Table~\ref{tab_col166}.

The \ion{O}{6} lines can be decomposed into three components at roughly 
--25, 0, and +17 \kms\, with b-values of 20, 8, and 8 \kms, respectively.
The 
peak optical depth occurs in the +17 \kms\ component, while the Ly$\alpha$ 
(and perhaps weak Ly$\beta$) is centered near 0 \kms.  Both species
have similar total velocity extents ($\pm50$ \kms).  We show the apparent
column density profile for the Ly$\alpha$ line in the bottom panel of 
Figure~\ref{acd16548}.  Integration of the Ly$\alpha$ profile yields a column density 
indistinguishable from that of the profile fit shown in Figure~\ref{lya_stack}.
The multi-component structure of the \ion{O}{6} lines indicates that 
the \ion{H}{1} is probably also multi-component in nature.

A small amount of \ion{N}{5} $\lambda1238.821$ absorption may also
be present in this absorber.  Both the apparent optical depth method
and a linear curve of growth applied to the equivalent width of this 
line yield the same column density: N(\ion{N}{5}) = 
$(6.03\pm2.82)\times10^{12}$ cm$^{-2}$.  This detection is somewhat 
tentative since there are other weak features in the STIS 
spectrum with similar equivalent widths (see \S5.16).  However, the
putative absorption does align well in velocity with the Ly$\alpha$
absorption and the zero velocity component of the \ion{O}{6} absorption.

\subsubsection{Ionization}

Comparison of the \ion{O}{6} and \ion{H}{1} N$_a$(v) profiles shows
an obvious change in ionization or metallicity as a function of velocity.
The simple profile decomposition of the \ion{O}{6} lines also favors a
 mix of ionization conditions traced by the \ion{O}{6} and \ion{H}{1}.  
The narrower structure within the \ion{O}{6} profiles near 0 and +17 
\kms\ traces gas at 
$T < (5-8)\times10^5$\,K.  
A single component fit to the \ion{H}{1} line yields b = $29.9\pm1.4$
\kms\ (Table~\ref{tab_summarylya}), which implies that the bulk of the \ion{H}{1} in the absorber 
must be at temperatures $T \lesssim 5.4\times10^4$\,K, which is too cool
to support collisional ionization of \ion{O}{6} in equilibrium situations.
In collisional ionization equilibrium, $f_{\rm H\,I}/f_{\rm O\,VI} > 10^3$ (
Sutherland \& Dopita 1993), and the expected ratio of 
N(\ion{H}{1})/N(\ion{O}{6}) far exceeds the ratio observed for the N$_a(v)$
profiles shown in Figure~\ref{acd16548} near zero \kms.  The gas must therefore
either be in a non-equilibrium situation or photoionized.

If the width of the broad negative velocity wing in the \ion{O}{6} lines is 
dominated by thermal broadening in hot gas, then the implied temperature
of the gas in the wing 
is $T \sim(3-5)\times10^5$\,K.   At these temperatures, hydrogen 
would have b $\sim70-90$ \kms, and therefore only a small portion of the 
observed \ion{H}{1} could be assigned to the hot \ion{O}{6} gas.  No more
than about $1.25\times10^{13}$ cm$^{-2}$, or about 50\% of the total \ion{H}{1}
column, can be attributed to gas with b $>70$ \kms.  If the hot gas is
centered near --25 \kms, then this estimate drops to $\lesssim25$\%.

A simple photoionization
model applied to the entire absorber, like the one described in the 
preceding section for the $z=0.13847$ 
absorber, requires a high ionization parameter and large cloud size
to explain the total
\ion{O}{6} and \ion{N}{5} column densities ($\log U \approx -0.5$,
$n_{\rm H} \sim 10^{-6}$ cm$^{-3}$, $L \sim 1$ Mpc).  A cloud this size would 
have a Hubble expansion broadening of $\sim70$ \kms, which is substantially
larger than the observe \ion{O}{6} Doppler width of $\sim30$ \kms.
 These constraints
are relaxed somewhat if collisional ionization also contributes to the 
production of the \ion{O}{6}.  We conclude that a combination of ionization
mechanisms may be required to produce the observed amount of  \ion{O}{6} in 
this absorber.

The absorption at this redshift is reminiscent of the absorption
seen at $z=0.14232$ along the PG\,0953+415 sight line (Tripp \& Savage
2000), for which similar conclusions regarding the ionization of that
system were reached (Savage et al.\ 2002). 
 The \ion{O}{6}/\ion{H}{1} and \ion{O}{6}/\ion{N}{5} 
column density ratios in the two systems are similar, as is the 
total \ion{H}{1} column density (within a factor of 2).  The \ion{H}{1}
absorber toward PG\,0953+415 has flanking Ly$\alpha$ lines, as does this one.
The Ly$\alpha$ line in the PG\,0953+415 absorber has b = $31\pm7$ \kms,
similar to the b = 30 \kms\ width in this system.  In both cases, the 
\ion{O}{6} lines also appear to have multi-component structure.  The 
multi-component \ion{O}{6} absorber at $z=0.1212$ toward H\,1821+643 (Tripp 
et al.\ 2001) also has many similar characteristics.
 
Observations of \ion{C}{4} would help to refine the velocity structure of 
the $z=0.16548$ absorption  and 
place stronger constraints on the ionization conditions.  For example,
if the gas is mostly photoionized by a hard ionizing spectrum, then 
then we would expect to see \ion{C}{4} in appreciable quantities,
N(\ion{C}{4}) $\gtrsim 10^{13}$ cm$^{-2}$.  If some of the gas is 
hot (T$ \ge 3\times10^{5}$\,K), then we would expect  N(\ion{C}{4}) 
$\lesssim 10^{13}$ cm$^{-2}$.  Having such information would also allow
a more direct comparison with the $z=0.14232$ absorber toward PG\,0953+415.


\subsection{The $z=0.16610$ Absorber}

\subsubsection{Column Densities and Kinematics}

\begin{figure}[ht!]
\figurenum{16}
\includegraphics{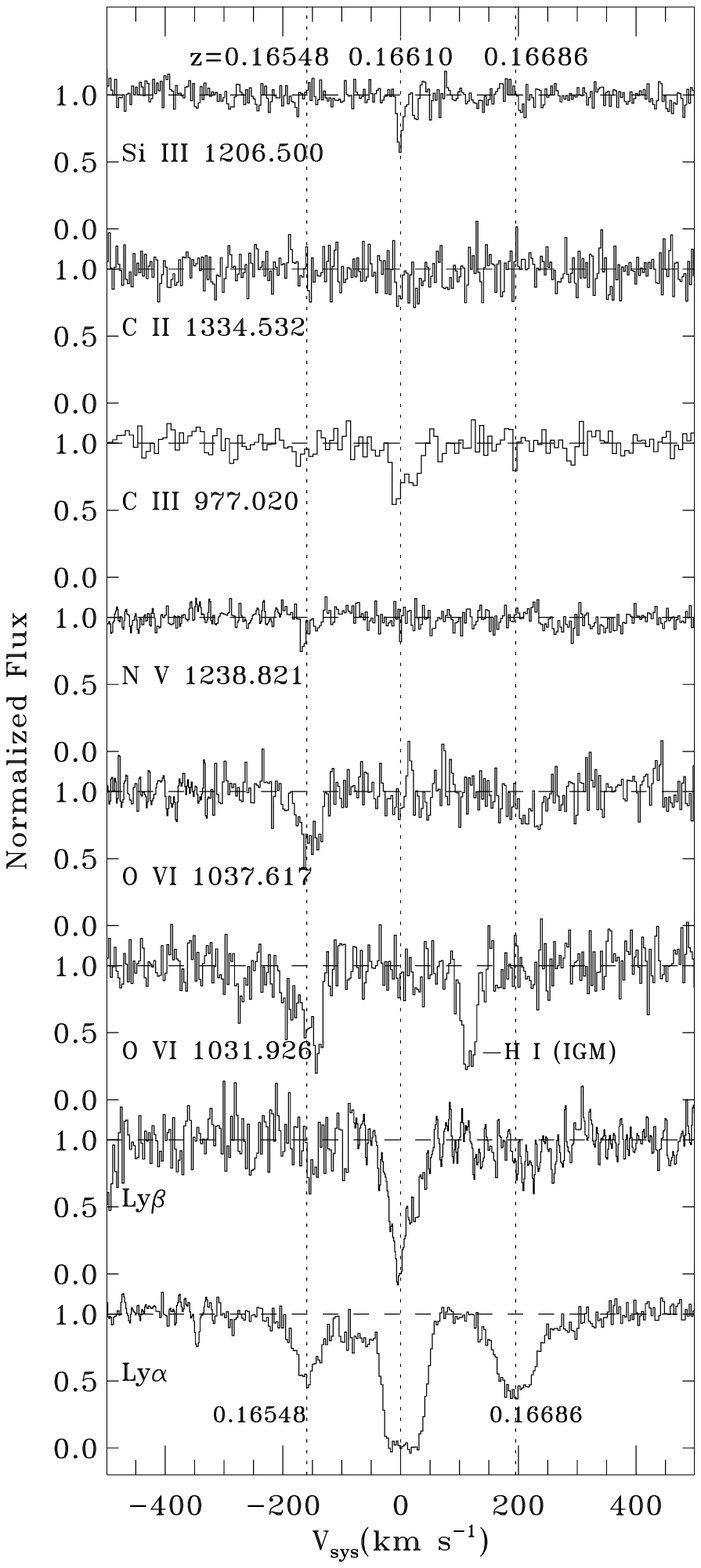}
\vspace{8.0in}
\caption{Absorption lines in the $z\approx0.166$ absorbers.  
Three Ly$\alpha$ absorbers are located within 400 \kms\
of each other at $z=0.16548$, 0.16610, and 0.16686.  These redshifts are 
indicated by the vertical dotted lines.\label{stack_166}}
\end{figure}

\begin{figure}[ht!]
\figurenum{18}
\includegraphics{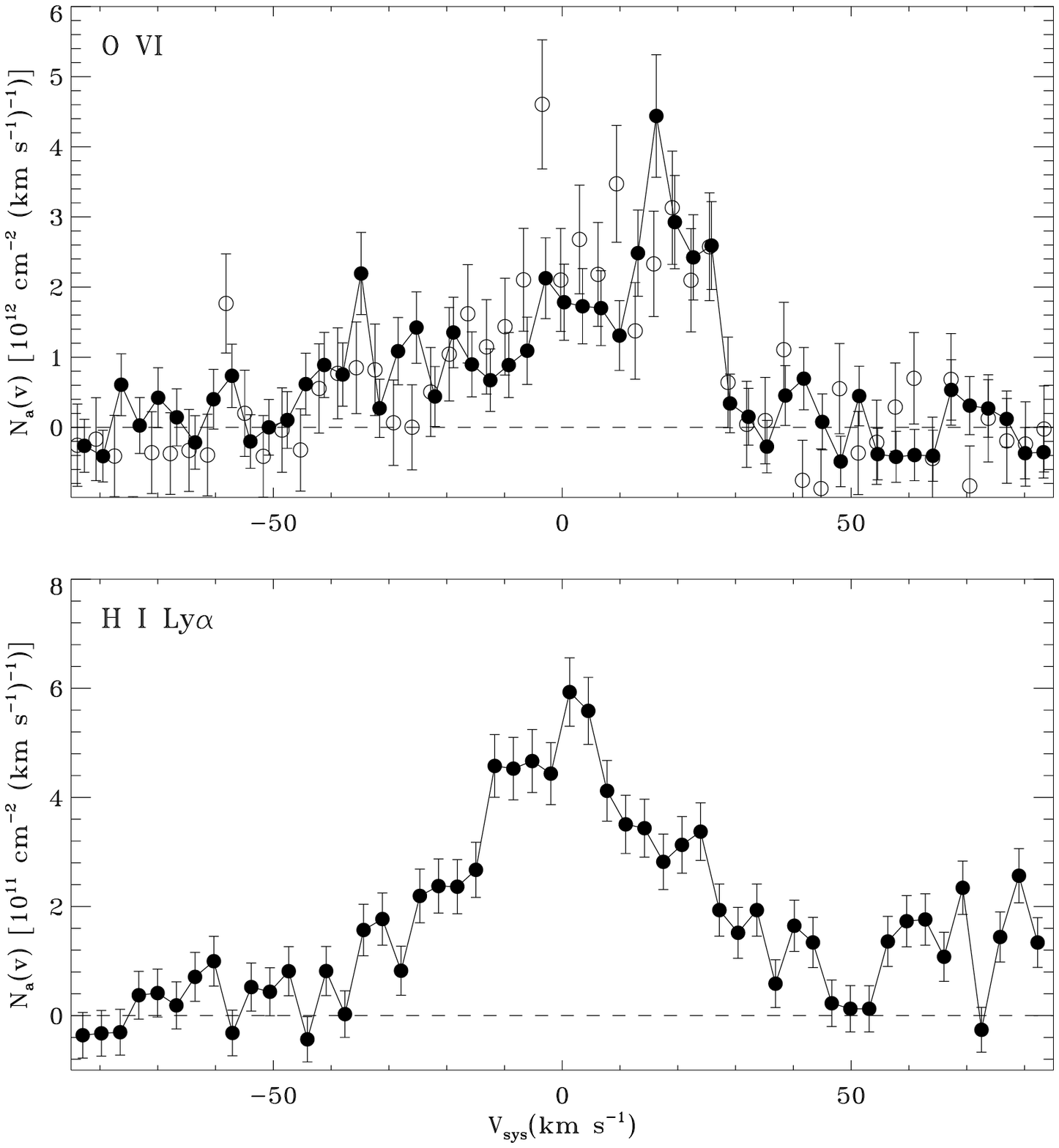}
\vspace{7.1in}
\caption{Apparent column density profiles for the two \ion{O}{6} lines and
 the \ion{H}{1} Ly$\alpha$ line in the 
$z=0.16548$ absorber constructed from the STIS absorption-line data 
shown in Figure~\ref{stack_166}. Error bars are $1\sigma$ estimates. {\it Top:} 
Data points for the $\lambda1031.926$ line are 
filled circles.  Data points for the $\lambda1037.617$ line are 
open circles.  To reduce confusion, the $\lambda1031.926$ data points have 
been connected with straight lines.  {\it Bottom: }The Ly$\alpha$ line.\label{acd16548}}
\end{figure}

The $z=0.16610$ absorber is the strongest of the three Ly$\alpha$ absorbers at 
$z\approx0.166$.
The \ion{Si}{3} $\lambda1206.500$ and \ion{C}{3} $\lambda977.020$ lines
have two components separated by $\approx 25$ \kms.  Both components are 
narrow (b $\lesssim 5 $\kms), implying that the gas is warm ($T \lesssim
2\times10^4$\,K).  The stronger 
component near $v_{sys} = 0$ \kms\ contains $\gtrsim70$\% of the 
total column density listed in Table~\ref{tab_col166}.  A small amount of \ion{C}{2} 
$\lambda1334.532$ absorption may be present (see Figure~\ref{stack_166}), but the
significance of this detection is limited to 2$\sigma$ confidence.

\subsubsection{Ionization}

\begin{figure}[ht!]
\figurenum{19}
\includegraphics{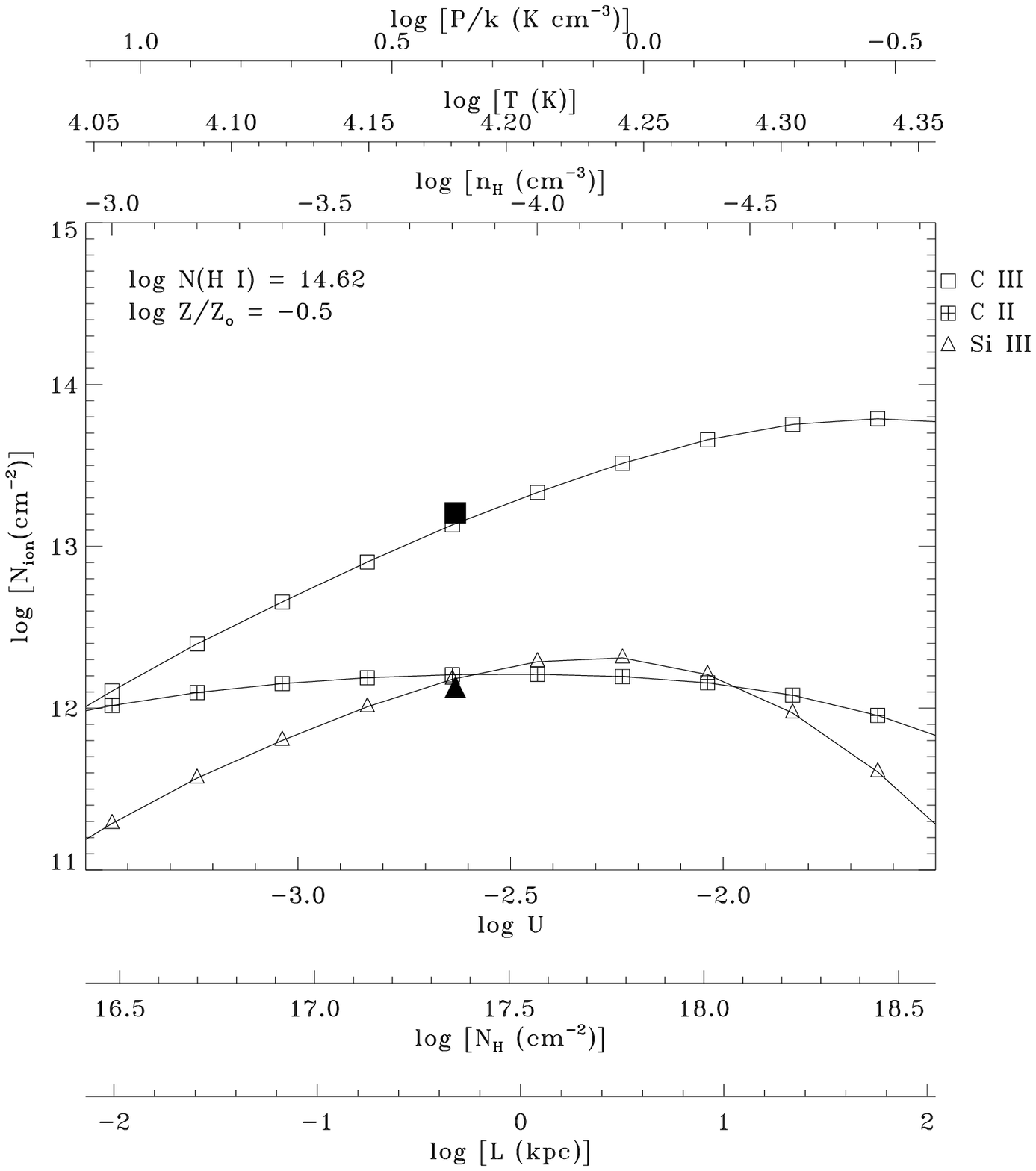}
\vspace{7.0in}
\caption{Predicted column densities for the photoionization model of the 
$z=0.16610$ system.  Relevant quantities for the physical conditions in
this plane-parallel, uniform density slab of gas (pressure, temperature,
density, ionization parameter, total hydrogen column density, and cloud thickness)
are plotted along the x~axes.  
Observed values for N(\ion{C}{3}) and N(\ion{Si}{3}) are shown with the filled
points.  The errors on the observed column densities are comparable to the 
size of the symbols. See text for details.\label{photo166}}
\end{figure}

We constructed {\tt CLOUDY} models with log N(\ion{H}{1}) = 14.62 for this 
absorber analogous to those described above.   The only significant 
constraints on the ionization parameter are the \ion{C}{3} and \ion{Si}{3}
column densities.  The ionization curves for these two species are shown
in Figure~\ref{photo166} for a model with 1/3 solar metallicity.  The total \ion{C}{3}
and \ion{Si}{3} column densities can be satisfied simultaneously in this 
model with no significant alteration of the relative abundance of 
C and Si for a very narrow range of ionization
parameters $\log U \sim -2.63$ ($n_{\rm H} \sim 1.6\times10^{-4}$ cm$^{-3}$).
The corresponding cloud thickness is 0.5 kpc and the total hydrogen column
is $2.2\times10^{17}$ cm$^{-2}$.
Alternatively, if the metallicity is solar, the allowed ionization parameter
decreases to $\log U \sim -2.94$ ($n_{\rm H} \sim 3.3\times10^{-4}$ cm$^{-3}$),
the cloud thickness decreases to $<100$ pc, and the total hydrogen column decreases
to $7.6\times10^{16}$ cm$^{-2}$.   Models with metallicities less than 
$\sim1/10$ solar have difficulties producing the observed amounts of 
\ion{Si}{3} at all values of $U$.  In all of these models, the predicted amount of \ion{C}{2}
is roughly an order of magnitude less than the amount listed in Table~\ref{tab_col166}.
This discrepancy can be removed if the value in Table~\ref{tab_col166} is considered to be
an upper limit.  Such an assumption seems reasonable given the low 
significance of the detection.  Higher quality data for the \ion{C}{2}
$\lambda1334.532$ line would provide a stronger constraint on the \ion{C}{2}
column density.

Incorporating dust in the models would reduce the abundance of silicon
relative to carbon and would lead to larger discrepancies with the observed 
ratio of \ion{Si}{3}/\ion{C}{3}.  Increasing the Si/C ratio above the solar
ratio of 0.14 would allow for a 
larger range of allowable ionization parameter overlap for \ion{Si}{3} and 
\ion{C}{3}, with lower inferred ionization parameters.  A modification of 
this type was employed by Tripp et al.\ (2002) to explain the heavy element
abundances in a Ly$\alpha$ absorber in the Virgo Cluster.  Supersolar 
Si/C enrichment is possible with Type~II supernovae, and while the absorption
we see does not strictly require such enrichment, it is more readily 
explained if some enrichment has occurred.

The \ion{O}{6} column density limit for this absorber is consistent with 
the ionization properties inferred from the lower ionization species.  In 
principle, observation of \ion{C}{4} in this absorber at an 
observed wavelength of 1805.35\,\AA\ would provide
additional constraints on the ionization of the gas.  For the parameters
adopted, we would expect a value of log N(\ion{C}{4}) $<12.5$.  Data of the 
type obtained for this study will be needed since the expected observed
equivalent widths should be only $\sim20$\,m\AA.

\subsection{The $z=0.16686$ Absorber}

This absorber is observed only in \ion{H}{1} Ly$\alpha$.  It has 
an \ion{H}{1} column density a factor of two greater than that of 
the $z=0.16548$ absorber, but has an \ion{O}{6} column density a factor 
of at least 4 less.  The weak absorption wing at $+65 \le v_{sys} \le +145$
\kms\ (see note 16 in Table~\ref{tab_summarylya})
is of unknown origin.  Its peak optical depth is too low to draw
meaningful conclusions about its intrinsic width.  
The primary absorber has an overall width (b = $38.5\pm1.2$ \kms) that is 
consistent with a temperature $T \lesssim 9\times10^5$\,K.  Solar 
abundance gas in collisional ionization equilibrium at this temperature has 
no appreciable \ion{O}{6} (Sutherland \& Dopita 1993). If the hydrogen 
is collisionally ionized at the high end of this temperature 
range, the absorber may contain very large amounts of ionized hydrogen,
as discussed in \S9.2.

\section{Weak \ion{O}{6} Absorption Systems}

We have identified three weak \ion{O}{6} absorbers along the PG\,1116+215
sight line.  These absorbers at $z=0.04125$, $0.05928$, and $0.06244$ 
are detected only in \ion{H}{1} Ly$\alpha$ and \ion{O}{6} absorption
(see \S5 for an overview).  Velocity plots of the Ly$\alpha$ and
\ion{O}{6} profiles for these absorbers can be found in Figures~\ref{weak_o6_04125},
\ref{weak_o6_05895}, and \ref{weak_o6_06244}.
We briefly consider the 
physical conditions in each of these systems below.

\subsection{The $z=0.04125$ Absorber}

\begin{figure}[ht!]
\figurenum{8}
\includegraphics{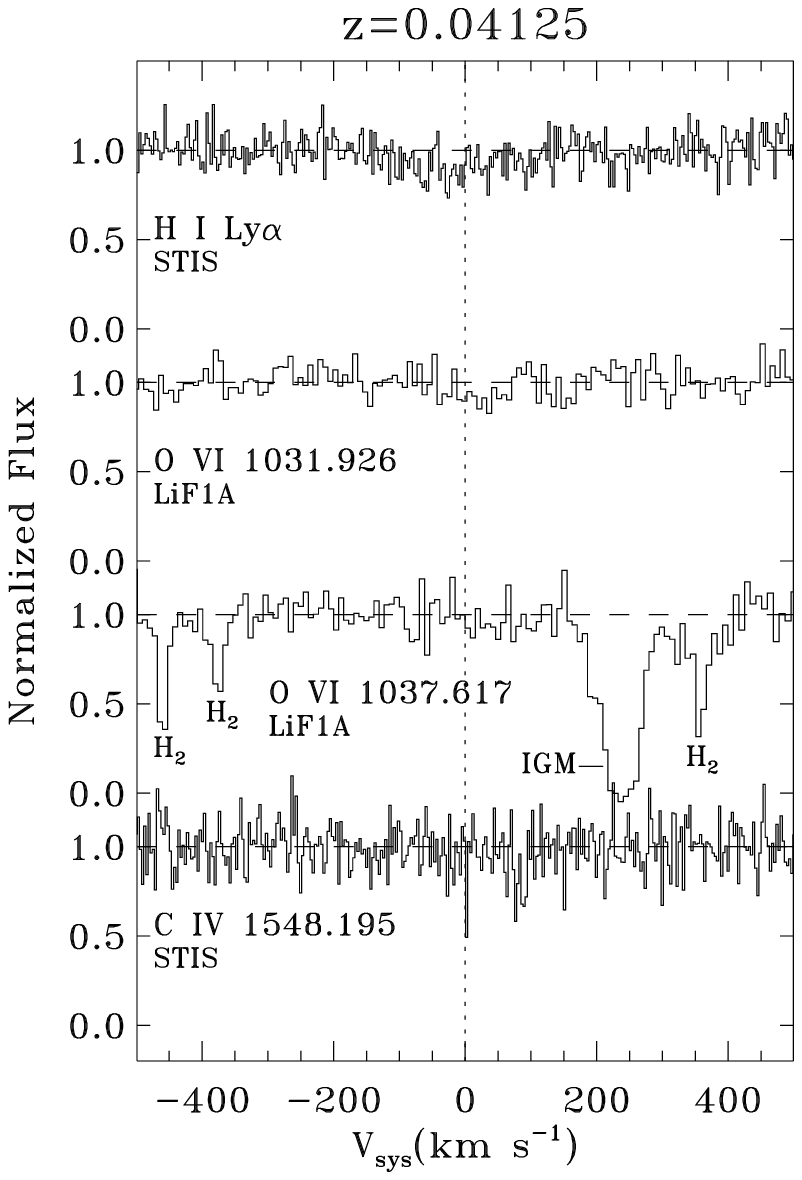}
\vspace{7.0in}
\caption{Normalized absorption profiles for the $z=0.04125$ absorption-line system.
The vertical dashed line indicates the systemic velocity of the Ly$\alpha$ line.  \label{weak_o6_04125}}
\end{figure}

For the weak \ion{O}{6} $\lambda1031.926$ absorption in this system we measure 
$W_\lambda = 28\pm10$\,m\AA.
The width of the line (b $= 35\pm15$ \kms) implies that the 
gas associated with the \ion{O}{6} is hot [$T \approx (0.4-1.9)\times10^6$\,K]
if the line width is broadened by thermal effects alone.  The great
width of the Ly$\alpha$ line (b $= 105\pm18$ \kms) is consistent with 
the presence of hot gas.  However, given the modest detection 
significance of the \ion{O}{6} feature ($< 3\sigma$), it is not 
possible to determine the precise relationship of the \ion{H}{1}
and \ion{O}{6} (see Figure~\ref{weak_o6_04125}).  The value of N(\ion{O}{6}) $\approx 2\times10^{13}$
cm$^{-2}$ in this absorber (Table~\ref{tab_summarylya}) implies an ionized hydrogen
column density of N(H$^+$) $\gtrsim 2\times10^{17} (Z/Z_\odot)^{-1}$ 
cm$^{-2}$ (Eq.\ 4).  This limit is well below, but consistent with,
the much higher 
H$^+$ column derived from the large ionization correction based solely
on the width of the broad Ly$\alpha$ line.  If the gas is in
collisional ionization equilibrium at a temperature 
$T \sim (0.5-1.0)\times10^6$\,K as implied by the \ion{H}{1} line width,
the metallicity of the gas derived from the ratio N(\ion{H}{1})/N(\ion{O}{6})
= $0.8\pm0.4$ is roughly 1/6 to 1/20 solar.

\subsection{The $z=0.05928$ Absorber}
\begin{figure}[ht!]
\figurenum{9}
\includegraphics{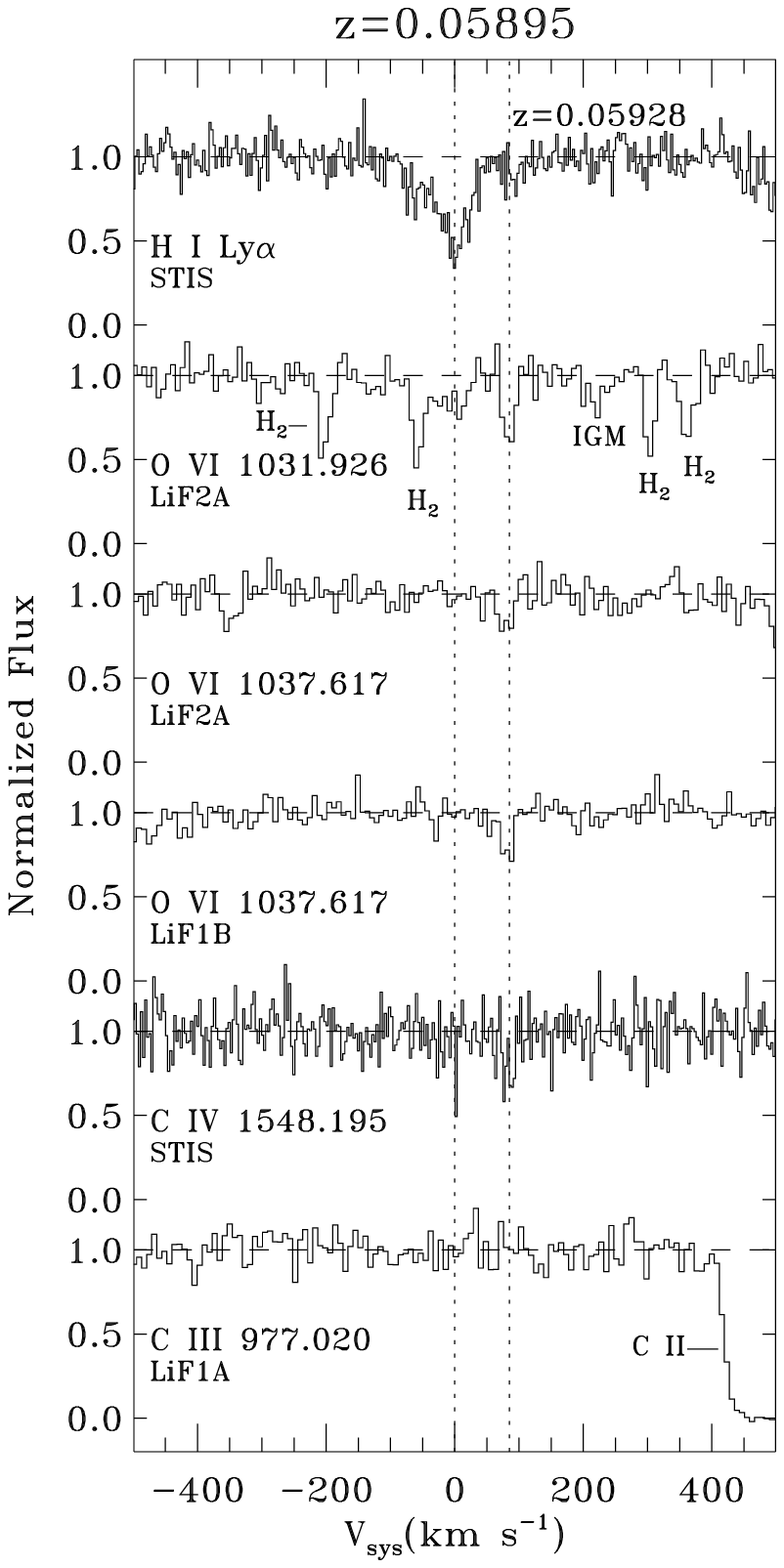}
\vspace{7.0in}
\caption{Normalized absorption profiles for the $z=0.05895, 0.05928$ absorption-line system.
The velocity scale is centered on the systemic velocity of the Ly$\alpha$ absorber at
$z=0.05895$.  
The second vertical line indicates the velocity of the system at $z=0.05928$.\label{weak_o6_05895}}
\end{figure}

The Ly$\alpha$ and \ion{O}{6} absorptions in this absorber occur 
$\sim80-90$ \kms\ redward of the strong Ly$\alpha$ absorption at 
$z=0.05895$.    Normalized absorption profiles are shown in Figure~\ref{weak_o6_05895}.
Both 
the Ly$\alpha$ and \ion{O}{6} absorptions are weak 
and narrow (b $\lesssim 10$ \kms).
The \ion{O}{6} $\lambda1031.926$ line has $W_\lambda = 26\pm7$\,m\AA,
and the $\lambda1037.617$ line has equivalent widths of $16\pm6$\,m\AA\
(LiF2A) and $20\pm6$\,m\AA\ (LiF1B).  Converting these equivalent
widths into  \ion{O}{6} column densities and averaging yields N(\ion{O}{6})
= $(2.45\pm0.78)\times10^{13}$ cm$^{-2}$.
 The widths of the \ion{O}{6} (b $< 10$ \kms) and \ion{H}{1}
(b $\approx 10$ \kms) imply temperatures $T < 10^5$\,K and $T \sim 6000$\,K, 
respectively.  The temperatures indicate that the \ion{O}{6} in this system
is produced by photoionization in warm gas rather than collisional ionization
in hot gas.  The amount of H$^+$ associated
with the \ion{O}{6} is $\sim10^5$ times greater than the amount of \ion{H}{1}
measured.  Weak \ion{C}{4} absorption may also be present in this 
system, which in conjunction with the absence of \ion{C}{3} 
(see Figure~\ref{weak_o6_05895}) confirms that the system is highly ionized.

\subsection{The $z=0.06244$ Absorber}
\begin{figure}[ht!]
\figurenum{10}
\includegraphics{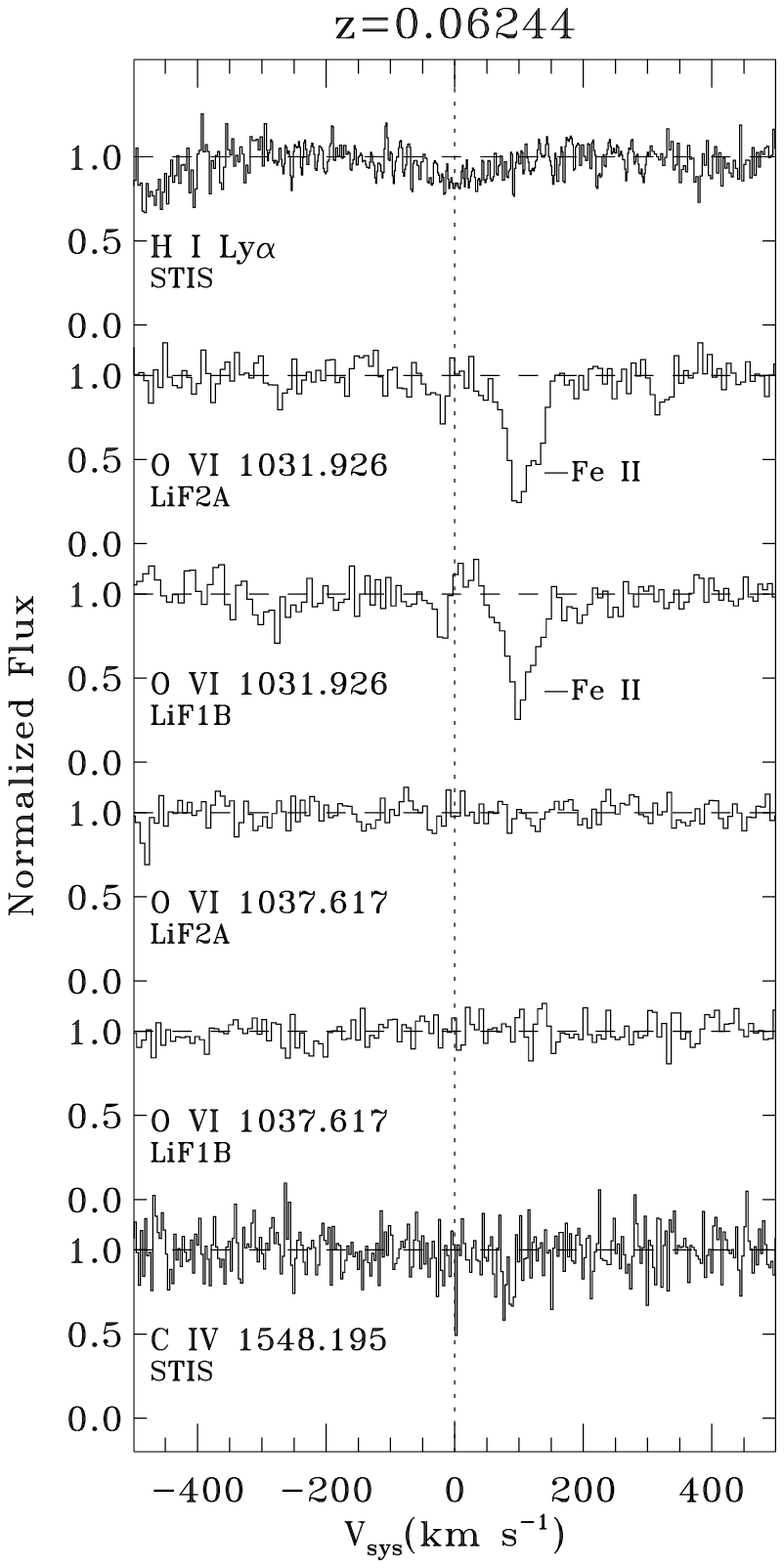}
\vspace{7.0in}
\caption{Normalized absorption profiles for the $z=0.06244$ absorption-line system.
The vertical dashed line indicates the systemic velocity of the Ly$\alpha$ line.  \label{weak_o6_06244}}
\end{figure}

This absorber consists of both broad \ion{H}{1} 
Ly$\alpha$ and relatively 
narrow \ion{O}{6} $\lambda1031.926$ lines.  
We estimate an observed \ion{O}{6} line strength of $W_\lambda = 19\pm8$\,m\AA\
(LiF1B) and  $W_\lambda = 18\pm7$\,m\AA\ (LiF2A).  Assuming a linear curve
of growth yields N(\ion{O}{6}) = $(1.39\pm0.56)\times10^{13}$ cm$^{-2}$.
The  \ion{O}{6} line width of b $\approx 8\pm7$ \kms\
translates into a temperature $T < 2.2\times10^5$\,K at 
the 1$\sigma$ upper line width confidence estimate (b = 15 \kms).  
The uncertainty on the
\ion{O}{6} width is large because the line is weak and 
barely resolved by FUSE.  
The significance of the 20 \kms\
kinematical offset between the \ion{O}{6}
and \ion{H}{1} centroids (see Figure~\ref{weak_o6_06244}) is difficult to assess since 
the \ion{H}{1}
line is so broad.  However, the reasonable agreement indicates that
much of the Ly$\alpha$ line width could be due to thermal broadening
in hot gas -- approximately 60 \kms\ of the 77 \kms\ line width could 
be accounted for by thermal broadening of hot gas 
directly associated with the \ion{O}{6}.  

In collisional ionization equilibrium, the value of N(\ion{H}{1})/N(\ion{O}{6})
= $1.1\pm0.5$ observed for this absorber implies $T \approx (1.8-2.0)\times10^5$\,K
for a solar abundance plasma.  The  fraction of hydrogen expected to be in
the form of \ion{H}{1} at
this temperature is $(3-4)\times10^{-6}$, so the amount of H$^+$ associated with the 
\ion{O}{6} is $\sim4\times10^{18}$ cm$^{-2}$.  This is only a few times less
than the value of N(H$^+$) $\sim1.2\times10^{19}$ cm$^{-2}$ derived assuming 
that the entire \ion{H}{1} line width is thermally broadened by a single 
temperature gas (\S9.2). A higher temperature solution at $T > 7\times10^5$\,K
is probably excluded by the \ion{H}{1} line width.
The conclusion that the \ion{O}{6} is associated with only a portion of the 
\ion{H}{1} line width does not exclude the possibility of a 
hotter component, however, since the detectability of weak \ion{O}{6} at 
higher temperatures is more difficult at these low equivalent width levels.

The combination of broad Ly$\alpha$ and narrow \ion{O}{6} in this absorber
is important, as there have been few such cases found in the low-redshift
IGM.  The absorber resembles that of the $z=0.31978$ system toward 
PG\,1259+593 (Richter et al.\ 2004)
in several respects.  Both systems have broad Ly$\alpha$ 
and narrow \ion{O}{6} $\lambda1031.926$ lines.  The ratio of line 
widths in both systems suggests that a substantial portion of the Ly$\alpha$ 
line width could be caused by thermal broadening at high temperatures,
in which case the amount of related ionized (H$^+$) gas must be very large.
Both systems show a 
slight offset of the \ion{O}{6} to the negative velocity side of the 
Ly$\alpha$ centroid (although the significance of this offset is unclear).
Neither system is detected in other metal lines; for the $z=0.31978$ system
it was possible to search for \ion{Ne}{8} and \ion{O}{4}, which led 
to a temperature constraint of $2\times10^5 < T < 6\times10^5$\,K 
(see Richter et al.\ 2004).  The $z=0.06244$ system is considerably 
weaker than the $z=0.31978$ system [ N(\ion{H}{1}) $\approx 1.5\times10^{13}$
cm$^{-2}$ vs.  $\approx 10^{14}$ cm$^{-2}$ ],
and has proportionally more \ion{O}{6} relative to \ion{H}{1} than
the $z=0.31978$ absorber 
[ N(\ion{H}{1})/N(\ion{O}{6}) = $1.1\pm0.5$ vs. $3.5\pm0.6$ ].

\section{Ly$\alpha$ Absorbers Toward PG\,1116+215}

\subsection{General Properties}

Some basic information about the 26 Ly$\alpha$ absorbers along the sight line
is contained in Table~\ref{tab_summarylya}.  All of the Ly$\alpha$ absorbers are detected 
at $\gtrsim3\sigma$ confidence, and all have observed equivalent widths 
$W_\lambda \gtrsim 32$\,m\AA\  (or $W_r \gtrsim 30$\,m\AA).  We find 
$<$b$> \approx 39\pm30$ \kms, with a median value of $\bar{\rm b} \approx 31$
\kms.  The mean value is weighted heavily by several broad Ly$\alpha$ 
lines, which may consist of multiple components (see below).  The mean
value is consistent with the average of $<$b$> = 38.0\pm15.7$ \kms\ 
found for the Ly$\alpha$ absorbers along many 
sight lines by Penton, Shull, \& Stocke (2000).  

For the 25 Ly$\alpha$ absorbers with $W_r \gtrsim 30$\,m\AA\ toward 
PG\,1116+215\footnotemark\, we find (d$N$/d$z$)$_{{\rm Ly}\alpha} = 
154\pm18$ for absorbers over an 
unblocked redshift path $\Delta X_{{\rm Ly}\alpha} = 0.162$.  
The error on this value reflects both the uncertainty in the 
blocking correction and the possibility that we have miscounted by 
three the number of absorbers in our census of Ly$\alpha$ absorbers along the 
sight line.  For example, this estimate includes the six 
absorbers at $z\approx0.166$ and $z=0.17340-0.17360$, some of 
which  could  possibly be associated
with the quasar host galaxy.  It also accounts for the possibility
that we may have missed a few features at arbitrary redshift 
near the 30\,m\AA\ equivalent cutoff limit.  For example, we have not included
the weak absorption on the positive velocity side of the $z=0.16686$ 
Ly$\alpha$ line in this census. 
If we omit the six Ly$\alpha$ absorbers within 5000 \kms\ of 
PG\,1116+215, we find (d$N$/d$z$)$_{{\rm Ly}\alpha}^\prime = 133$
for an unblocked
redshift path of $\Delta X_{{\rm Ly}\alpha}^\prime = 0.143$.
The estimate of (d$N$/d$z$)$_{{\rm Ly}\alpha}$ for the PG\,1116+215 
sight line is slightly smaller than the value of 
(d$N$/d$z$)$_{{\rm Ly}\alpha} = 190\pm28$ found by 
Richter et al.\ (2004) for absorbers with rest-frame equivalent widths 
$W_r \ge 30$\,m\AA\ along the PG\,1259+593 sight line.  The unblocked 
redshift 
path at this equivalent width limit for that sight line is 
$\Delta X_{{\rm Ly}\alpha} = 
0.247$.  Values of (d$N$/d$z$)$_{{\rm Ly}\alpha}$
for both sight lines are roughly consistent with the value of 
(d$N$/d$z$)$_{{\rm Ly}\alpha} \approx 225\pm27$ found in the HST/GHRS 
Ly$\alpha$ survey by Penton et al.\ (2000) for absorbers with a similar
equivalent width cutoff.

\begin{figure}[ht!]
\figurenum{20}
\includegraphics{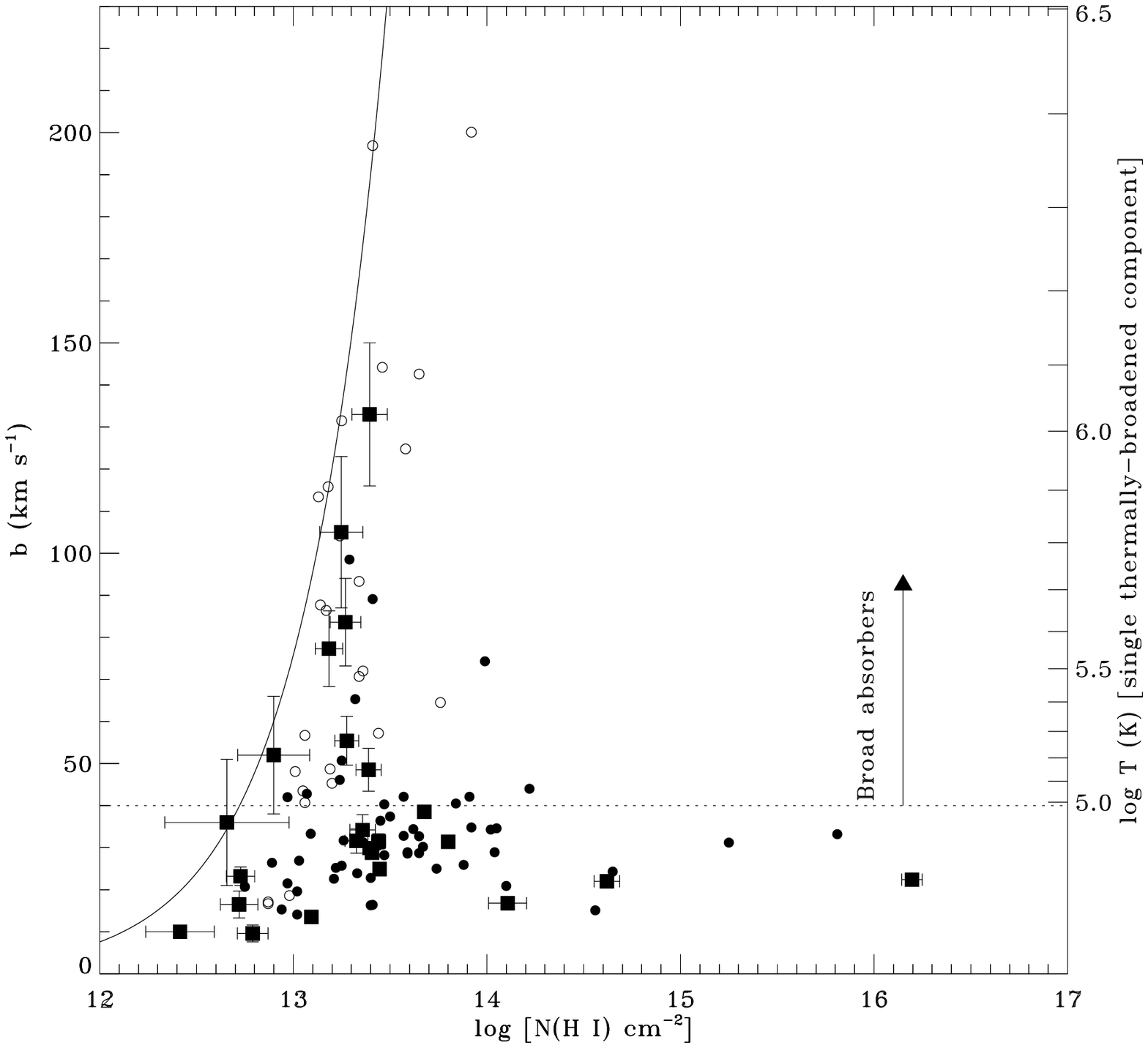}
\vspace{5.7in}
\caption{Line width versus \ion{H}{1} column density for the Ly$\alpha$ 
absorbers identified along the PG\,1116+215 and PG\,1259+593 sight lines.
The corresponding temperature for a single, thermally-broadened
line is shown on the right side of the plot.
The PG\,1116+215 data are shown as filled squares with $1\sigma$
error bars; in some cases the error bars are smaller than the symbol size.
The PG\,1259+593 data from Richter et al.\ (2004) are shown as circles;
open circles indicate less reliable values. The solid curve indicates the
relation between b-value and column density for a Gaussian 
line with a central optical depth of 10\%.  Weak, broad Ly$\alpha$ 
lines in the region to the left of this curve would be difficult to 
detect in the data used for this study.  \label{bvsn}}
\end{figure}

In Figure~\ref{bvsn} we plot the width of the Ly$\alpha$ absorbers 
as a function of their \ion{H}{1} column density.  The data points,
shown as filled squares, have $1\sigma$ error bars attached; in
some cases, these errors are smaller than the symbol size.  We also
plot the points for the PG\,1259+593 sight line (Richter et al.\ 2004) 
for the systems with reliably determined values of b and N (filled circles)
and those with less certain values (open circles).  The smooth 
solid curve is the relationship between b and N for  Gaussian 
lines with central optical depths of 10\%. The absence of points to 
the left of this line is probably a selection effect caused by the 
difficulties in continuum placement for broad weak lines at the $S/N$ 
of the data in the two studies. 
 There are distinct regions
of this figure populated by the Ly$\alpha$ absorbers along both 
sight lines. It is clear  that broad (${\rm b} \gtrsim 40$ \kms)
absorbers are present along both sight lines at 
a statistically significant level.  Furthermore, the broad absorbers 
also tend to be weak (i.e., low N(\ion{H}{1})), suggesting that they may trace hot gas,
as we discuss below.

\footnotetext{The number of absorbers does not include the weak 
($W_r \approx 14$\,m\AA) absorber at $z=0.05928$ or the weak 
($W_r \approx 18$\,m\AA) absorber at $z=0.08632$.  It also treats the two
 components of the 
$z=0.05895$ system as a single absorber because the profile decomposition
of the $z=0.05895$ system is not unique.}

\subsection{Broad Ly$\alpha$ Lines}

Eight of the intergalactic 
Ly$\alpha$ absorbers identified along the sight line 
have measured widths that exceed $\sim40$ \kms.  A width of $\gtrsim40$ \kms\
 corresponds to 
a temperature $T \gtrsim 10^5$\,K if the line is broadened solely 
by thermal processes.  The 8 broad Ly$\alpha$ lines include the absorbers
at $z=0.01635$, $0.04125$, $0.06072$, $0.06244$, $0.08587$,
$0.09279$, $0.13370$, and 
possibly $0.16686$.  Table~\ref{tab_broadlya} summarizes the properties of these
absorbers.  We have included the $z=0.16686$ absorber since
it is only marginally narrower than the 40 \kms\ cutoff 
(b = $38.5\pm1.2$ \kms). Single-component fits to the broad absorbers are 
shown in Figure~\ref{lya_stack}.   
The $z=0.03223$ system is not included in this 
subset of broad absorbers since it clearly consists of an ensemble of 
narrower components (see Figure~\ref{lya_stack}). 
A few other absorbers have sufficiently
large line width errors that they might also qualify, but we have 
not included these in the discussion that follows. 

All of the broad absorbers have observed equivalent widths $W_\lambda
\gtrsim 80$\,m\AA\ ($W_r \gtrsim 74$\,m\AA), except the absorber at 
$z=0.08587$, which has $W_\lambda = 39\pm10$ ($W_r =36\pm9$\,m\AA).
 All of the broad absorbers have 
N(\ion{H}{1}) $\lesssim 3\times10^{13}$ cm$^{-2}$ except  the 
$z=0.16680$ system, which has N(\ion{H}{1}) $\approx 4.75\times10^{13}$ 
cm$^{-2}$. These broad absorbers are relatively rare and difficult
to detect except in high quality datasets.  For example,  Penton et al.\
(2004) identify only 7 broad Ly$\alpha$ absorbers with $W_r > 100$\,m\AA\
in their sample of 109 absorbers
along 15 sight lines covering a 
total redshift path of 0.770 (and one of these is the PG\,1116+215 absorber 
at $z=0.04125$).

The best candidate tracers of the warm-hot IGM are the broadest lines.
The broadest Ly$\alpha$ lines also tend to be those with the small maximum optical
depths.  As a result, the absorbers at $z=0.04125$ (b $\approx 105$ \kms),
$z=0.06244$ (b $\approx 77$ \kms), $z=0.09279$ (b $\approx 133$ \kms),
and $z=0.13370$ (b $\approx 84$ \kms) have substantial width
uncertainties due to the placement of the continuum level.  Even so, 
the great breadth of these lines indicates that the gas must be hot 
[$T \sim (0.3-1.0)\times10^6$\,K] if the lines consist of single 
components that are not broadened substantially by gas flows or turbulent 
gas motions.  
The detections of \ion{O}{6} in two of the broad 
absorbers ($z=0.04125$ and $0.06244$) increases the likelihood that 
hot gas is present at these redshifts (see \S8).

Taking the 8 broad Ly$\alpha$ absorbers toward PG\,1116+215
and the unblocked redshift path of 
$\Delta X_{{\rm Ly}\alpha} = 0.162$ derived in \S3, we find a 
broad Ly$\alpha$ number density per unit redshift of d$N$/d$z \approx 50$
for absorbers with $W_r > 30$\,m\AA.
If we adopt the formalism of Richter et al.\ (2004) for 
estimating the cosmological mass density of the broad Ly$\alpha$ 
absorbers, we can express $\Omega_b$(BLy$\alpha$) as a function of $\Delta X$ 
and the measured \ion{H}{1} column density:

\begin{equation}
\Omega_b({\rm BLy}\alpha) = \frac{\mu_H~m_{\rm H}~H_0}{\rho_c~c~\Delta X} 
~\sum{f_{\rm H}(T_i)~{\rm N}_i{\rm (H\,I)}} \approx 
1.667\times10^{-23}~\Delta X^{-1}~\sum{f_{\rm H}(T_i)~{\rm N}_i{\rm (H\,I)}}
\end{equation}

\noindent
where $m_{\rm H} = 1.673\times10^{-24}$ gm 
is the atomic mass of hydrogen, $\mu_H = 1.3$ corrects 
the mass for the presence of 
helium, $H_0$ = 75 \kms\ Mpc$^{-1}$ is the Hubble constant, and 
$\rho_c = 3H_0^2/8\pi G = 1\times10^{-29}$ gm~cm$^{-3}$ is the 
current critical density.  In this 
expression, the sum over index $i$
is a measure of the total hydrogen column density
in the broad absorbers, with $f_{\rm H}(T_i)$ being the conversion factor
between \ion{H}{1} and total H given by Sutherland \& Dopita (1993)
for temperatures $T \sim 10^5 -10^7$\,K:

\begin{equation}
\log f_{\rm H}(T_i) \approx -13.9 + 5.4 \log T - 0.33 (\log T)^2.
\end{equation}

Assuming the broad Ly$\alpha$ absorbers are purely thermally
broadened, we can use the familiar expression 
b(\kms) = $0.129 \sqrt{T(\rm K)}$ for hydrogen together with the 
values of b and N(\ion{H}{1}) listed in Table~~\ref{tab_summarylya} to estimate 
$\Omega_b({\rm BLy}\alpha) = 0.020\,h^{-1}_{75}$.  This baryonic density is 
enormous -- nearly half of the baryonic mass in the universe
predicted by measurements of 
the deuterium abundances at high redshift 
(Kirkman et al.\ 2003 and references therein)
and by measurements of the cosmic microwave background  (Spergel et al.\ 2003).
The two primary uncertainties in such an estimate lie in the assumption of a 
single-component structure for the absorbers and the large ionization 
corrections required for the high temperatures implied by the large 
Doppler widths.  The first of these is difficult to judge from the existing
data because the single-component model fits shown in 
Figure~\ref{lya_stack} provide
reasonable approximations to the observed profiles.  Multi-component structure
could be present, but it is not required to fit the data as it is 
for some of the other absorption systems along the sight line (e.g., 
$z=0.03223$, $0.05895$). The observed data do not preclude the possibility 
that the lines are intrinsically broad.
The ionization corrections depend strongly on the 
temperature of the gas, and thus also on
the measured profile widths.  The atomic 
physics entering this calculation are known well, though non-equilibrium 
situations could change the corrections.   The 
corrections are large for the collisional ionization equilibrium case,
especially for lines with b~$\gtrsim 100$ \kms.
For example,
$f_{\rm H} = [1.6\times10^5, 1.9\times10^6, 4.3\times10^6]$ for 
b = [50, 100, 130] \kms\ and $T = [1.5\times10^5, 6.0\times10^5, 1.0\times10^6]$\,K.

Another possible complication, especially for 
the broadest lines with the lowest maximum optical depths, is that 
fluctuations in the QSO continuum or blends of lines from gas associated 
with the QSO could mimic some of the broad absorption features attributed
to the IGM.  We believe both of these possibilities are unlikely for the 
PG\,1116+215 sight line. No prominent resonance
lines at the redshifts
of the QSO or the possible associated absorbers at $z=0.17340-0.17360$ 
fall at the wavelengths of the broad Ly$\alpha$ absorbers.  
Furthermore, the QSO continuum is remarkably smooth longward of the 
associated Ly$\alpha$ absorption (i.e., at $\lambda \gtrsim 1430$\,\AA)
where metal-lines in the quasar spectrum might be expected to cause
depressions in the continuum.  Still, there are some minor
undulations present at shorter wavelengths, and the possibility that
these are indicative of some unknown source of local continuum 
fluctuations cannot be ruled out with the present data alone.

Tripp et al.\ (1998) reported Ly$\alpha$ detections in their low-resolution,
high-$S/N$
GHRS spectrum of PG\,1116+215 for 5 of the 8
broad Ly$\alpha$ absorbers detected in our STIS spectrum.  At least
two of the three absorbers that were not identified ($z=0.06244$, 
$z=0.08587$, and $z=0.13370$) are present as very weak 
features in the GHRS spectrum (see Figure~\ref{compfig}), but had insufficient
statistical significance to be included in the line list of Tripp et al.\ (1998).  The detection of at least 7 of the 8 broad absorbers
in the two sets of spectra taken more than 3 years apart indicates
that the features are long-lived and not transient depressions in the 
quasar continuum.

A more conservative approach to calculating 
$\Omega_b({\rm BLy}\alpha)$ can be taken by restricting attention 
to only those cases where 
40 $\lesssim$ b $\lesssim$ 100 \kms.  This limits the sample 
of broad Ly$\alpha$ candidates to 6 absorbers and implies 
d$N$/d$z$ $\approx 37$.  This number per unit redshift
is  higher than  the 
value of 23 found by Richter
et al.\ (2004) for the 8 broad Ly$\alpha$ absorbers with $40\le$~b~$\le 100$
\kms\ 
toward PG\,1259+593; the weakest broad absorber included in their
estimate has $W_r = 45$\,m\AA, so the two samples have similar selection
criteria.
Here, we find $\Omega_b({\rm BLy}\alpha) \approx 0.0046\,h^{-1}_{75}$, 
which is still very 
large.  Richter et al.\ (2004) found a value of 
$\Omega_b({\rm BLy}\alpha) \approx  0.0031\,h^{-1}_{75}$ for the broad
Ly$\alpha$ absorbers
along the PG\,1259+593 sight line.  The baryon estimates for the 
broad Ly$\alpha$ absorbers along the two sight lines are 
comparable to the amount baryonic mass in stars and gas inside
galaxies -- $\Omega_b({\rm *, gas}) \approx  0.0032\,h^{-1}_{75}$
(Fukugita 2003; Fukugita et al.\ 1998).

\section{Absorber-Galaxy Relationships}

The high quality of the STIS and FUSE observations of PG1116+215
provide an unusual opportunity to study  weak intergalactic absorption
lines at low redshifts. These sensitive data allow 
studies of the relationships  (or lack thereof) 
between the absorption-line systems and
nearby galaxies, galaxy groups, and galaxy clusters.
Galaxy redshifts in the immediate vicinity of PG1116+215 have been
measured by Ellingson, Green, \& Yee (1991) and Tripp et al.\ (1998).  
Ellingson et al.\ (1991)
report redshifts of galaxies as faint
as $r = 21$, but their observations are confined to galaxies within
2.5\arcmin\ of the sight line.  The Tripp et al.\ (1998) survey does not go as deep
but covers a much larger field (including objects within $\sim$50\arcmin\
from the QSO). Tripp et al. estimate that their survey is $\sim78$\%
complete for objects with $B_J \leq$ 19 within 30\arcmin\ of the QSO.
The redshifts from Ellingson et al.\ (1991) are all at or beyond the
QSO redshift.

\begin{figure}[ht!]
\figurenum{21}
\includegraphics{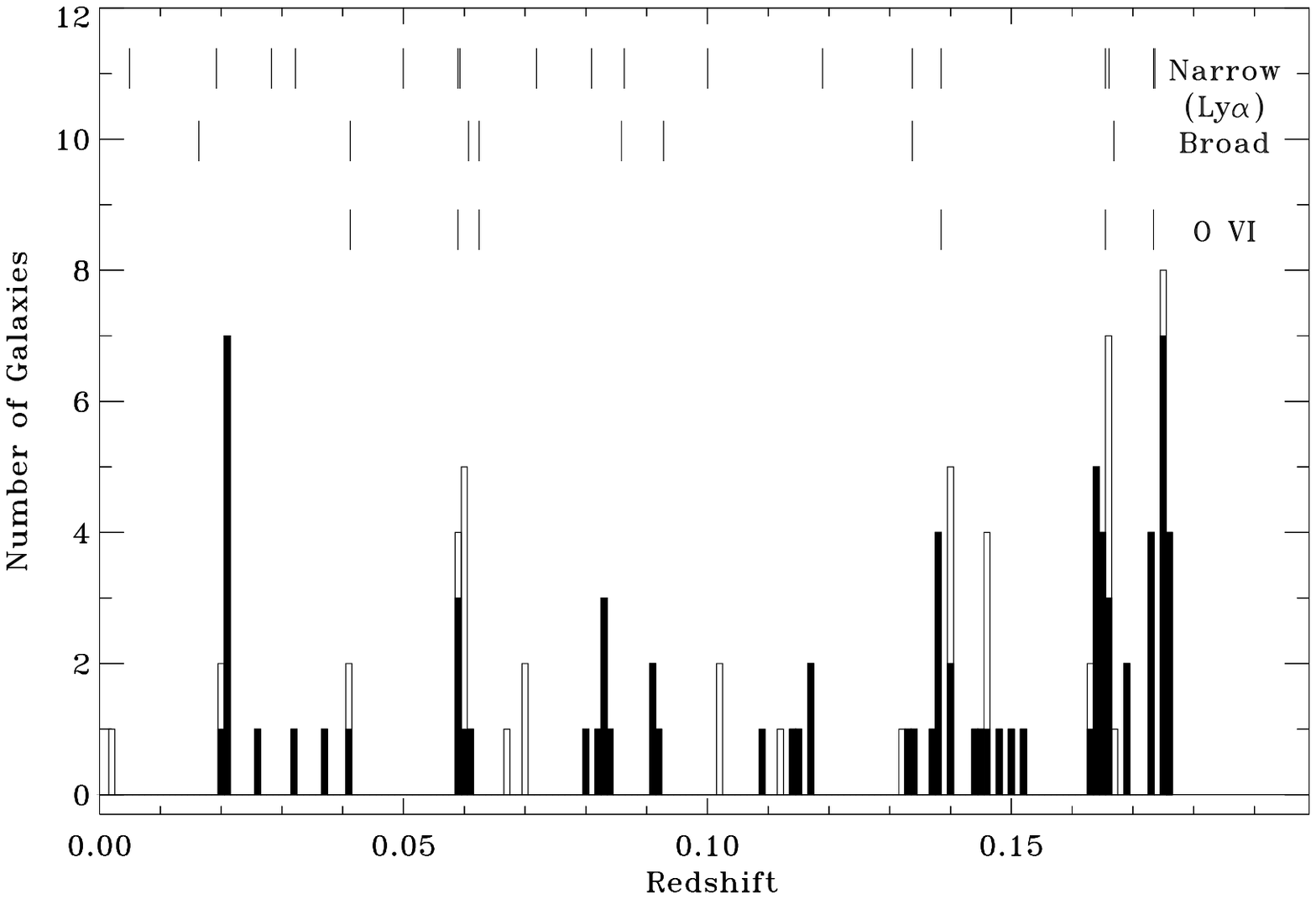}
\vspace{5.5in}
\caption{Histogram of 103 galaxy redshifts within 50\arcmin\ of
PG\,1116+215 at $z < 0.1765$.  The histogram bins have a width
$\Delta z = 0.001$.  Galaxies within 30\arcmin\ are highlighted with 
filled bins.  The redshifts of the Ly$\alpha$ and \ion{O}{6} 
absorption systems observed along the sight line are indicated by vertical
tick marks above the histogram.  The Ly$\alpha$ absorbers are separated into
narrow (${\rm b} \lesssim 40$ \kms) and broad (${\rm b} \gtrsim 40$ \kms) 
absorbers.  The tick marks for the $z=0.05895,0.05928$ and $z=0.17340,0.17360$
pairs of lines blend together.\label{galaxyhist}}
\end{figure}

\begin{figure}
\figurenum{22}
\epsscale{0.8}
\plotone{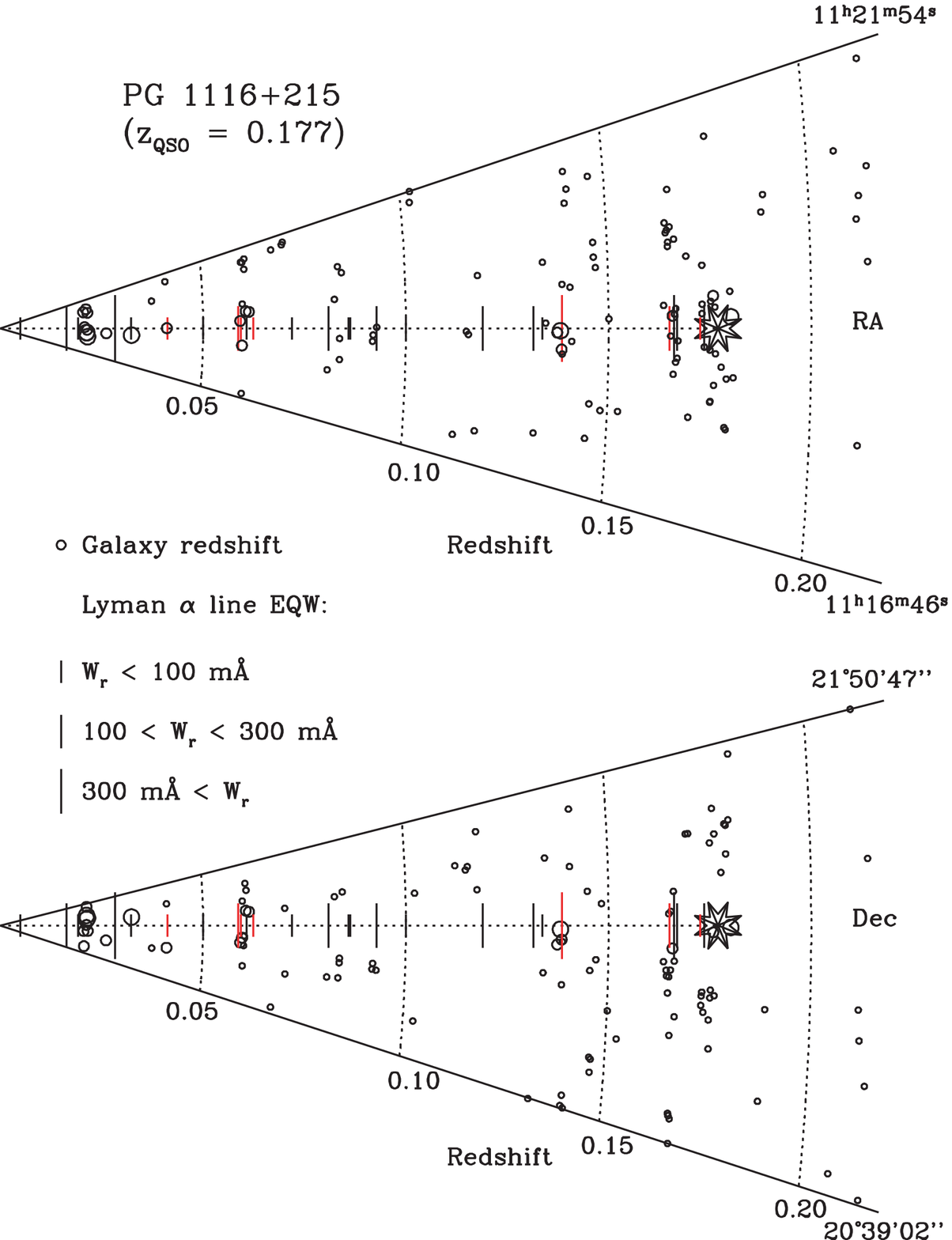}
\caption{Comparison of the galaxy distribution within 50\arcmin\ of
PG\,1116+215 to the redshifts of the Ly$\alpha$ and \ion{O}{6} absorbers
measured in the STIS and FUSE spectra presented in this
paper. Galaxies are indicated with open circles in a right ascension 
slice (top) and a declination slice (bottom). The size of the circle indicates
proximity to the sight line; the largest circles show galaxies with
projected distance $\rho \leq 500$ kpc, intermediate-size circles
represent galaxies at 500 $< \rho \leq 1$ Mpc, and the smallest
circles indicate galaxies at $\rho > 1$ Mpc. Redshifts of Ly$\alpha$
lines are plotted with vertical lines on the sight line, and the
length of the vertical line reflects the \ion{H}{1} Ly$\alpha$ 
equivalent width as
indicated in the key.  The redshifts of the Ly$\alpha$ absorbers
with corresponding \ion{O}{6} are highlighted in red.\label{galaxywedge}}
\end{figure}

Figures~\ref{galaxyhist} and \ref{galaxywedge} compare the redshift
distributions of the Ly$\alpha$ and \ion{O}{6} absorbers (including
the weaker \ion{O}{6} lines identified at $z = 0.04125$, 0.05895,
0.05928, and 0.06244) to the distribution of the galaxies along the
line-of-sight to PG1116+215. Figure~\ref{galaxyhist} presents
histograms of the number of galaxies within 50\arcmin\ and
30\arcmin\ of the sight line as a function of redshift.
Figure~\ref{galaxywedge} shows the locations of the galaxies in right
ascension and declination slices versus galaxy redshift (circles), as well
as the redshifts of the Ly$\alpha$ and \ion{O}{6} absorption
lines. The vertical lines of varying size plotted on the sight line in
each slice of Figure~\ref{galaxywedge} show the Ly$\alpha$ redshifts;
the line size indicates the Ly$\alpha$ equivalent width following the
legend in the figure. The redshifts of the Ly$\alpha$ absorbers 
containing \ion{O}{6} are highlighted in red.
The largest circles in
Figure~\ref{galaxywedge} represent galaxies within 500 kpc in
projection from the sight line,\footnote{For ease of comparison with
projected distances reported in Tripp et al. (1998), including
additional galaxies near absorbers of interest not listed in
Table~\ref{nearestgal}, for the calculation of projected distances we
assume the same cosmological parameters adopted by Tripp et
al. (1998): H$_{0}$ = 75 km s$^{-1}$ Mpc$^{-1}$ and $q_{0}$ = 0.}
intermediate-size circles indicate galaxies with projected distances
($\rho$) ranging from 500 kpc to 1 Mpc, and small circles show
galaxies at $\rho >$ 1 Mpc. For the \ion{O}{6} absorbers,
Table~\ref{nearestgal} lists the galaxy with the smallest projected
distance to the sight line and smallest velocity difference from the
absorber.

Tripp et al.\ (1998) have shown that Ly$\alpha$ absorption-line systems
are significantly correlated with galaxies in the direction of
PG1116+215. Visual inspection of Figures~\ref{galaxyhist} -
\ref{galaxywedge} and Table~\ref{nearestgal} suggests that the
\ion{O}{6} absorbers are strongly correlated with galaxies as
well. For each of the seven \ion{O}{6} absorption lines identified in
the PG1116+215 spectrum, we find at least one galaxy within $\rho
\leq$ 750 kpc and $\Delta v \leq 600$ km s$^{-1}$; six out of the
seven \ion{O}{6} systems have $\Delta v \leq$ 320 km s$^{-1}$. In
addition, most of the \ion{O}{6} absorbers are found close to the
peaks in the galaxy distribution.  It is interesting that no
\ion{O}{6} is positively identified near the prominent group of
galaxies at $z \approx$ 0.0212; however, \ion{O}{6} at this redshift
falls in a region of the FUSE spectrum that is heavily blocked
by unrelated lines (see Figure~\ref{fusespec}), so it remains possible that
\ion{O}{6} is simply hidden by blending in this case.

We have considered the possibility that our visual impression is just
coincidental and that the \ion{O}{6} absorbers are actually randomly
distributed with respect to the galaxies in the direction of
PG\,11116+215. To quantitatively assess this probability, we performed
the following simple Monte Carlo statistical test.  We carried
out a large number of simulations in which we randomly
distributed a number of absorbers along the sight line equal to the
observed number of \ion{O}{6} systems, and then we determined the
number of instances in which the simulations by chance showed the same
number of absorbers as observed with $\rho \leq$ 750 kpc and $\Delta v
\leq$ 350 \kms. Since associated absorption systems with
$z \approx z_{\rm QSO}$ sometimes show strong evidence that
the gas is close to the central engine of the QSO (e.g., Hamann \&
Ferland 1999; Yuan et al.\ 2002; Ganguly et al.\ 2003) and therefore is
unrelated to the intervening absorbers, we excluded the two \ion{O}{6}
absorbers observed toward PG1116+215 within 5000 \kms\ of the
QSO redshift from our first Monte Carlo analysis, and we set the
maximum redshift for the random simulations to 5000 \kms\ less
than $z_{\rm QSO}$. We also treated the \ion{O}{6} lines at 
$z =$ 0.05895 and 0.05928 as a single ``absorber'' since these lines
are separated by only $\sim$90 \kms\ and therefore could
easily arise in a single absorbering entity (e.g., a single
galaxy).\footnote{Indeed, our criteria for matching galaxies and
\ion{O}{6} absorbers in Table~\ref{nearestgal} results in both of
these \ion{O}{6} lines being assigned to the same galaxy. It should be
noted, though, that many of the \ion{O}{6} lines in
Table~\ref{nearestgal} have other galaxies at comparable but slightly
larger $\rho$ and/or $\Delta v$, and it is not always clear which 
single galaxy, if any, should be matched with an absorber.  This
caveat applies to the \ion{O}{6} lines at $z$ = 0.05895,
0.05928 (see Table~2 in Tripp et al.\ 1998), and these lines could each
be associated with different galaxies.} Finally, we required all of
the random absorbers in the simulations to have $z \geq 0.02430$
 because below that redshift, the FUSE spectrum is
strongly blanketed with unrelated lines (which could easily hide weak
\ion{O}{6} systems) and the physical radius covered by the galaxy
redshift surveys is limited.  With these exclusions, we have four
intervening \ion{O}{6} absorbers in the PG\,1116+215 spectrum (at
$z$ = 0.04125, 0.059, 0.06244, and 0.13847). After running a
total of $4 \times 10^{5}$ Monte Carlo trials, we found 81 trials in
which the randomly distributed systems showed the same number of
absorbers within $\rho =$ 750 kpc and $\Delta v$ = 350 km s$^{-1}$ as
the actually observed number. This indicates that the probability that
the observed intervening \ion{O}{6} absorbers toward PG\,1116+215 are
randomly distributed with respect to 
the galaxies is 81/($4 \times 10^{5}) = 2.0
\times 10^{-4}$. We conclude that it is highly unlikely that the
intervening \ion{O}{6} absorbers are randomly distributed with respect
to galaxies in the direction of PG1116+215.

However, it is striking how well the two ``associated'' \ion{O}{6} lines
toward PG\,1116+215 line up with two of the most pronounced peaks in the
galaxy distribution, so one might reasonably 
argue that the \ion{O}{6} absorbers at $z=0.16548$ and $z=0.17340$, despite
their proximity to the QSO redshift, are also
intervening. Consequently we have also carried out a Monte Carlo
calculation in which the \ion{O}{6} lines within 5000 km s$^{-1}$ of
$z_{\rm QSO}$ were included in the count, and the random distributions
were allowed to extend all of the way to the QSO redshift. In this
second set of Monte Carlo trials we found that the probability that
the \ion{O}{6} lines are randomly distributed was even slightly
smaller, $1.3 \times 10^{-4}$. Therefore, our conclusion holds with
this set of assumptions as well, and as we shall see in the next section
our conclusions about baryonic content of the absorbers do not 
depend strongly on whether the velocity interval near the QSO is 
included in our calculations of the redshift path or $\Omega_b$(\ion{O}{6}).

Finally, we note that most of the Ly$\alpha$ absorbers along the sight line 
also occur at redshifts where there are peaks in the galaxy redshift 
distribution.  Notable exceptions include the somce of the 
closer Ly$\alpha$ absorbers at $z=0.00493$, $0.01635$, and $0.04996$.  At 
these lower redshifts, the galaxy redshift surveys cover a smaller volume of 
space than at higher redshifts, so the absence of galaxies at these redshifts 
will need to be tested more thoroughly.  The generally good correlation of
both Ly$\alpha$ and \ion{O}{6} 
with galaxy redshifts is a strong indication that the absorbers are 
truly intervening material rather than material ejected from QSOs 
at high velocities (e.g., Richards et al. 1999).

\section{Baryonic Content of the Warm-Hot Intergalactic Medium}

The detection of broad Ly$\alpha$ lines and \ion{O}{6} in the IGM
demonstrates that there may be multiple baryon reservoirs 
in the low-redshift universe.  Previous investigations have found that 
(d$N$/d$z$)$_{\rm O\,VI} \approx 14\pm^9_6$ for absorbers with 
$W_r > 50$\,m\AA\ over an unblocked redshift path of 0.43 
(Savage et al.\, 2002, and references therein).  Toward PG\,1116+215, 
we expect two and detect two \ion{O}{6}
absorbers with $W_r > 50$\,m\AA\ over an unblocked redshift interval of 
$\Delta X_{\rm O\,{\sc VI}} = 0.117$, which corresponds to 
(d$N$/d$z$)$_{\rm O\,VI} \approx 17$.
Adding in the weaker \ion{O}{6} absorbers along the sight line
increases (d$N$/d$z$)$_{\rm O\,VI}$ to $\sim60$ for an equivalent width 
threshold $W_r \ge 30$\,m\AA. The relatively high $S/N$ observations of 
H\,1821+643 (Tripp et al.\ 2000) similarly indicate that d$N$/d$z$ 
increases significantly with increasing sensitivity to weak lines.

The baryonic content of the \ion{O}{6} systems is given by

\begin{equation}
\Omega_b({\rm O\,VI}) = \frac{\mu_H~m_{\rm H}~H_0}{\rho_c~c}~\sum~\frac{{\rm N}_i{\rm (O\,VI)}}{f_{_i{\rm O\,VI}}~\Delta X_i~({\rm O/H})_i}~.
\end{equation}

\noindent
In this equation, $\Omega_b({\rm O\,VI})$ is inversely proportional to the 
ionization fraction of \ion{O}{6} and the metallicity of the gas.
Savage et al.\ (2002) derived a 
value of $\Omega_b$(\ion{O}{6}) $>0.002h^{-1}_{75}$ assuming $f_{\rm O\,VI} <
0.2$ and a metallicity of 0.1 solar for systems with $W_r > 50$\,m\AA.  This 
estimate is comparable to the baryonic mass contained in stars and gas inside 
galaxies, $\Omega_b({\rm *, gas}) \approx  0.0032\,h^{-1}_{75}$
(Fukugita 2003; Fukugita et al.\ 1998), and may rival the amount contained
in the broad Ly$\alpha$ absorbers (see \S9.2).  
While we 
cannot rule out a metallicity of 0.1 solar for the two stronger \ion{O}{6}
absorbers toward
PG\,1116+215, the observational constraints are more readily satisfied 
for metallicities greater than $\sim0.5$ solar.  As a result, the 
baryonic estimate might decrease slightly if the \ion{O}{6} systems along
other sight lines have similar metallicities.  We note that Savage et al.\
(2002) favor a metallicity of $\sim0.4$ solar for the $z=0.06807$ \ion{O}{6}
absorber toward PG\,0953+415.
The ionization fraction of \ion{O}{6}
rarely exceeds 0.2 (see Tripp \& Savage 2000;
Sembach et al.\ 2003), so the above estimate for
$\Omega_b$(\ion{O}{6}) also has a strong dependence on the value of 
$f_{\rm O\,VI}$.

A summary of the baryonic content of the broad Ly$\alpha$ and \ion{O}{6}
absorbers can be found in Table~\ref{tab_baryonsummary}.  For both types of 
absorbers, we list values of $\Omega_b$ and $\Omega_b^\prime$.  The 
baryonic content does not depend strongly on whether or not the 5000 \kms\
interval nearest PG\,1116+215 is included in the calculations.
For the Ly$\alpha$ absorbers, we list the values of $\Omega_b$(BLy$\alpha$) 
for all of the broad lines observed (${\rm b} \gtrsim 40$\,\kms)
as well as the values for the more restricted line width interval of
$40 \lesssim {\rm b} \lesssim 100$\,\kms\ considered in \S9.2.
\ion{O}{6} values for two limiting equivalent widths ($W_r \ge
30$ and 50\,m\AA) are given.  We also tabulate an average value 
of $\Omega_b$(\ion{O}{6}) $\gtrsim 0.0027 h_{75}^{-1}$ for 13 absorbers
with $W_r \ge 50$\,m\AA\ along 6 lines of sight assuming a metallicity 
of 0.1 solar and an \ion{O}{6} ionization fraction $f_{\rm O\,VI} <0.2$.
References for the other lines of sight used in this summary 
can be found in the notes for the table.

Different types of \ion{O}{6} systems may have different physical
 properties.  The two primary \ion{O}{6} absorbers toward
PG\,1116+215 at $z=0.13847$ and $z=0.16548$ clearly have very
different ionization structures.  Photoionization may play a role in the 
production of \ion{O}{6} in these systems, though it cannot be the 
primary ionization mechanism if the \ion{O}{6} and lower ionization 
stages observed are co-spatial as implied by the similarities in their
kinematics.  
The \ion{O}{6} absorption associated
with the broad Ly$\alpha$ absorbers at $z=0.04125$ and $0.06244$ 
may trace yet another type of environment.  For these systems, the Ly$\alpha$
and \ion{O}{6} line shapes are consistent with thermal broadening in 
hot gas.  It is particularly interesting that the \ion{O}{6} line widths in
the $z=0.06244$ system and in the $z=0.31978$ absorber toward PG\,1259+593
are narrow and in roughly the right proportion to the \ion{H}{1} line
widths to be due to thermal broadening in hot, collisionally ionized
gas.  If instead the lines had been produced by photoinization, the
cloud sizes required to reproduce the \ion{O}{6} columns 
would have been large ($\gtrsim0.5$ Mpc), and the lines 
would have been broadened considerably by the Hubble flow.
 The detection of \ion{H}{1} in high temperature gas has profound 
implications for the baryonic content of the gas since the amount of 
related ionized gas implied is very large indeed (see \S9.2). 

Obtaining high quality data of the type used in this study for other sight 
lines is essential for revealing weak features like the \ion{O}{6} 
absorption in these broad Ly$\alpha$ systems.
Quantifying the baryonic content of the \ion{O}{6} and broad Ly$\alpha$
absorbers more precisely will require estimates of the \ion{O}{6}
ionization fractions ($f_i$) and gas metallicities (O/H) 
in the different types
of systems rather than assuming common values for all cases.  
There are two measurements that are particularly
valuable for determinations of the \ion{O}{6} absorber properties 
and constraining the value of $(f_{\rm O\,VI})~({\rm O/H})$.
The first is 
an accurate assessment of the strength and kinematics of
\ion{C}{4} absorption in these systems.  The ratio N(\ion{O}{6})/N(\ion{C}{4})
is a powerful diagnostic of the primary ionization mechanism for the 
highly ionized gas.  When the kinematical information in the profiles 
indicate that a direct comparison of column densities of the two species is 
meaningful, the ratio can be used to discriminate between collisional
ionization and photoionization.  For PG\,1116+215, the existing HST/STIS E230M 
spectrum does not 
cover the \ion{C}{4} absorption associated with the two strong \ion{O}{6}
absorbers at $z=0.13847$ and $z=0.16548$.  The STIS/E140M covers some of the 
lower redshift absorbers  but the data are not of sufficient
$S/N$ to detect the \ion{C}{4} lines in most cases.  In the one case where 
\ion{C}{4} is possibly detected (e.g., $z=0.05928$), the comparable strengths 
of the \ion{O}{6} and \ion{C}{4} lines supports  photoionization as the
preferred ionization mechanism (see \S8.2).
The second type of measurement is an 
observation of the extreme-ultraviolet lines of \ion{O}{3} $\lambda832.927$ and
\ion{O}{4} $\lambda787.711$, and 
\ion{O}{5} $\lambda629.730$
for those \ion{O}{6} absorbers at redshifts sufficiently high to place the 
lines in the wavelength range covered by  FUSE 
($\lambda > 912$\,\AA, $z \gtrsim 0.2-0.4$).
Having access to several stages of ionization may allow determinations of 
both the ionization and metallicity of the gas.  Additional extreme-ultraviolet
diagnostics such as \ion{Ne}{8} $\lambda\lambda770.409, 780.324$
and \ion{Mg}{10} $\lambda\lambda609.790, 624.950$ are also valuable 
in assessing whether the gas is collisionally ionized at high temperatures
(see Savage et al. 2004).
Unfortunately, none of the \ion{O}{6}
absorbers toward PG\,1116+215 occur at redshifts high enough to observe
the EUV lines with the FUSE LiF channels or with STIS.

Many of the low-redshift \ion{O}{6} absorption systems are clearly multi-phase
in nature. This conclusion holds for the  $z=0.13847$ absorber toward
PG\,1116+215 as well as absorbers toward H\,1821+643, (Tripp et al.\ 2000),
PG\,1259+593 (Richter et al.\ 2004), and possibly PG\,0953+415 
(Savage et al.\ 2002).  Multiple ionization  mechanisms are required
to explain the observed amounts of \ion{O}{6} and 
lower ionization stages in the gas.  The PG\,1116+215 $z=0.13847$ absorber
is particularly interesting in this regard since it has the highest 
\ion{H}{1} content of the low-redshift \ion{O}{6} absorbers studied 
to date and yet has an \ion{O}{6} column density comparable to those
of many of the other systems.  As mentioned in \S6.3, its ionization
properties bears some 
resemblance to those of ionized high-velocity clouds near the Milky Way.
Gnat \& Sternberg (2004) have modeled the ionization pattern expected for 
photoionized dark matter dominated mini-halos in the Local Group 
(see also Kepner et al.\ 1999) and 
have suggested that some of the ionized high
velocity gas near the Milky Way could occur in the outer envelopes of low 
surface brightness dwarf galaxies.  Their model does not produce significant 
amounts of \ion{O}{6}, which must still be produced by other means.
Locally, the \ion{O}{6} could occur in warm-hot gas interfaces caused by
interactions of the high-velocity gas with a hot ($T\sim 10^6$\,K) 
low-density ($n\lesssim 10^{-4}$ cm$^{-3}$) Galactic corona or 
Local Group medium (Sembach et al.\ 2003).  
Similar interactions of warm and hot gases are 
expected to occur as the IGM evolves, especially in regions where galaxies
have formed.

We conclude by noting that our results demonstrate that broad Ly$\alpha$ 
and \ion{O}{6} absorbers could be substantial 
reservoirs of baryons in the low-redshift universe. In some cases, 
these may be one and the same.   Narrow 
Ly$\alpha$ absorbers also contribute significantly to the total 
baryon budget. For example, 
Penton et al.\ (2004) have estimated that the baryonic content of the 
Ly$\alpha$ clouds in the low-redshift universe is $\approx40\pm10$\%
of the total baryonic mass, under the assumption that the Ly$\alpha$
clouds are photoionized.  Understanding the ionization of the Ly$\alpha$ 
and \ion{O}{6} clouds will lead to a more rigorous accounting of the baryon
content of the clouds.  Comsological hydrodynamic simulations
predict that stronger Ly$\alpha$ absorbers should be more strongly 
correlated with galaxies than weaker absorbers, and that there should be both
photoionized and collisionally ionized systems (Dav\'e et al.\ 1999).  
Quantifying these statements both observationally and theoretically will 
be a major challenge for astronomers in the coming years.

\section{Summary}

The results of our HST/STIS and FUSE study of the Ly$\alpha$ 
and \ion{O}{6} absorption systems along the PG\,1116+215 sight line 
are as follows.

\noindent
1) We detect 25 Ly$\alpha$ absorbers along the sight line 
at rest-frame equivalent widths $W_r \gtrsim 30$\,m\AA, yielding
(d$N$/d$z$)$_{{\rm Ly}\alpha} = 
154\pm18$  over an 
unblocked redshift path $\Delta X_{{\rm Ly}\alpha} = 0.162$.   
We also detect two 
weak Ly$\alpha$ features with $W_r =15-20$\,m\AA.  Most, if not
all, of these are intergalactic systems.  The metal-line
absorption systems at $z\approx0.166$ and $z = 0.17340-0.17360$ could 
be associated with the quasar, but we believe that it is just as likely 
that they 
are truly intervening systems since they occur at redshifts corresponding
to peaks in the galaxy redshift distribution along the sight line.

\noindent
2) Some of the Ly$\alpha$ absorbers have broad line widths (b $\gtrsim
40$ \kms).  The detection of narrow \ion{O}{6} absorption in at least 
one broad Ly$\alpha$ absorber along the sight line ($z=0.06244$) supports 
the idea that thermal broadening in hot gas could account for much of the 
breadth of the \ion{H}{1} lines.
If the \ion{H}{1} line widths are dominated by thermal broadening at 
$T > 10^5$\,K, the amount of baryonic material in these 
absorbers is enormous because the ionization corrections required to account 
for the observed \ion{H}{1} columns are large.  

\noindent
3) We find d$N$/d$z$ $\approx 50$ for broad Ly$\alpha$ absorbers with 
$W_r \gtrsim 30$\,m\AA\ and ${\rm b} \ge 40$ \kms.  This number drops to 
d$N$/d$z$ $\approx 37$ if the line widths are restricted to 
$40 \le {\rm b} \le 100$ \kms.
As much as half the baryonic mass in the low-redshift universe could be 
contained in broad collisionally-ionized Ly$\alpha$ systems.

\noindent
4) We detect \ion{O}{6} absorption in several of the Ly$\alpha$ clouds along 
the sight line.  Two detections at $z=0.13847$ and $z=0.16548$ are 
confirmed by the presence of other ions at these redshifts, while the 
detections at $z=0.04125$, $0.05895$, $0.05928$, and $0.06244$ are based upon 
the Ly$\alpha$ and \ion{O}{6} detections alone.  We find 
(d$N$/d$z$)$_{\rm O\,VI} \approx 17$ for \ion{O}{6} absorbers
line with
$W_r > 50$\,m\AA\ toward PG\,1116+215.  

\noindent
5) Compiling the information available for 13 low-redshift \ion{O}{6} absorbers
with $W_r \ge 50$\,m\AA\ along 5 sight lines yields 
(d$N$/d$z$)$_{\rm O\,VI} \approx 14$ and $\Omega_b$(\ion{O}{6}) $\gtrsim0.0027 h_{75}^{-1}$, assuming a metallicity of 0.1 solar and an \ion{O}{6} 
ionization fraction $f_{\rm O\,VI} \le 0.2$.
The implied baryonic content of the \ion{O}{6} absorbers is comparable to 
the baryonic mass contained in stars and gas inside galaxies.  

\noindent
6) We detect metal-line absorption in the $z=0.13847$ system.  Species 
detected include \ion{H}{1}, \ion{C}{2-III}, \ion{N}{2-III}, \ion{N}{5},
\ion{O}{1}, \ion{O}{6}, and \ion{Si}{2-IV}.
The numerous Lyman-series lines and weak Lyman limit allow us to 
estimate N(\ion{H}{1}) = $(1.57\pm^{0.18}_{0.14})\times10^{16}$ 
cm$^{-2}$ in this
system.  Investigation of the ionization of this system indicates
that the absorber is multi-phase in nature. The presence of \ion{O}{6} 
likely requires collisional ionization, while the ionization pattern 
seen in the lower ionization stages can be explained in part by 
photoionization in a low-density plasma.  Regardless of the ionization 
mechanism, the absorber is composed mainly of ionized gas.

\noindent
7) We detect a group of Ly$\alpha$ lines near $z=0.166$.  Three absorbers
within a velocity interval of 500 \kms\ contribute to the majority of the
overall Ly$\alpha$ absorption.  Additional weak Ly$\alpha$ components are 
also present in this velocity interval.

\noindent
8) The $z=0.16548$ Ly$\alpha$ absorber contains \ion{O}{6} and possibly
\ion{N}{5}.  The \ion{O}{6} absorption appears to consist of at least
three components, two of which are narrow (b $< 10$ \kms).  
If the gas is photoionized, a high ionization parameter is required 
($\log U \approx -0.5$, $n_{\rm H} \sim 10^{-6}$ cm$^{-3}$) and the cloud must
be large ($L \sim 1$ Mpc).  If collisional processes also contribute
to the ionization of the cloud, these constraints can be relaxed.  This 
absorber resembles the $z=0.14232$ absorber toward PG\,0953+415.  
Observations of \ion{C}{4} would help to solidify statements 
about the ionization properties of this absorber.

\noindent
9) The $z=0.16610$ Ly$\alpha$ absorber contains \ion{C}{2}, \ion{C}{3},
and \ion{Si}{2}.  Its ionization properties are marginally consistent with 
photoionization in a moderate ionization plasma ($\log U \approx -2.63$,
$n_{\rm H} \sim 1.6\times10^{-4}$ cm$^{-3}$).  Increasing the relative 
abundance of silicon to carbon compared to the solar ratio would improve 
the agreement of the observed and predicted column densities of 
\ion{C}{2} and \ion{Si}{2}.  

\noindent
10) We find a strong correlation of the redshifts of the Ly$\alpha$ and
\ion{O}{6} absorbers with the redshifts of galaxies along the sight line.
Monte-Carlo simulations indicate that the  probability of a random distribution
of \ion{O}{6} absorbers is $<2\times10^{-4}$.  None of the intervening
\ion{O}{6} absorbers is nearer than $\sim100\,h_{75}^{-1}$ kpc to a known
galaxy, and most are farther than $\sim600\,h_{75}^{-1}$.  This  suggests
that the \ion{O}{6} arises in intragroup gas rather than in the extended
halos of large galaxies.

\noindent
11) A few Ly$\alpha$ clouds along the sight line lie at redshifts where 
no galaxies have yet been detected ($z=0.00493$, $0.01635$, $0.04996$).  
Deeper galaxy searches at these and other redshifts would help to further
elucidate the relationship (or lack thereof) of the absorbers to galaxies.

\acknowledgments
We thank the FUSE and HST mission operations teams for their dedicated 
efforts to keep spectroscopic resources in space available to the 
astronomical community.
Partial financial support has been provided to KRS 
by NASA contract NAS5-32985 and to TMT by NASA through
Long Term Space Astrophysics grant NNG04GG73G.
PR is supported by the Deutsche 
Forschungsgemeinschaft.

\clearpage
\newpage

\clearpage
\newpage
\begin{figure}[ht!]
\figurenum{2}
\includegraphics{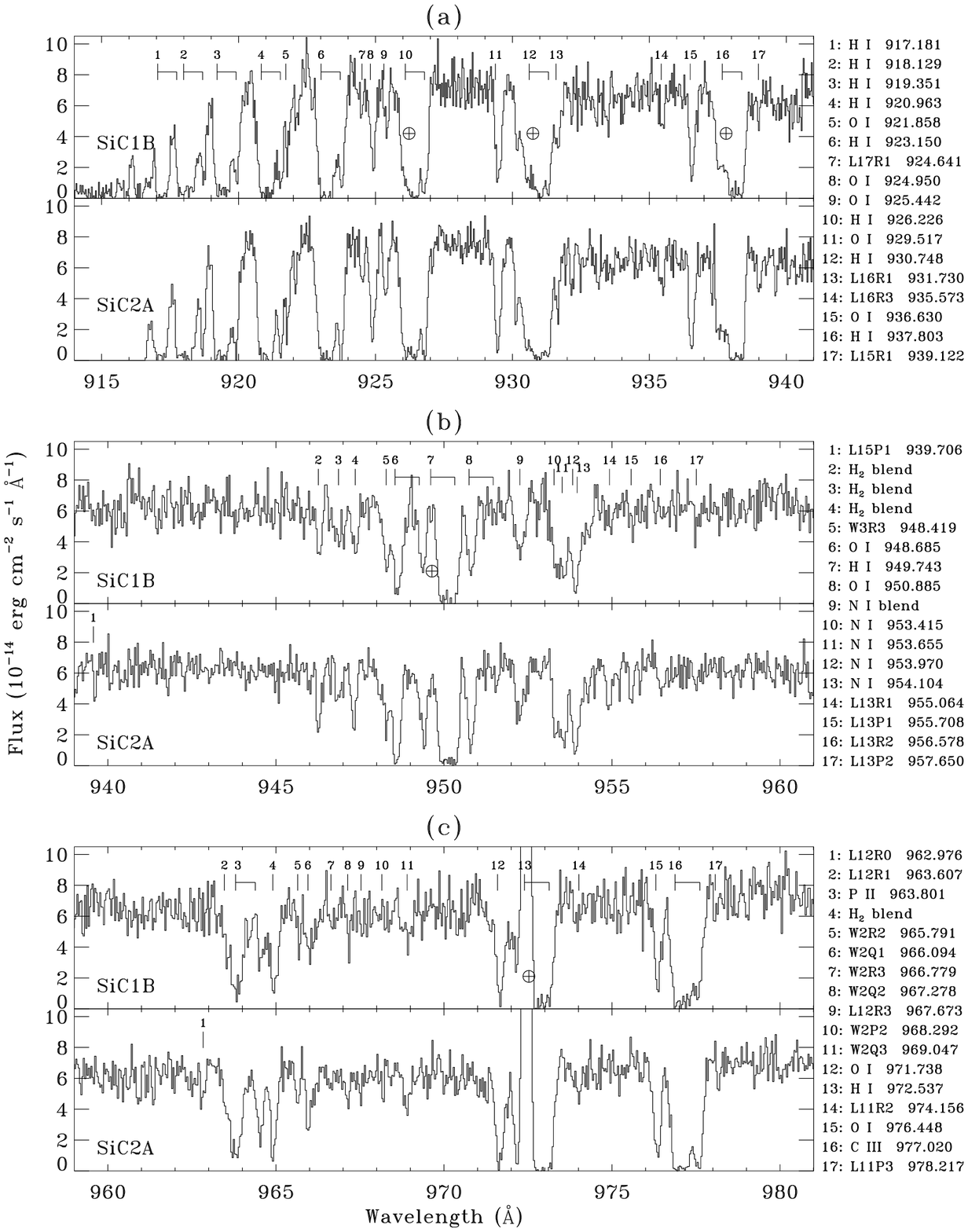}
\vspace{8.75in}
\caption{See caption at end of figure.\label{fusespec}}
\end{figure}

\clearpage
\newpage
\begin{figure}[ht!]
\figurenum{2 (continued)}
\includegraphics{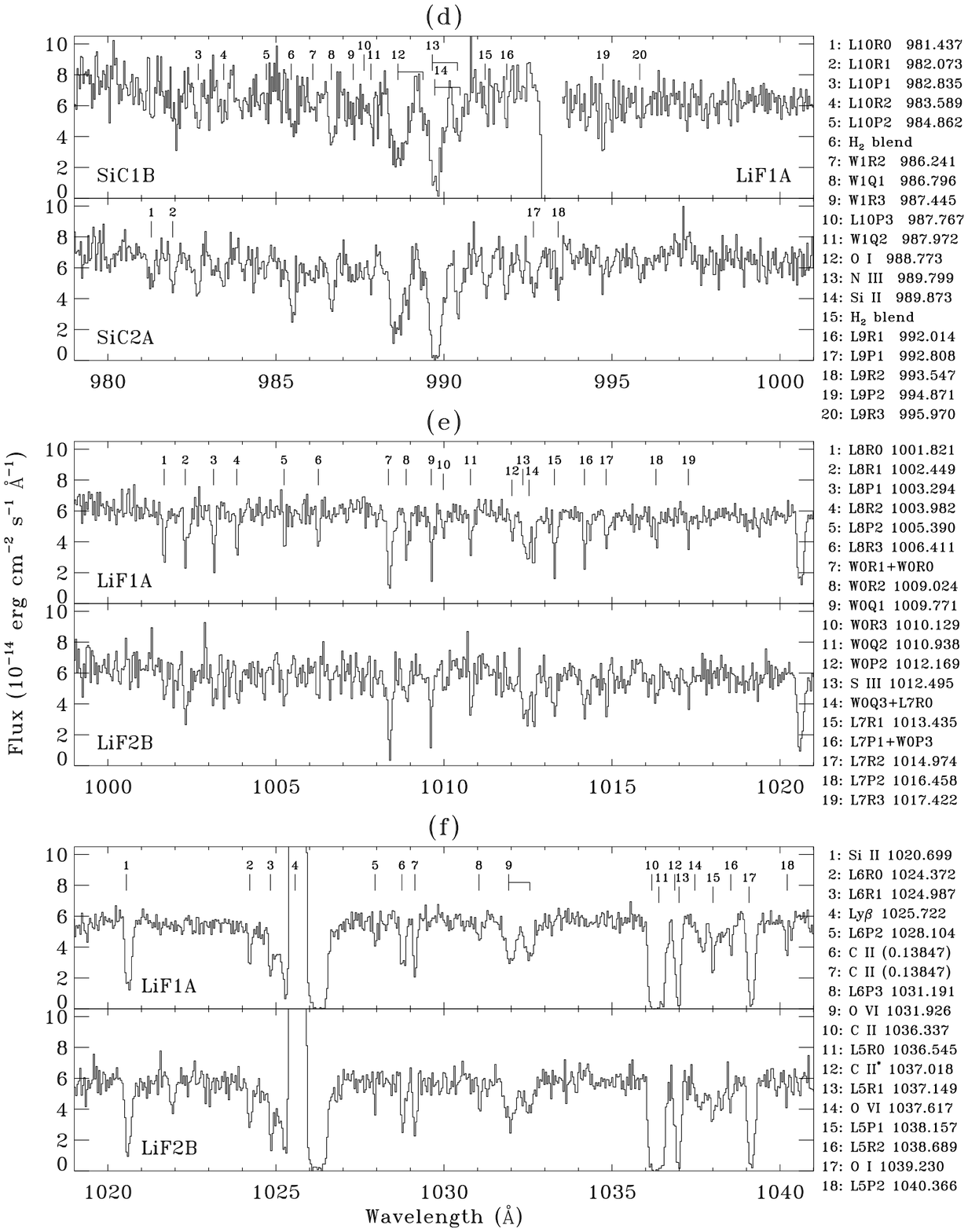}
\vspace{8.75in}
\caption{See caption at end of figure.}
\end{figure}

\clearpage
\newpage
\begin{figure}[ht!]
\figurenum{2 (continued)}
\includegraphics{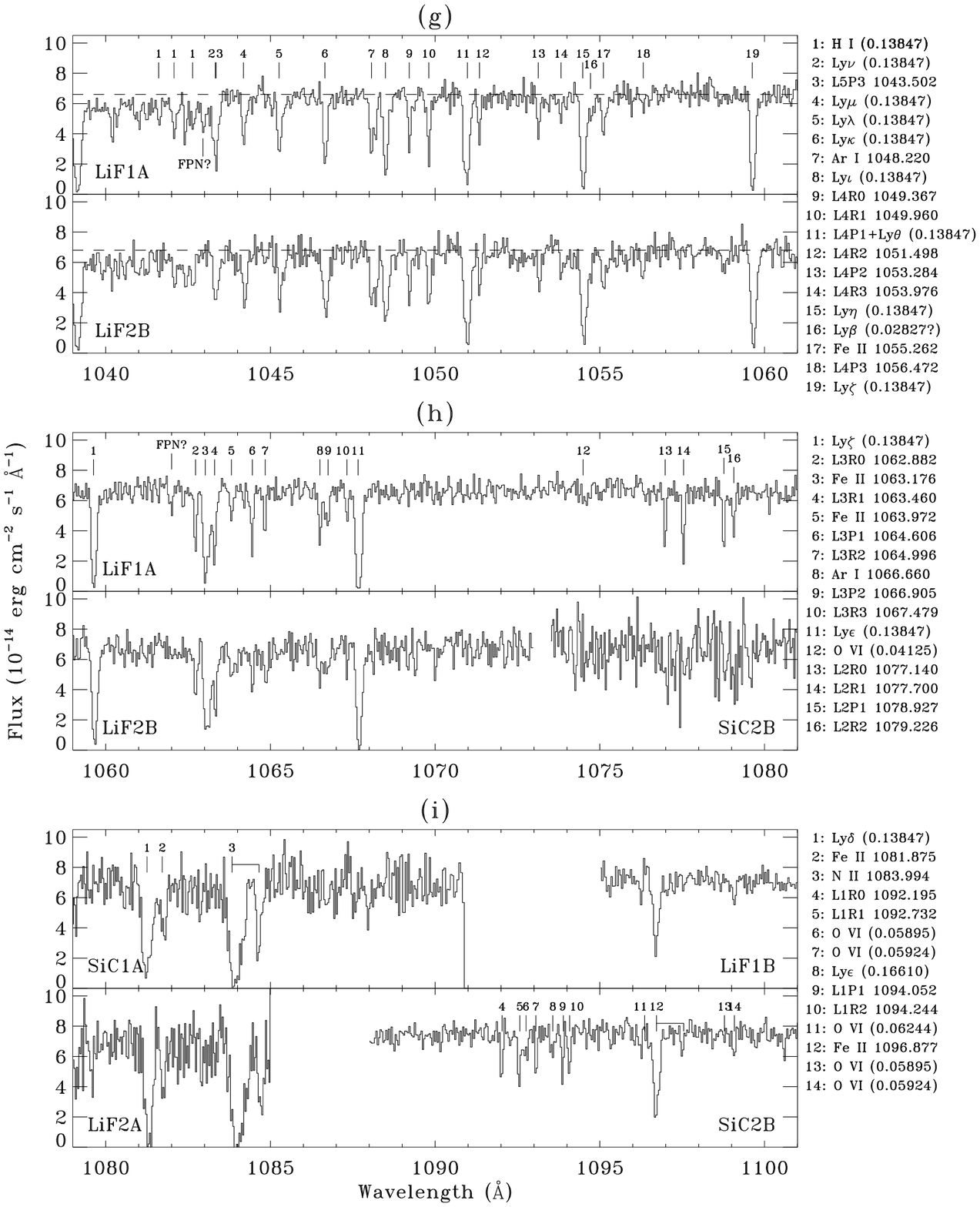}
\vspace{8.75in}
\caption{See caption at end of figure.}
\end{figure}

\clearpage
\newpage
\begin{figure}[ht!]
\figurenum{2 (continued)}
\includegraphics{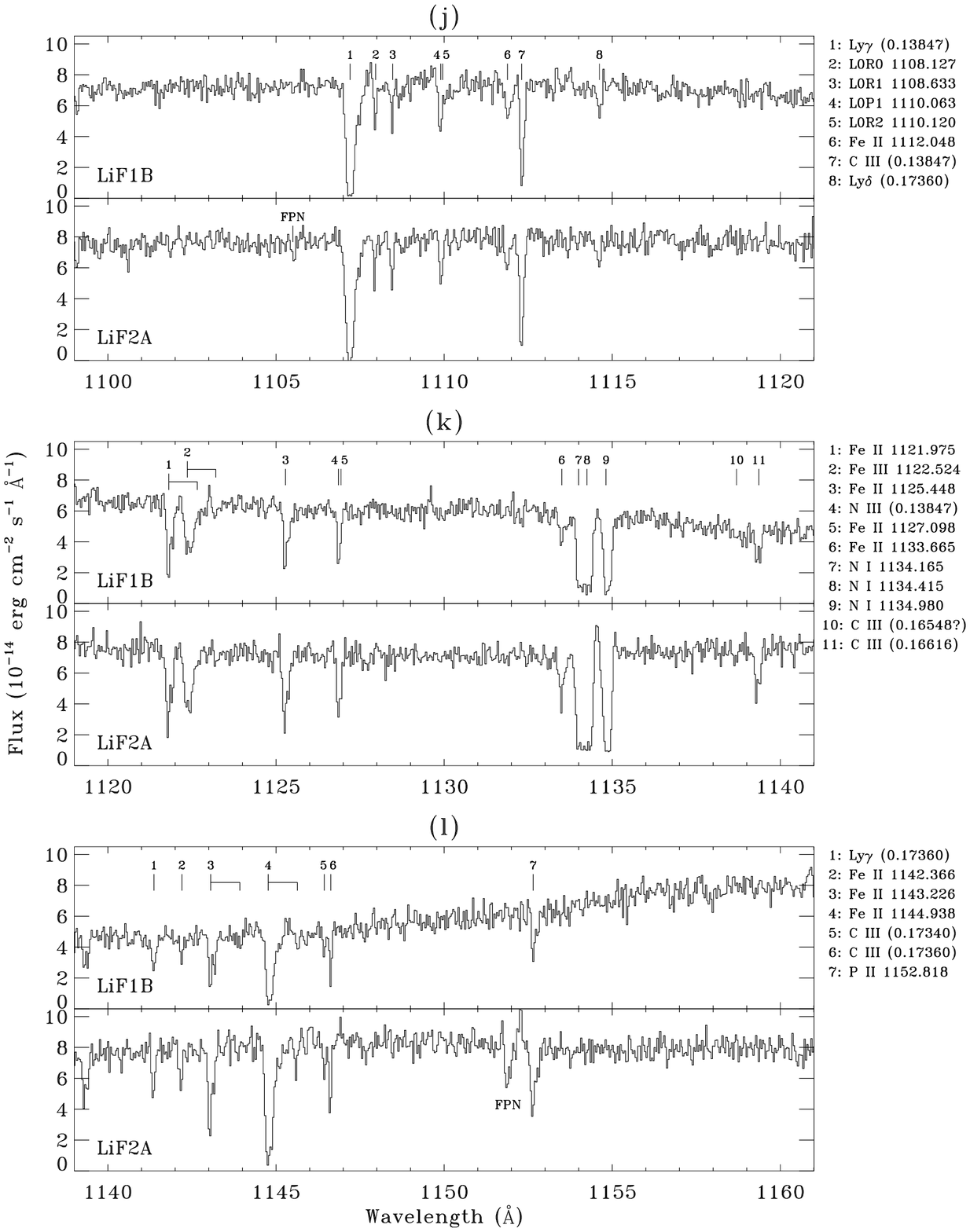}
\vspace{8.75in}
\caption{See caption at end of figure.}
\end{figure}

\clearpage
\newpage
\begin{figure}[ht!]
\figurenum{2 (continued)}
\includegraphics{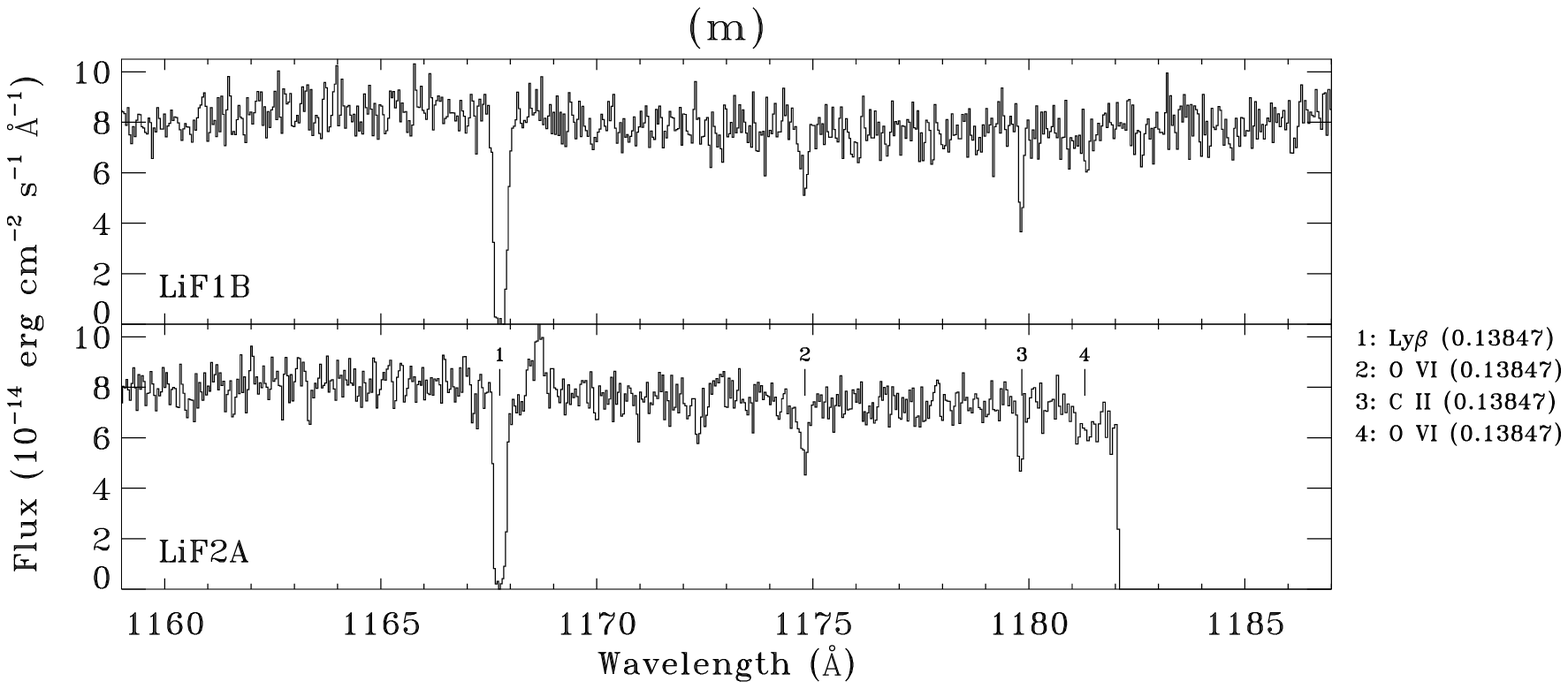}
\vspace{3in}
\caption{Fully reduced FUSE spectra of PG\,1116+215 as a function of 
heliocentric wavelength between 915 and 1187\,\AA.
Data from two channels are shown at most wavelengths, with SiC data 
shown at $\lambda \lesssim 1000$\,\AA, and LiF data shown at $\lambda
\gtrsim 1000$\,\AA.  SiC data are also shown in the LiF coverage gap near 1070--1090\,\AA.  The detector segments shown are identified in the lower 
corners of each panel.
The data have a spectral resolution of 
$\approx 20-25$ \kms\ (FWHM).  Line detections are denoted by 
tick marks above the spectra.  Line identifications for these detections are listed at the right hand side
of each panel.   Redshifts (in parentheses) are indicated for intergalactic
lines. Rest wavelengths are indicated for interstellar 
lines.  Interstellar molecular hydrogen lines are labeled according to 
their rotational and vibrational levels as described in the text. 
In cases where the +184 \kms\ high-velocity interstellar feature is present, an
offset tick mark is attached to the primary tick mark at the rest 
wavelength of the line.
Several fixed-pattern noise (FPN) and unidentified (UID) features are 
also labeled.  The dashed horizontal line in panel~g illustrates
the continuum placement longward of the Lyman-limit system at $z=0.13847$.}
\end{figure}

\clearpage
\newpage

\begin{figure}[ht!]
\figurenum{3}
\includegraphics{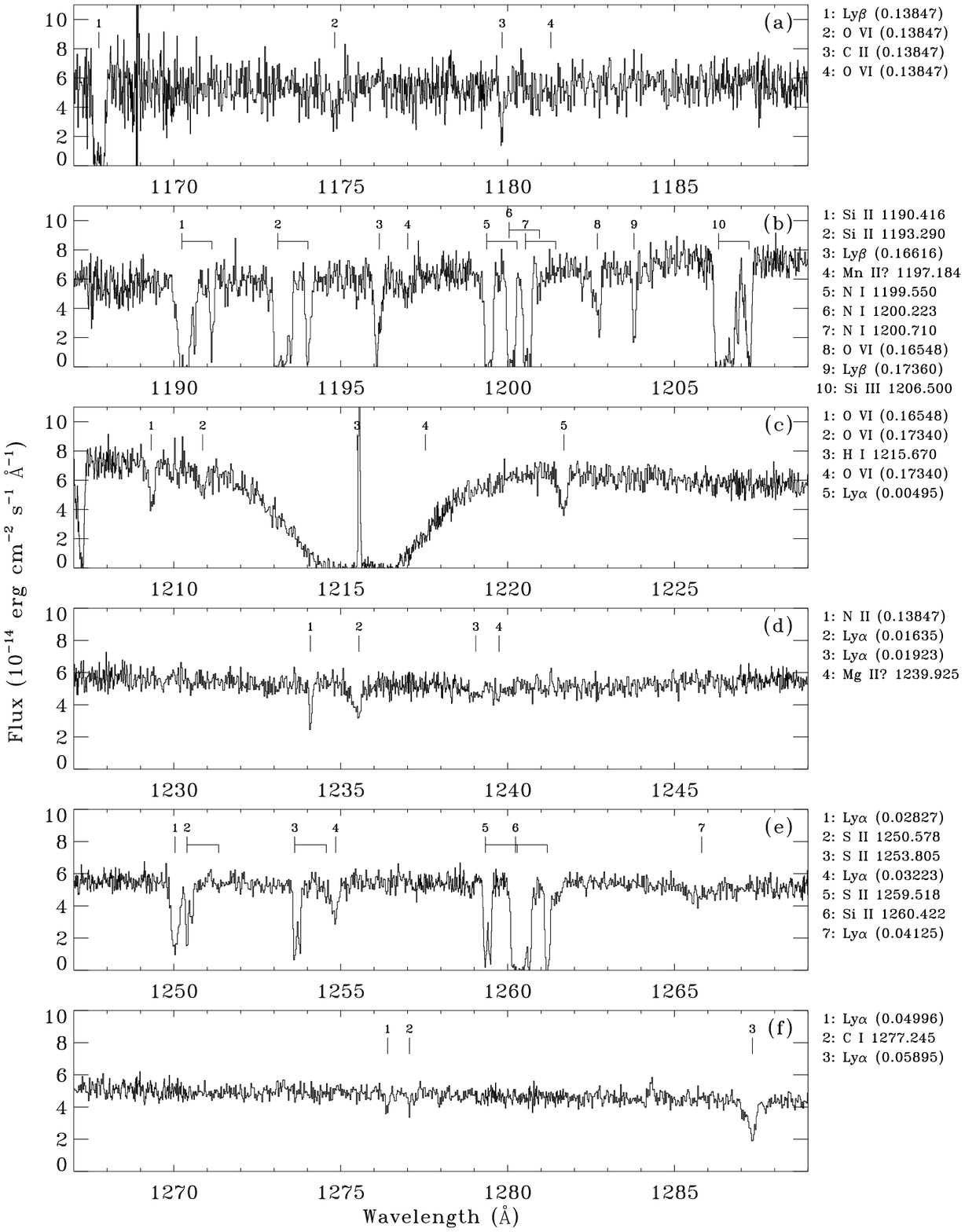}
\vspace{8.9in}
\caption{See caption at end of figure.\label{e140mspec}}
\end{figure}

\begin{figure}[ht!]
\figurenum{3 (continued)}
\includegraphics{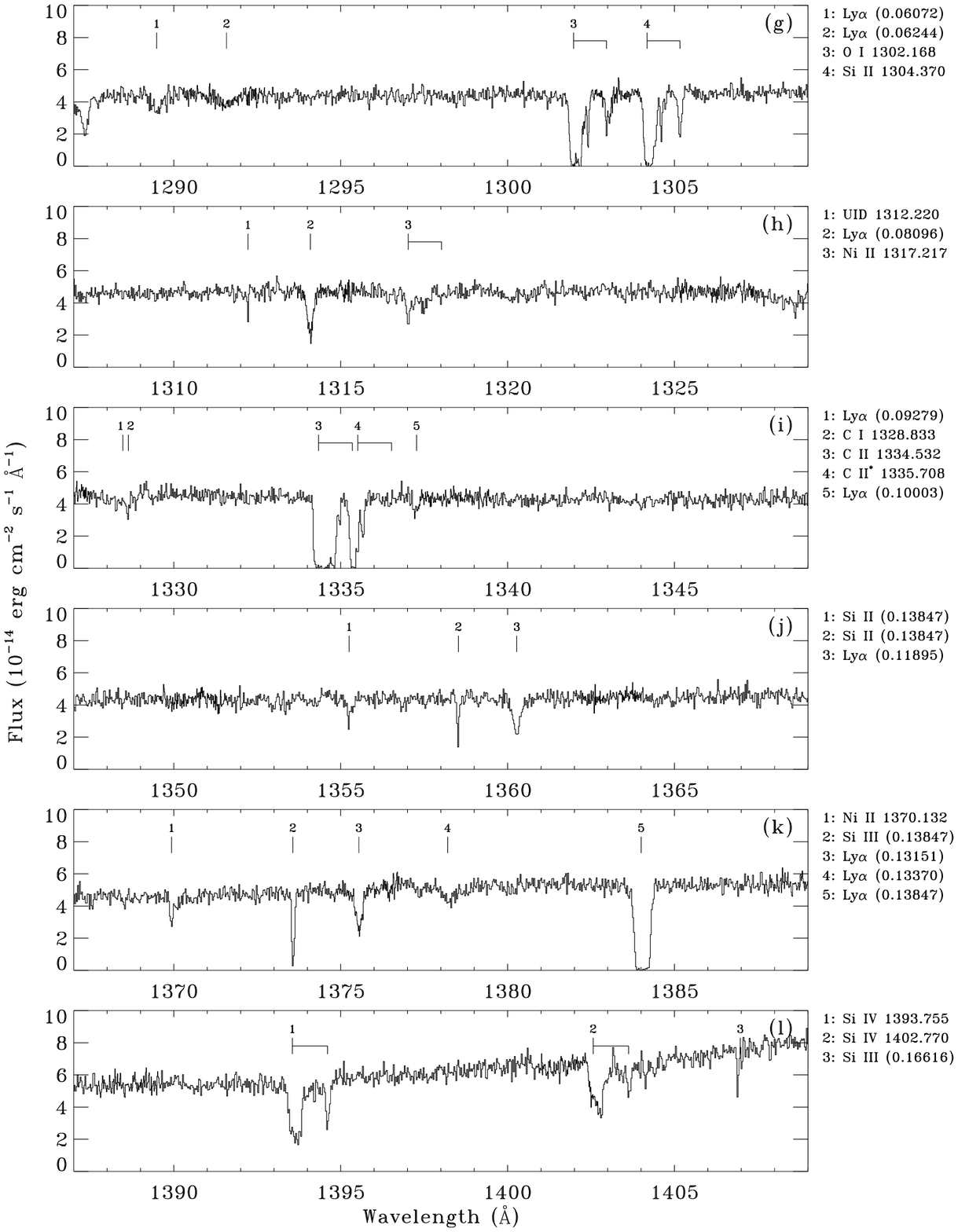}
\vspace{8.9in}
\caption{See caption at end of figure.}
\end{figure}

\begin{figure}[ht!]
\figurenum{3 (continued)}
\includegraphics{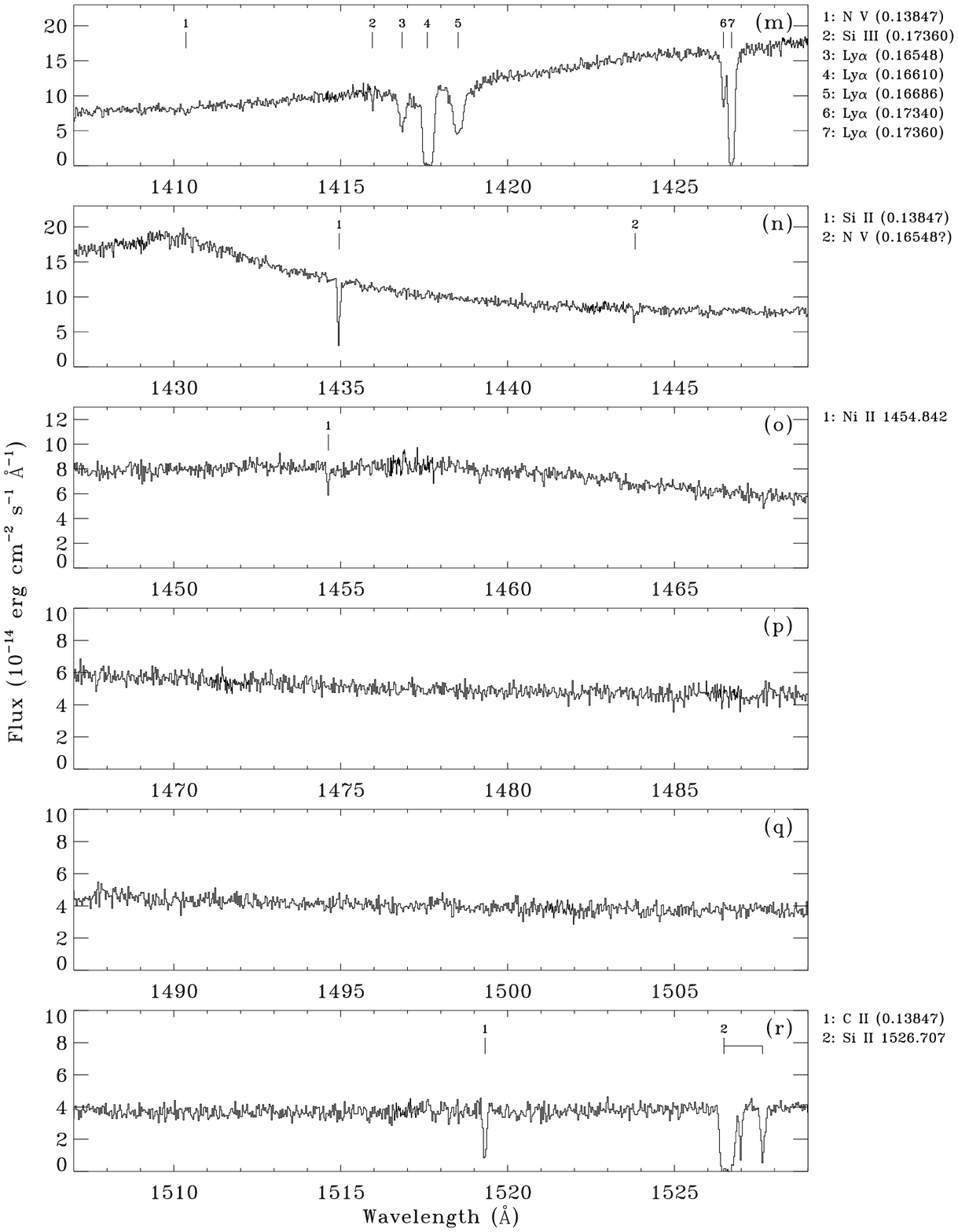}
\vspace{8.9in}
\caption{See caption at end of figure.}
\end{figure}

\begin{figure}[ht!]
\figurenum{3 (continued)}
\includegraphics{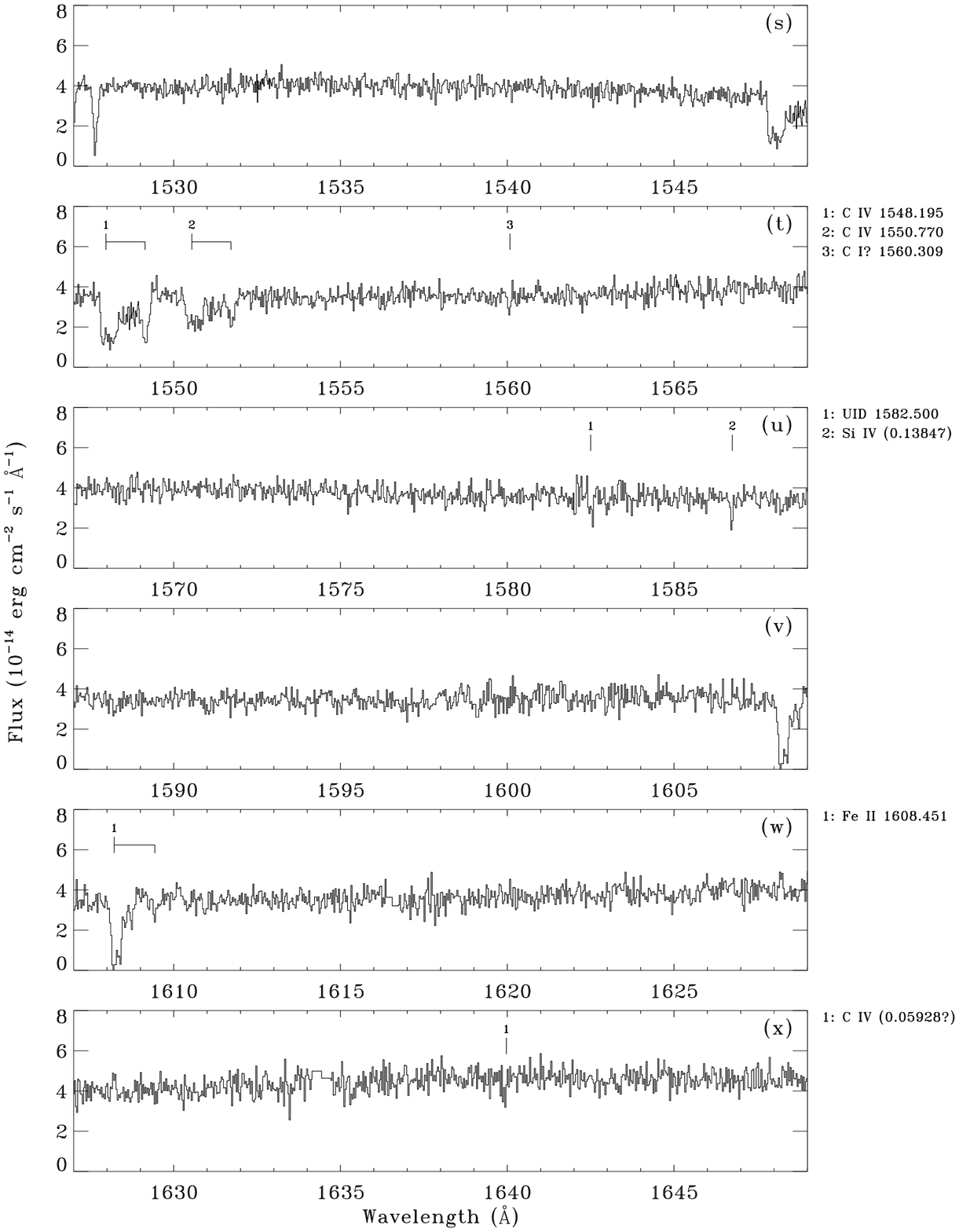}
\vspace{8.9in}
\caption{See caption on next page.}
\end{figure}

\begin{figure}[ht!]
\figurenum{3 (continued)}
\includegraphics{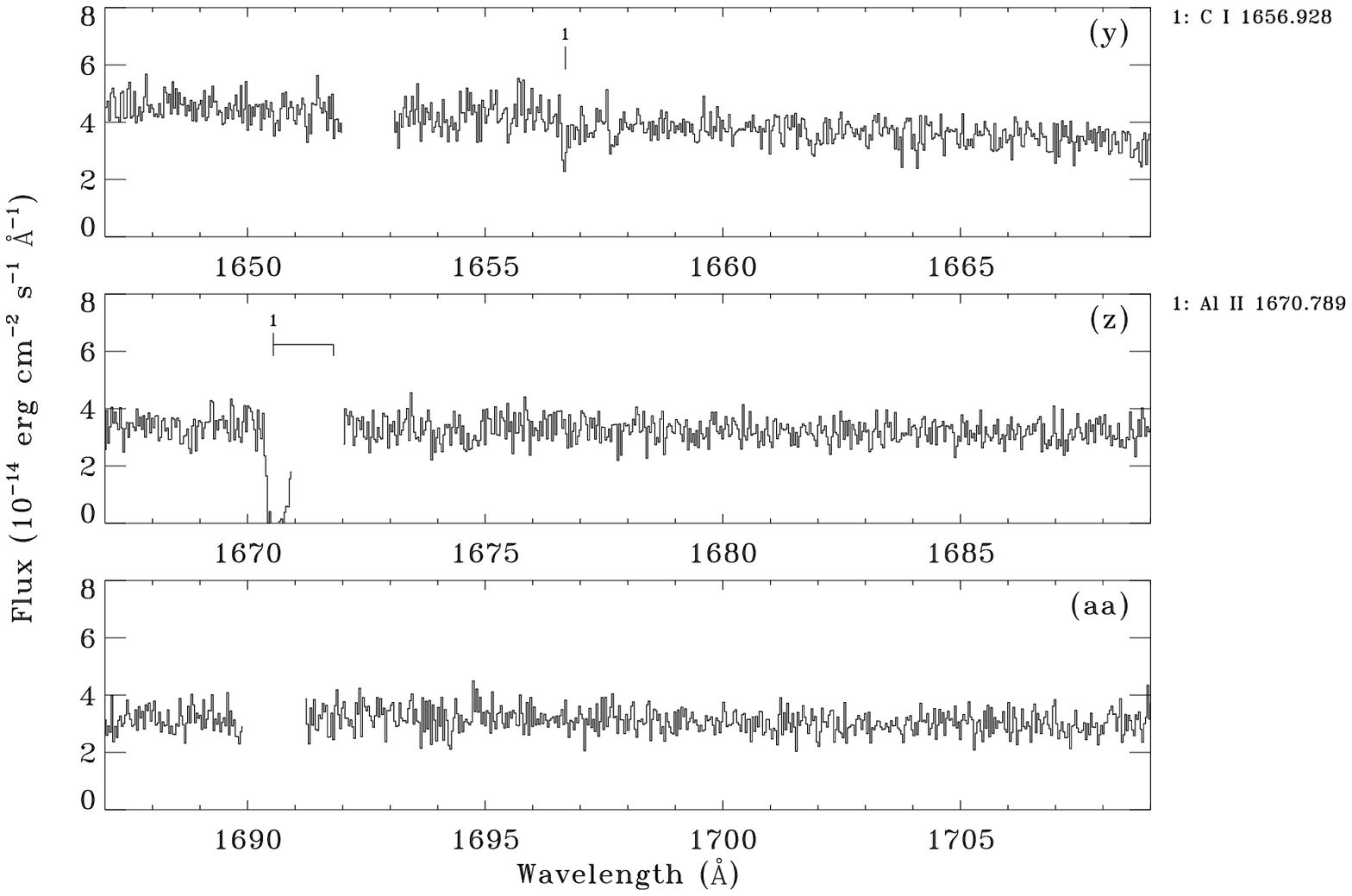}
\vspace{5.0in}
\caption{Fully reduced HST/STIS E140M spectra of 
PG\,1116+215 as a function of heliocentric wavelength
 between 1167 and 1709\,\AA.  The data have a 2-pixel (FWHM) spectral 
resolution of $\approx 6.5$ \kms.  
We have binned the data into two pixel bins for clarity, but all
measurements were conducted on the fully sampled data.  
Labels are similar to those in Figure~\ref{fusespec}. }
\end{figure}

\clearpage
\newpage
\begin{figure}[ht!]
\figurenum{4}
\includegraphics{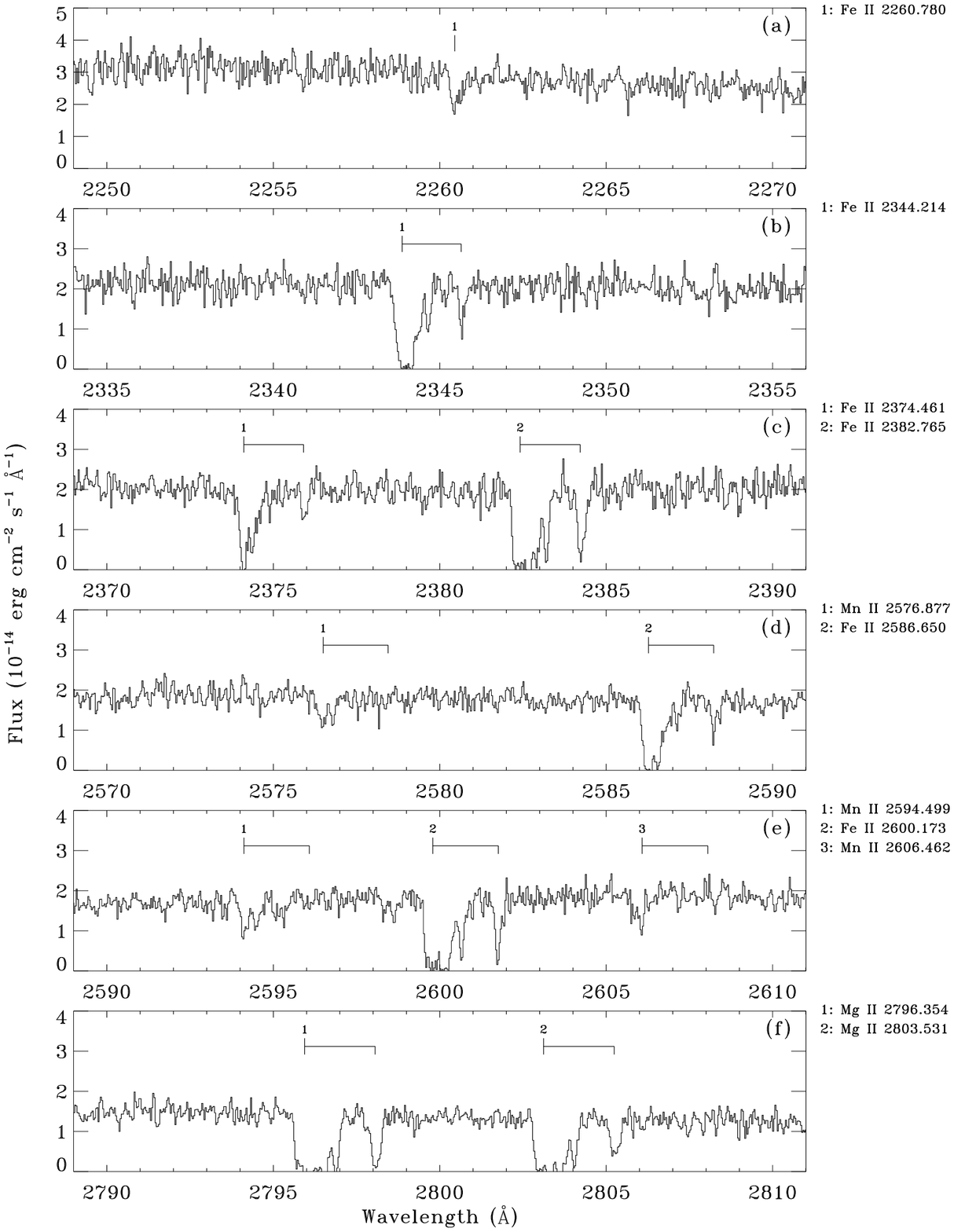}
\vspace{8.9in}
\caption{See caption on next page.\label{e230mspec}}
\end{figure}
\clearpage
\newpage
\noindent
Fig.\,\ref{e230mspec}. --- Portions of the fully reduced HST/STIS E230M 
spectra of 
PG\,1116+215 as a function of heliocentric wavelength
between 2250 and 2851\,\AA.  The data have a 2-pixel (FWHM) spectral 
resolution of $\approx 10$ \kms.  The unbinned data are plotted at the 
nominal sampling interval of 
$\approx5$ \kms.
Labels are similar to those in Figure~\ref{fusespec}.
\clearpage
\newpage

\begin{figure}[ht!]
\figurenum{5}
\includegraphics{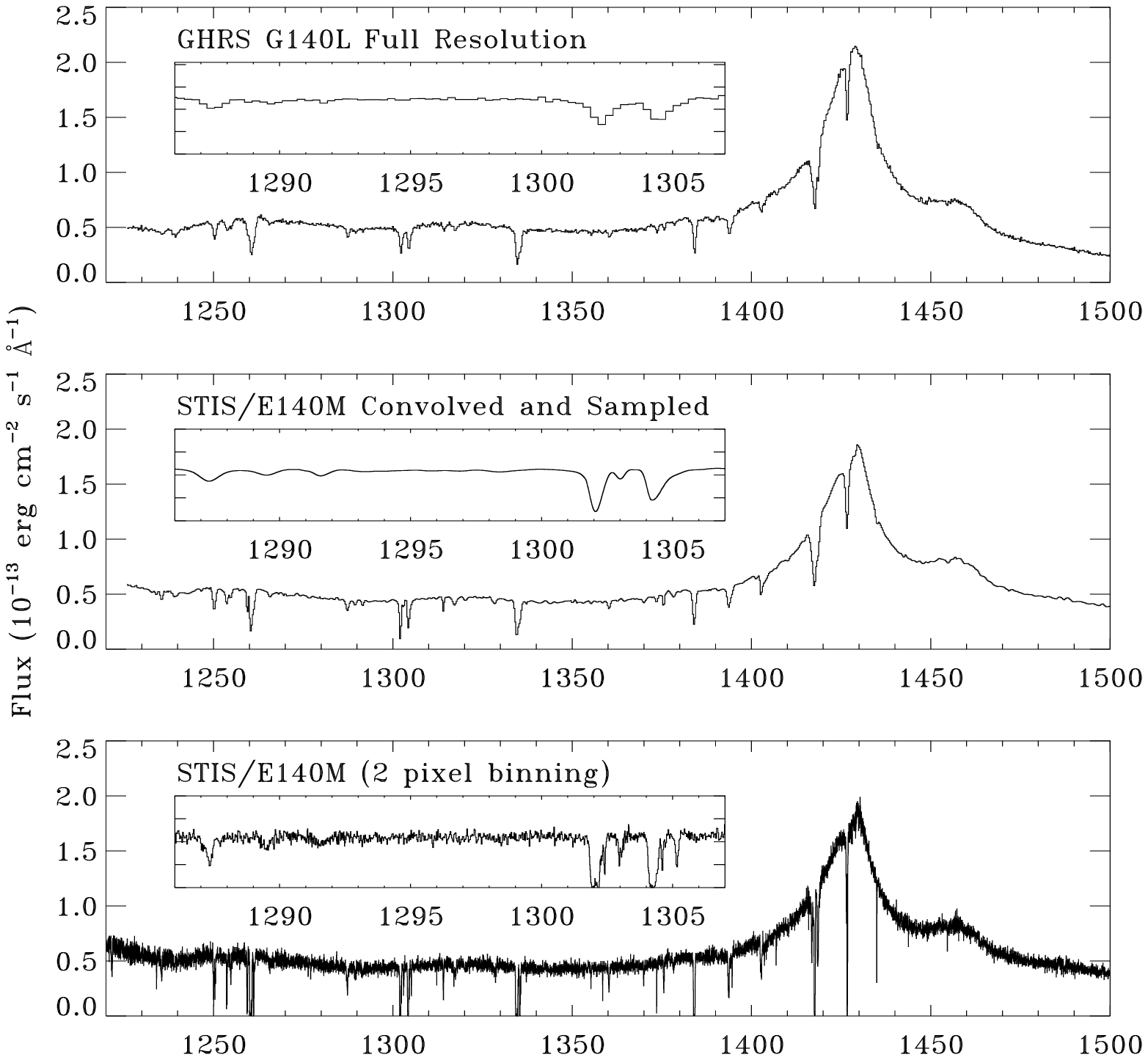}
\vspace{6.0in}
\caption{A comparison of the HST/STIS E140M and GHRS G140L spectra 
of PG\,1116+215 between 1220 and 1500\,\AA.  {\it Top:} 
GHRS G140L spectra obtained with FP-SPLIT procedures by Tripp
 et al. (1998).  The spectra have a total exposure time of 3.5 hours
and a spectral resolution of $\approx 160$ \kms\ (FWHM).  {\it Middle:}
STIS E140M spectra of PG\,1116+215 convolved with a Gaussian smoothing
kernel having FWHM = 160 \kms\ and resampled to the GHRS wavelength vector.  
The resulting spectrum looks
very similar to the GHRS G140L spectrum.  {\it Bottom:} Full-resolution
STIS E140M spectra of PG\,1116+215 binned by 2 pixels.  
Expanded regions of this plot can be found in Figure~\ref{e140mspec}. \label{compfig}  }
\end{figure}

\begin{figure}[ht!]
\figurenum{6}
\includegraphics{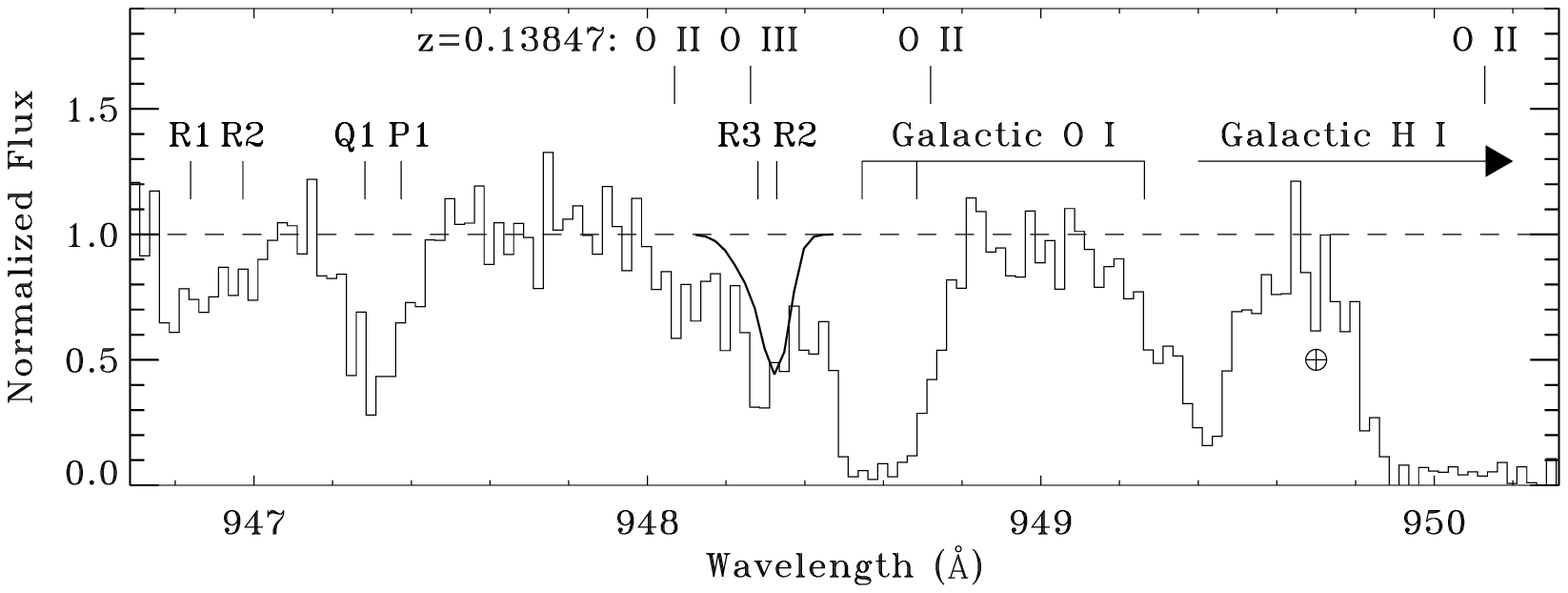}
\vspace{2.5in}
\caption{A portion of the FUSE SiC2 spectrum of PG\,1116+215 in the 
wavelength region covering the redshifted lines of \ion{O}{2} 
$\lambda\lambda832.757, 833.329, 834.466$ and \ion{O}{3} $\lambda832.927$ 
in the $z=0.13847$ metal-line system.  The 
heliocentric wavelengths of 
the redshifted oxygen lines are indicated with tick marks at the top of the 
figure.  Galactic lines of \ion{H}{1} $\lambda949.743$, \ion{O}{1} 
$\lambda948.468$, and several H$_2$ lines in the Lyman series 14-0 
and Werner series 3-0 vibrational bands are labeled
immediately above the spectrum.
For the Galactic \ion{O}{1} line, the 
three ticks indicate components at negative intermediate velocity
($-44$ \kms), zero velocity, and high positive velocity (+184 \kms).  
The H$_2$ lines indicated include 
$\lambda\lambda946.978$ (14-0) R(1), 947.111 (3-0) R(2), 947.421 (3-0) Q(1),
947.513 (14-0) P(1), 948.419 (3-0) R(3), and 948.468 (14-0) R(2), all of which 
occur in the $-44$ \kms\ interstellar component along the sight line.  The
expected absorption due to the two interstellar H$_2$ 
lines near 948.3\,\AA\ is shown as a heavy smooth curve overplotted on 
the spectrum.
Terrestrial airglow emission between 949.5 and 949.9\,\AA\ is indicated
with a crossed circle below the spectrum.  
\label{o3fig}}
\end{figure}

\clearpage
\newpage

\begin{figure}[ht!]
\figurenum{7}
\includegraphics{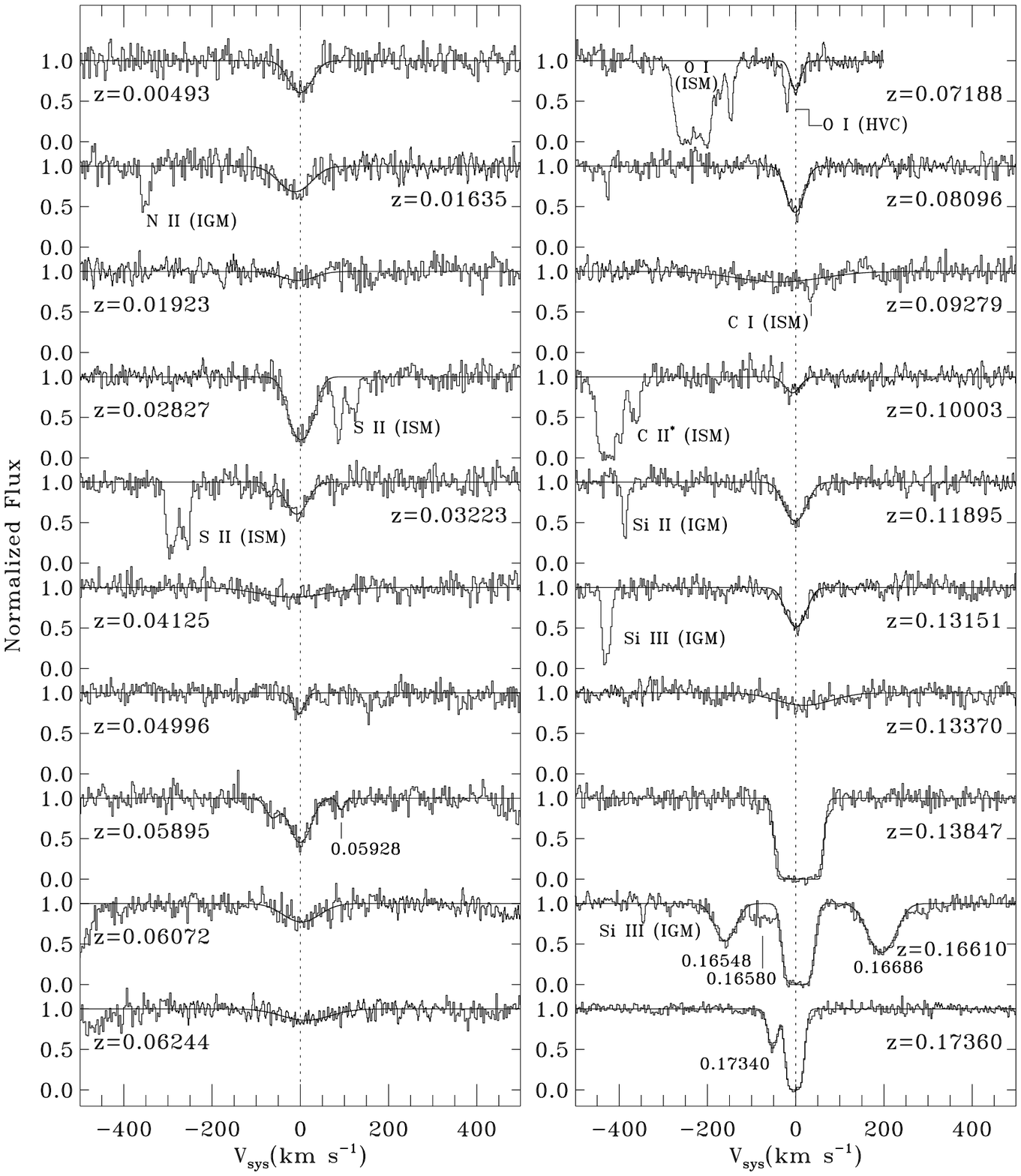}
\vspace{8.0in}
\caption{See caption on next page.\label{lya_stack}}
\end{figure}
\clearpage
\newpage
\noindent
Fig.\,\ref{lya_stack}. --- Ly$\alpha$ absorbers toward PG\,1116+215.  
The redshifts of 
 Ly$\alpha$ features are indicated for the main absorption and for the 
``satellite'' absorbers at $z=0.05928$, 0.16548, 0.16580,
0.16686, and 0.17340.  
Other ISM and IGM features are identified
by species.  A portion of the spectrum containing
Galactic  \ion{Si}{2} $\lambda1304.370$ 
absorption  near the $z=0.07188$ absorber 
has been omitted for clarity.  Gaussian components fit to each intervening 
Ly$\alpha$ line are overplotted
as smooth curves.  See text for details.

\clearpage
\newpage

\clearpage
\newpage

\clearpage
\newpage

\clearpage
\newpage

\clearpage
\newpage

\clearpage
\newpage

\clearpage
\newpage

\begin{figure}[ht!]
\figurenum{14}
\includegraphics{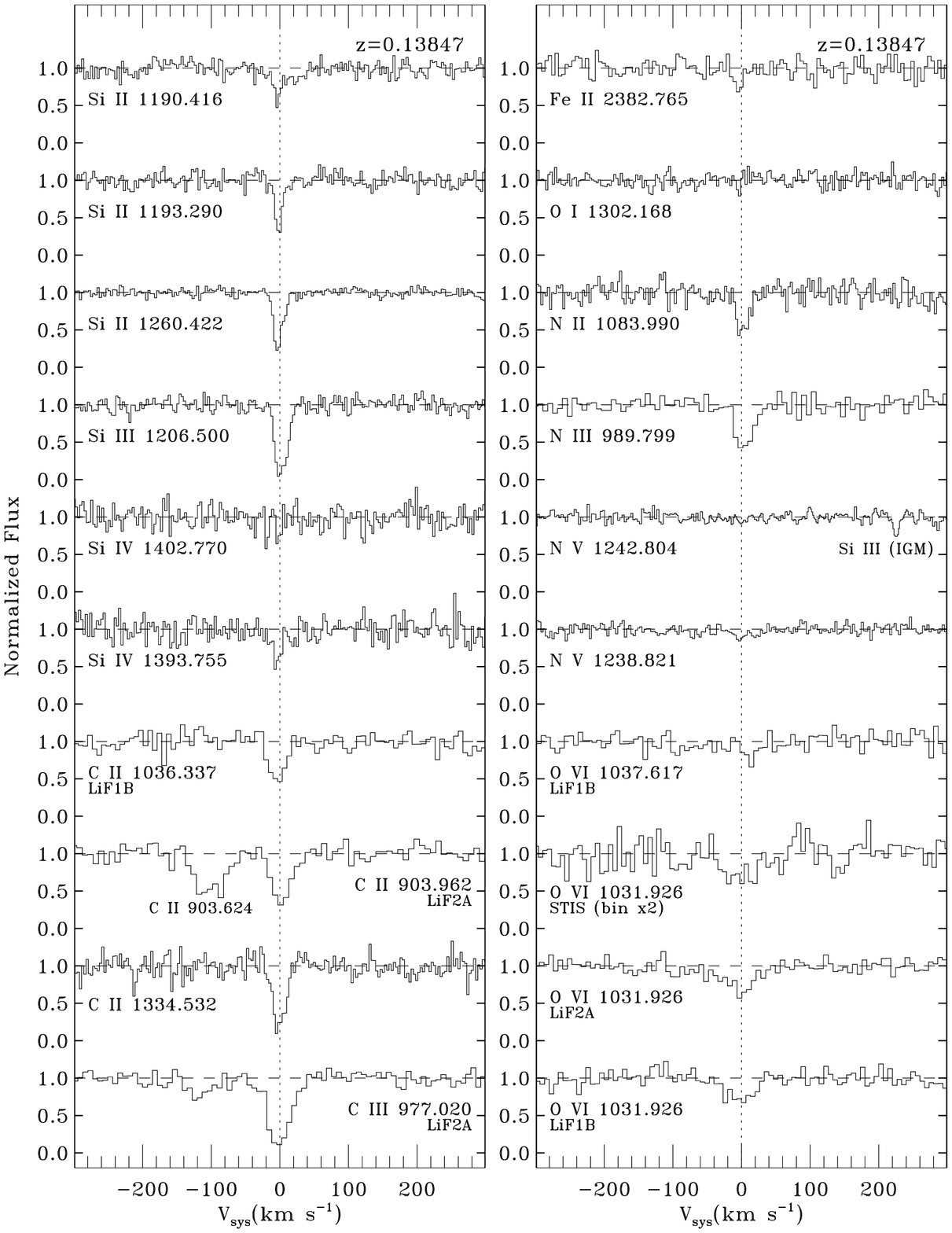}
\vspace{8.5in}
\caption{See caption on next page\label{stack_ion_13847}}
\end{figure}

\clearpage
\newpage
\noindent
Fig.\,\ref{stack_ion_13847}. --- Continuum-normalized metal lines in the $z=0.13847$ absorber.  
Both FUSE
and STIS data are shown.  
The FUSE data are from the SiC2 or LiF1 channels.  The three detections of 
\ion{O}{6} $\lambda1031.926$ are shown at the bottom of the right hand
panel.  The STIS data for the \ion{O}{6} line has been binned into 2-pixel
samples since the spectrum is noisy at these wavelengths (see Figure~\ref{e140mspec}).

\clearpage
\newpage

\clearpage
\newpage

\clearpage
\newpage

\begin{figure}[ht!]
\figurenum{17}
\includegraphics{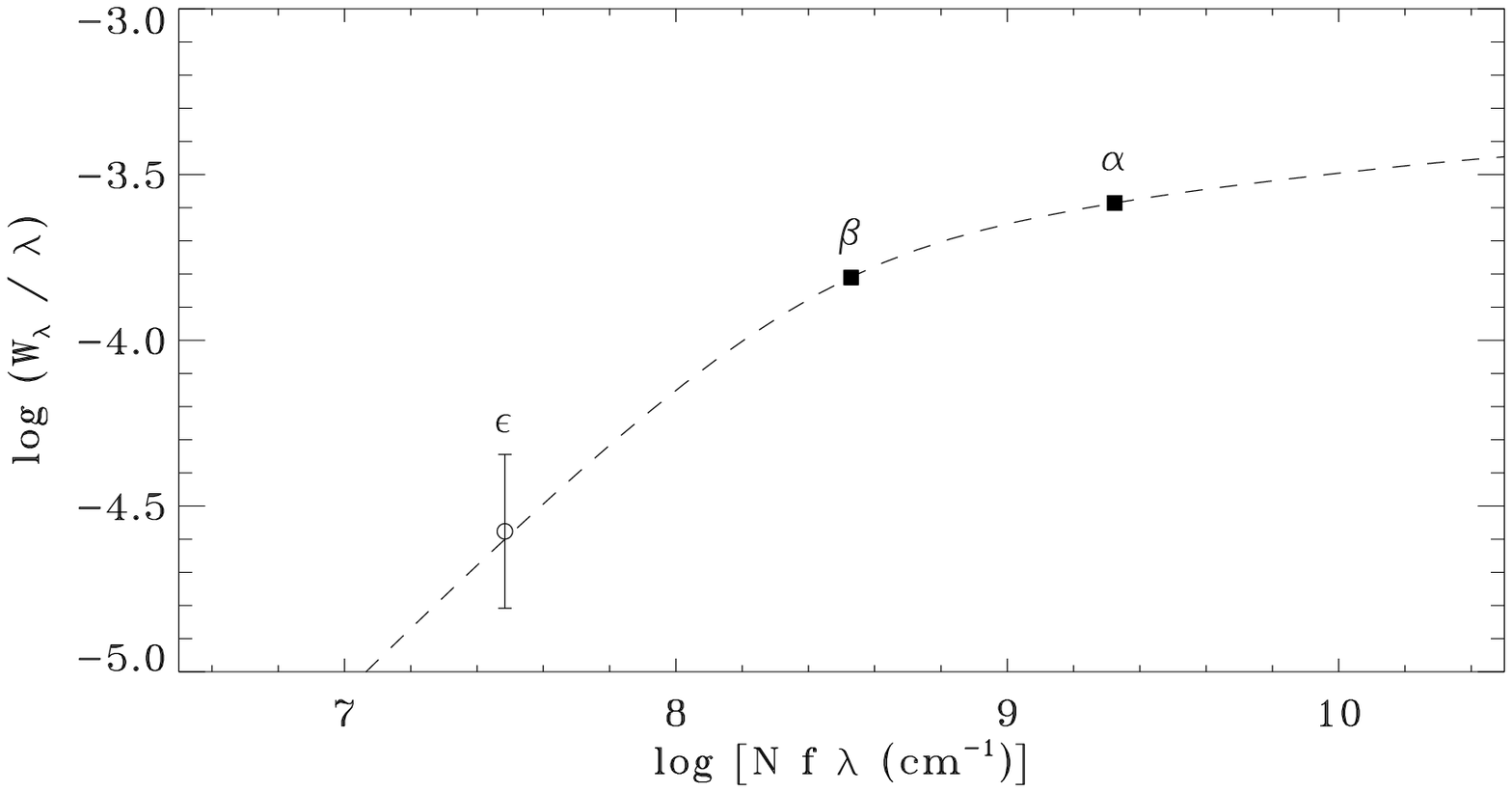}
\vspace{4.0in}
\caption{Single-component curve of growth for the \ion{H}{1} Lyman-series
lines in the $z=0.16610$ absorber.
 The STIS  Ly$\alpha$ and Ly$\beta$ measurements are shown as filled 
square data points.  The FUSE LiF2 measurement for the Ly$\epsilon$ line 
is represented by the open circle.
Error bars are $1\sigma$ estimates.
A small residual amount of Ly$\alpha$ absorption ($\sim$60\,m\AA) was not 
included in the fit since it is outside the velocity range of the Ly$\beta$
absorption (see text and Figure~\ref{lya_stack}).
The best fit COG has N(\ion{H}{1}) = $(4.17\pm^{0.59}_{0.48})\times10^{14}$
cm$^{-2}$ (log N = $14.62\pm^{0.06}_{0.05}$) and b = $22.0\pm0.9$ \kms.\label{cog16610}}
\end{figure}
\clearpage
\newpage

\clearpage
\newpage

\clearpage
\newpage

\clearpage
\newpage

\clearpage
\newpage

\clearpage
\newpage

\begin{deluxetable}{cccccccc}
\tablewidth{0pt} 
\tablecaption{FUSE Observations of PG\,1116+215\tablenotemark{a} \label{tab_fuse}}
\tablehead{Dataset & Date & \# Exp. 
& \multicolumn{2}{c}{\underline{LiF1 / SiC1}}
& \multicolumn{2}{c}{\underline{LiF2 / SiC2}} & Aperture \\
                   & (U.T. Start) &   & T$_{\rm tot}$  & T$_{\rm ngt}$  
& T$_{\rm tot}$  & T$_{\rm ngt}$\\
                   & &      & (ksec) & (ksec) & (ksec) & (ksec)}
\startdata
P1013101 & 2000-Apr-28 & \phn7 & 11.0 & 11.0 & 11.0 & 11.0 & $30\arcsec\times30\arcsec$  \\
P1013102 & 2001-Apr-22 & \phn5 & 11.1 & \phn9.4 & 11.1 & \phn9.4 & $30\arcsec\times30\arcsec$\\
P1013103 & 2001-Apr-22 & \phn6 & \phn8.4 & \phn6.3 & \phn8.4 & \phn6.3 & $30\arcsec\times30\arcsec$ \\
P1013104 & 2001-Apr-23 & \phn5 & 11.2 & \phn9.5 & 11.2 & \phn9.5 & $30\arcsec\times30\arcsec$\\
P1013105 & 2001-Apr-23 & 20 & 35.0 & 27.8 & 35.1 & 27.8 & $30\arcsec\times30\arcsec$\\
\enddata
\tablenotetext{a}{Note -- 
Entries in this table include the dataset identification, 
U.T. date at the start of the observation, number of exposures in the 
observation, exposure times (total and night-only) for the LiF1/SiC1 channels 
and LiF2/SiC2 channels, and 
aperture used (same size for all four channels).  Exposure times are totals
after screening for valid data with event bursts removed.}

\end{deluxetable}

\clearpage
\newpage
\begin{deluxetable}{ccccccc}
\tabletypesize{\small}
\tablewidth{0pt} 
\tablecaption{HST/STIS E140M and E230M Observations of PG\,1116+215\tablenotemark{a} \label{tab_stis}}
\tablehead{Program & Grating/Tilt & Dataset & Date & \# Exp.& T$_{\rm tot}$ & Slit\\
                   &         & &  (U.T. Start)   &   & (ksec) }
\startdata
GO-8097 & E140M/1425 & O5A301010--O5A301020 & 2000-May-20 & 9 & 12.6  & $0.2\arcsec\times0.06\arcsec$\\
GO-8097 & E140M/1425 & O5A302010--O5A302020 & 2000-May-15 & 5 & \phn7.3  & $0.2\arcsec\times0.06\arcsec$\\
GO-8165 & E140M/1425 & O5E701010--O5E701070 & 2000-Jun-22 & 7 & 10.0 & $0.2\arcsec\times0.06\arcsec$\\
GO-8165 & E140M/1425 &  O5E702010--O5E702070  & 2000-Jun-30 & 7 & 10.0 & $0.2\arcsec\times0.06\arcsec$\\
GO-8097 &  E230M/2415 &  O5A302030--O5A302050 & 2000-May-15 & 4 & \phn5.6 & $0.2\arcsec\times0.06\arcsec$\\
\enddata
\tablenotetext{a}{Note -- Entries in this table include the Guest Observer progam number,  grating and central wavelength, dataset identifications, U.T. date at the start of the first observation in the series, number of exposures in the series,
total exposure time, and slit used for the 
observation.}
\end{deluxetable}

\clearpage
\newpage
\begin{deluxetable}{lcrcl}
\tabletypesize{\footnotesize}
\tablewidth{490pt} 
\tablecaption{Interstellar Line Cross Reference for Lines \label{tab_ism}}
\tablehead{Species & $\lambda$(\AA)\tablenotemark{a}   & $\log f\lambda$\tablenotemark{a} & ~~~~~ID\tablenotemark{b}~~~~~ & Comments }
\startdata
\ion{H}{1} & \phn917.181 & $-0.178$& 2a-1 & Additional \ion{H}{1} blanketing at $\lambda < 917$\,\AA\\
\ion{H}{1} & \phn918.129 & $-0.072$ & 2a-2 & HVC component at +184 \kms\\
\ion{H}{1} & \phn919.351 & 0.043 & 2a-3 & HVC component at +184 \kms\\
\ion{H}{1} & \phn920.963 & 0.170 & 2a-4 & HVC component at +184 \kms\\
\ion{O}{1} & \phn921.857 & $-0.001$ & 2a-5 \\
\ion{H}{1} & \phn923.150 & 0.135 & 2a-6 & HVC component at +184 \kms\\
H$_2$ L17R7 & \phn924.641 & 0.574 & 2a-7\\
\ion{O}{1} & \phn924.950 &  0.155 & 2a-8\\
\ion{O}{1} & \phn925.446 & $-0.484$ & 2a-9\\
\ion{H}{1} & \phn926.226 & 0.294 & 2a-10  & HVC component at +184 \kms\\
\ion{O}{1} & \phn929.517 & 0.329 & 2a-11\\
\ion{H}{1} & \phn930.748 & 0.475 & 2a-12 & HVC component at +184 \kms\\
H$_2$ L16R1 & \phn931.730 & 0.850 & 2a-13\\
H$_2$ L16R3 & \phn935.573 & 0.789 & 2a-14\\
\ion{O}{1} & \phn936.630 &  0.534 & 2a-15\\
\ion{H}{1} & \phn937.803 & 0.688 & 2a-16& HVC component at +184 \kms\\
H$_2$ L15R1 & \phn939.122 & 0.793 & 2a-17\\
H$_2$ L15P1 & \phn939.706 & 0.437 & 2b-1\\
H$_2$ W3R1 & \phn946.384 & 1.082 & 2b-2 & Blend with H$_2$ W3R0\\
H$_2$ W3R0 &  \phn946.423 & 1.757 & 2b-2 & Blend with H$_2$ W3R1\\
H$_2$ L14R1 & \phn946.978 & 1.271 & 2b-3 & Blend with H$_2$ W3R2\\
H$_2$ W3R2 &  \phn947.111 & 1.103 & 2b-3 & Blend with H$_2$ L14R1\\
H$_2$ W3Q1 & \phn947.421 & 1.406 & 2b-4 & Blend with H$_2$ L14P1\\
H$_2$ L14P1 & \phn947.513 & 0.521 & 2b-4 & Blend with H$_2$ W3Q1\\
H$_2$ W3R3  & \phn948.419 & 1.085 & 2b-5 & \\
\ion{O}{1} & \phn948.685 & 0.778 & 2b-6 & HVC component at +184 \kms?\\
\ion{H}{1} & \phn949.743 & 0.946 & 2b-7 & HVC component at +184 \kms\\
\ion{O}{1} & \phn950.885 & 0.176& 2b-8 & HVC component at +184 \kms\\
\ion{N}{1} & \phn952.303 & 0.338 & 2b-9 & Blend with \ion{N}{1} $\lambda\lambda952.415,952.532$ \\
\ion{N}{1} & \phn953.415 & 1.091& 2b-10\\
\ion{N}{1} & \phn953.655 & 1.372& 2b-11\\
\ion{N}{1} & \phn953.970 & 1.499& 2b-12\\
\ion{N}{1} & \phn954.104 & 0.582& 2b-13\\
H$_2$ L13R1 & \phn955.064 & 0.974 & 2b-14 \\
H$_2$ L13P1 & \phn955.708 & 0.608 & 2b-15 \\
H$_2$ L13R2 & \phn956.578 & 0.945 & 2b-16 \\
H$_2$ L13P2 & \phn957.650 & 0.661 & 2b-17 \\
H$_2$ L12R0 & \phn962.976 & 1.109 & 2c-1 \\
H$_2$ L12R1 & \phn963.607 & 0.841 & 2c-2 \\
\ion{P}{2} & \phn963.801 & 3.148 & 2c-3 &  HVC component at +184 \kms\\
H$_2$ W2R0 & \phn964.981 & 0.870 & 2c-4 & Blend with H$_2$ L12R2 and W2R1\\
H$_2$ L12R2 & \phn965.045 & 0.189 & 2c-4 & Blend with H$_2$ W2R0 and W2R1\\
H$_2$ W2R1 & \phn965.062 & 1.528 & 2c-4 & Blend with H$_2$ W2R0 and L12R2 \\
H$_2$ W2R2 & \phn965.791 & 1.481 & 2c-5 \\
H$_2$ W2Q1 & \phn966.094 & 1.529 & 2c-6 \\
H$_2$ W2R3 & \phn966.779 & 0.864 & 2c-7 \\
H$_2$ W2Q2 & \phn967.278 & 1.530 & 2c-8 \\
H$_2$ L12R3 & \phn966.673 & 1.356 & 2c-9 \\
H$_2$ W2P2 & \phn968.292 & 0.847 & 2c-10 \\
H$_2$ W2Q3 & \phn969.047 & 1.533 & 2c-11 \\
\ion{O}{1} & \phn971.738 & 1.052 & 2c-12\\
\ion{H}{1} & \phn972.537 & 1.450 & 2c-13\\
H$_2$ L11R2 & \phn974.156 & 1.108 & 2c-14 \\
\ion{O}{1} & \phn976.448 & 0.509 & 2c-15\\
\ion{C}{3} & \phn977.020 & 2.869 & 2c-16\\
H$_2$ L11P3 & \phn978.217 & 0.821 & 2c-17 \\
H$_2$ L10R0 & \phn981.437 & 1.310& 2d-1 \\
H$_2$ L10R1 & \phn982.073 & 1.130& 2d-2 \\
H$_2$ L10P1 & \phn982.835 & 0.822& 2d-3 \\
H$_2$ L10R2 & \phn983.589& 1.056& 2d-4 \\
H$_2$ L10P2 & \phn984.862 & 0.907& 2d-5 \\
H$_2$ W1R0 & \phn985.632 & 1.838& 2d-6 &  Blend with H$_2$ W1R1\\
H$_2$ W1R1 & \phn985.642 & 1.525& 2d-6 &  Blend with H$_2$ W1R0\\
H$_2$ W1R2 & \phn986.241 & 1.418& 2d-7 \\
H$_2$ W1Q1 & \phn986.796 & 1.557& 2d-8 \\
H$_2$ W1R3 & \phn987.445 & 1.415& 2d-9 \\
H$_2$ L10P3 & \phn987.767 &0.943 & 2d-10 \\
H$_2$ W1Q2 & \phn987.972 & 1.558& 2d-11 \\
\ion{O}{1} & \phn988.773 & 1.662 & 2d-12\\
\ion{N}{3} & \phn989.799 & 2.085 & 2d-13\\
\ion{Si}{2} & \phn989.873 & 2.228 & 2d-14\\
H$_2$ L9R0 & \phn991.376 & 1.413& 2d-15 &  Blend with H$_2$ W1P3\\
H$_2$ W1P3 & \phn991.378 & 1.060& 2d-15 &  Blend with H$_2$ L9R0\\
H$_2$ L9R1 & \phn992.014 & 1.257& 2d-16 \\
H$_2$ L9P1 & \phn992.808 & 0.893& 2d-17 \\
H$_2$ L9R2 & \phn993.547 & 1.227& 2d-18 \\
H$_2$ L9P2 & \phn994.871 & 0.934& 2d-19 \\
H$_2$ L9R3 & \phn995.970 & 1.219& 2d-20 \\
H$_2$ L8R0 & 1001.821 & 1.426 & 2e-1 & \\
H$_2$ L8R1 & 1002.449 & 1.264 & 2e-2 & \\
H$_2$ L8P1 & 1003.294 & 0.933 & 2e-3 & \\
H$_2$ L8R2 & 1003.982 & 1.225 & 2e-4 & \\
H$_2$ L8P2 & 1005.390 & 1.002 & 2e-5 & \\
H$_2$ L8R3 & 1006.411 & 1.202 & 2e-6 & \\
H$_2$ W0R1 & 1008.498 & 1.317 & 2e-7 & Blend with H$_2$ W0R0\\
H$_2$ W0R0 & 1008.552 & 1.646 & 2e-7 & Blend with H$_2$ W0R1\\
H$_2$ W0R2 & 1009.024 & 1.197 & 2e-8 & \\
H$_2$ W0Q1 & 1009.771 & 1.393 & 2e-9 & \\
H$_2$ W0R3 & 1010.129 & 1.143 & 2e-10 & $W_\lambda = 30$\,m\AA; used for comparison with H$_2$ W3R3 in Figure\,\ref{o3fig} \\
H$_2$ W0Q2 & 1010.938 & 1.394 & 2e-11 & \\
H$_2$ W0P2 & 1012.169 & 0.748 & 2e-12 & \\
\ion{S}{3} & 1012.495 & 1.647 & 2e-13 & \\
H$_2$ L7R0 & 1012.810 & 1.476 & 2e-14 & Blend with H$_2$ W0Q3 \\
H$_2$ W0Q3 & 1012.680 & 1.396 & 2e-14 & Blend with H$_2$ L7R0 \\
H$_2$ L7R1 & 1013.435 & 1.318 & 2e-15 & \\
H$_2$ L7P1 & 1014.326 & 0.948 & 2e-16 & \\
H$_2$ W0P3 & 1014.504 & 0.924 & 2e-16 & \\
H$_2$ L7R2 & 1014.974 & 1.286 & 2e-17 & \\
H$_2$ L7P2 & 1016.458 & 1.017 & 2e-18 & \\
H$_2$ L7R3 & 1017.422 & 1.271 & 2e-19 & \\
\ion{Si}{2} & 1020.699 & 1.234 & 2f-1 & \\
H$_2$ L6R0 & 1024.372 & 1.470 & 2f-2 & \\
H$_2$ L6R1 & 1024.987 & 1.305 & 2f-3 & \\
\ion{H}{1} Ly$\beta$ & 1025.722 & 1.909 & 2f-4 & $\oplus$ airglow emission present\\
H$_2$ L6P2 & 1028.104 & 1.032 & 2f-5 & \\
H$_2$ L6P3 & 1031.191 & 1.061 & 2f-8 & $W_\lambda = 30$\,m\AA; used for comparison with H$_2$ W3R3 in Figure\,\ref{o3fig}\\
\ion{O}{6} & 1031.926 & 2.136 & 2f-9 & HVC component at +184 \kms \\
\ion{C}{2} & 1036.337 & 2.088 & 2f-10 & HVC at +184 \kms\ blends with \ion{C}{2}$^*$ $\lambda1037.018$ \\
H$_2$ L5R0 & 1036.545 & 1.448 & 2f-11 & \\
\ion{C}{2}$^*$ & 1037.018 & 2.088 & 2f-12 &  Blend with \ion{C}{2} $\lambda1036.337$ HVC at +184 \kms \\
H$_2$ L5R1 & 1037.149 & 1.271 & 2f-13 & \\
\ion{O}{6} & 1037.617 & 1.834 & 2f-14 & HVC component at +184 \kms \\
H$_2$ L5P1 & 1038.157 & 0.952 & 2f-15 & \\
H$_2$ L5R2 & 1038.689 & 1.235 & 2f-16 & \\
\ion{O}{1} & 1039.230 & 0.974 & 2f-17 & \\
H$_2$ L5P2 & 1040.366 & 1.006 & 2f-18 & \\
H$_2$ L5P3 & 1043.502 & 1.049 & 2g-3 & Blended with \ion{H}{1} Ly$\nu$ at $z=0.13847$\\
\ion{Ar}{1} & 1048.220 & 2.440 & 2g-7 & \\
H$_2$ L4R0 & 1049.367 & 1.388 & 2g-9 & \\
H$_2$ L4R1 & 1049.960 & 1.210 & 2g-10 & \\
H$_2$ L4P1 & 1051.033 & 0.910 & 2g-11 & Blend with \ion{H}{1} Ly$\theta$ at $z = 0.13847$\\
H$_2$ L4R2 & 1051.498 & 1.165 & 2g-12 & $W_\lambda = 40$\,m\AA; used for comparison with H$_2$ L14R2 in Figure\,\ref{o3fig}\\
H$_2$ L4P2 & 1053.284 & 0.976 & 2g-13 & \\
H$_2$ L4R3 & 1053.976 & 1.146 & 2g-14 & \\
\ion{Fe}{2} & 1055.262 & 0.812 & 2g-17 & \\
H$_2$ L4P3 & 1056.472 & 1.004 & 2g-18 & Weak line\\
H$_2$ L3R0 & 1062.882 & 1.276 & 2h-2 & \\
\ion{Fe}{2} & 1063.176 & 1.765 & 2h-3 & \\
H$_2$ L3R1 & 1063.460 & 1.101 & 2h-4 & \\
\ion{Fe}{2} & 1063.972 & 0.704 & 2h-5 & \\
H$_2$ L3P1 & 1064.606 & 0.780 & 2h-6 & \\
H$_2$ L3R2 & 1064.996 & 1.057 & 2h-7 & $W_\lambda = 37$\,m\AA; used for comparison with H$_2$ L14R2 in Figure\,\ref{o3fig}\\
\ion{Ar}{1} & 1066.660 & 1.857 & 2h-8 & \\
H$_2$ L3P2 & 1066.905 & 0.877 & 2h-9 & \\
H$_2$ L3R3 & 1067.479 & 1.033 & 2h-10 & \\
H$_2$ L2R0 & 1077.140 & 1.092 & 2h-13 & \\
H$_2$ L2R1 & 1077.700 & 0.927 & 2h-14 & \\
H$_2$ L2P1 & 1078.927 & 0.624 & 2h-15 & \\
H$_2$ L2R2 & 1079.226 & 0.868 & 2h-16 & \\
\ion{Fe}{2} & 1081.875 & 1.134 & 2i-2 & \\
\ion{N}{2} & 1083.994 & 2.079 & 2i-3 & HVC component at +184 \kms \\
H$_2$ L1R0 & 1092.195 & 0.809 & 2i-4 & \\
H$_2$ L1R1 & 1092.732 & 0.627 & 2i-5 & \\
H$_2$ L1P1 & 1094.052 & 0.334 & 2i-9 & \\
H$_2$ L1R2 & 1094.244 & 0.553 & 2i-10 & \\
\ion{Fe}{2} & 1096.877 & 1.554 & 2i-12 & HVC component at +184 \kms~?\\
H$_2$ L0R0 & 1108.127 & 0.264 & 2j-2 & \\
H$_2$ L0R1 & 1108.633 & 0.076 & 2j-3 & \\
H$_2$ L0P1 & 1110.063 & --0.197 & 2j-4 & \\
H$_2$ L0R2 & 1110.120 & 0.014 & 2j-5 & \\
\ion{Fe}{2} & 1112.048 & 0.695 & 2j-6 & \\
\ion{Fe}{2} & 1121.975 & 1.512 & 2k-1 & \\
\ion{Fe}{3} & 1122.524 & 1.786 & 2k-2 & \\
\ion{Fe}{2} & 1125.448 & 1.244 & 2k-3 & \\
\ion{Fe}{2} & 1127.098 & 0.102 & 2k-5 & May blend with \ion{N}{3} at $z=0.13847$\\
\ion{Fe}{2} & 1133.665 & 0.728 & 2k-6 & \\
\ion{N}{1} & 1134.165 & 1.219 & 2k-7 & \\
\ion{N}{1} & 1134.415 & 1.512 & 2k-8 & \\
\ion{N}{1} & 1134.980 & 1.674 & 2k-9 & \\
\ion{Fe}{2} & 1142.366 & 0.661 & 2l-2 & \\
\ion{Fe}{2} & 1143.226 & 1.342 & 2l-3 & HVC component at +184 \kms~?\\
\ion{Fe}{2} & 1144.938 & 1.978  & 2l-4 & HVC component at +184 \kms\\
\ion{P}{2} & 1152.818 & 2.451 & 2l-7 & \\
\ion{Si}{2} & 1190.416 & 2.541 & 3b-1 & HVC component at +184 \kms \\
\ion{Si}{2} & 1193.290 & 2.842 & 3b-2 & Blend with \ion{S}{3} $\lambda1193.208$, HVC at +184 \kms \\
\ion{Mn}{2} & 1197.184 & 2.414 & 3b-4 & Tentative identification\\
\ion{N}{1} & 1199.550 & 2.199 & 3b-5 & \\
\ion{N}{1} & 1200.223 & 2.018 & 3b-6 & \\
\ion{N}{1} & 1200.710 & 1.715 & 3b-7 & \\
\ion{Si}{3} & 1206.500 & 3.293 & 3b-10 & HVC component at +184 \kms \\
\ion{H}{1} Ly$\alpha$ & 1215.670 & 2.704 & 3c-3 \ & $\oplus$ airglow emission present \\
\ion{Mg}{2} & 1239.925 & --0.106 & 3d-4 & Weak line, \ion{Mg}{2} $\lambda1240.395$ absent \\
\ion{S}{2} & 1250.584 & 0.832 & 3e-2 & \\
\ion{S}{2} & 1253.805 & 1.136 & 3e-3 & HVC at +184 \kms\ blends with Ly$\alpha$ at $z=0.03223$\\
\ion{S}{2} & 1259.518 & 1.320 & 3e-5 & HVC at +184 \kms\ blends with \ion{Si}{2} $\lambda1260.422$ \\
\ion{Si}{2} & 1260.422 & 3.171  & 3e-6 & \\
\ion{C}{1} & 1277.245 & 2.037 & 3f-2 \\
\ion{O}{1} & 1302.168 & 1.796 & 3g-3 & \\
\ion{Si}{2} & 1304.370 & 2.052 & 3g-4 & \\
\ion{Ni}{2} & 1317.217 & 2.284  & 3h-3 & HVC component at +184 \kms?\\
\ion{C}{1} & 1328.833 & 2.003 & 3i-2 & Weak line \\
\ion{C}{2} & 1334.532 & 2.234 & 3i-3 & HVC component blends with \ion{C}{2}$^* \lambda1335.708$ \\
\ion{C}{2}$^*$ & 1335.708 & 2.188 & 3i-4 & No HVC component visible \\
\ion{Ni}{2} & 1370.132 & 2.023  & 3k-1 & \\
\ion{Si}{4} & 1393.755\tablenotemark{c}& 2.854 & 3l-1 & HVC component at +184 \kms \\
\ion{Si}{4} & 1402.770\tablenotemark{c} & 2.552 & 3l-2 & HVC component at +184 \kms \\
\ion{Ni}{2} & 1454.842 & 1.672 & 3o-1 & Tentative identification \\
\ion{Si}{2} & 1526.707 & 2.307 & 3r-2 & HVC component at +184 \kms \\
\ion{C}{4} & 1548.195\tablenotemark{d} & 2.468 & 3t-1 & HVC component at +184 \kms \\
\ion{C}{4} & 1550.770\tablenotemark{d} & 2.167 & 3t-2 & HVC component  at +184 \kms \\
\ion{Fe}{2} & 1608.451 & 1.968 & 3w-1 & HVC component at +184 \kms \\
\ion{C}{1}  & 1656.928 & 2.392 & 3y-1 & Weak line\\
\ion{Al}{2} & 1670.789 & 3.463 & 3z-1& HVC  at +184 \kms\ falls between orders\\
\ion{Fe}{2} & 2260.780 & 0.742 & 4a-1\\
\ion{Fe}{2} & 2344.214 & 2.427 & 4b-1& HVC component  at +184 \kms \\
\ion{Fe}{2} & 2374.461 & 1.889 & 4c-1& HVC component  at +184 \kms \\
\ion{Fe}{2} & 2382.765 & 2.882 & 4c-2& HVC component  at +184 \kms \\
\ion{Mn}{2} & 2576.877 & 2.969 & 4d-1\\
\ion{Fe}{2} & 2586.650 & 2.252 & 4d-2& HVC component  at +184 \kms\\
\ion{Mn}{2} & 2594.499 & 2.860 & 4e-1\\
\ion{Fe}{2} & 2600.173 & 2.793 & 4e-2& HVC component  at +184 \kms\\
\ion{Mn}{3} & 2606.462 & 2.712 & 4e-3& HVC component  at +184 \kms\\
\ion{Mg}{2} & 2796.354 & 3.234 & 4f-1& HVC component  at +184 \kms\\
\ion{Mg}{2} & 2803.531 & 2.933 & 4f-2& HVC component  at +184 \kms\\
\enddata
\tablenotetext{a}{Wavelengths and $f$-values are from Morton (2003), except as noted.}
\tablenotetext{b}{The listed values correspond to the line 
identification numbers of each interstellar feature shown in Figures~\ref{fusespec},
\ref{e140mspec}, and \ref{e230mspec}.}
\tablenotetext{c}{\ion{Si}{4} wavelength from Morton (1991).  Morton (2004) prefers 
\ion{Si}{4} wavelengths of 1393.760 and 1402.773\,\AA.}
\tablenotetext{d}{\ion{C}{4} wavelength from Morton (1991).  Morton (2004) prefers \ion{C}{4} wavelengths of 1548.202 and 1550.781\,\AA.}
\tablenotetext{e}{Line is not shown in Figure~\ref{e140mspec} since it occurs at the edge of a
STIS echelle order and is only partially detected.}

\end{deluxetable}

\clearpage
\newpage
\begin{deluxetable}{ccccccccc}
\tabletypesize{\scriptsize}
\tablewidth{0pt} 
\tablecaption{Ly$\alpha$ Absorber Summary\tablenotemark{a} \label{tab_summarylya}}
\tablehead{$\lambda$ & $z$  & $W_\lambda$(Ly$\alpha$) & b(\ion{H}{1}) & N(\ion{H}{1})\tablenotemark{b} &  b(\ion{O}{6}) & N(\ion{O}{6})\tablenotemark{c} & $\frac{N({\rm H\,I})}{N({\rm O\,VI})}$ & Note\tablenotemark{d} \\
(\AA) &&  (m\AA) &(\kms) & (cm$^{-2}$) &(\kms) & (cm$^{-2}$) }
\startdata
\medskip
1221.66 & 0.00493  & $\phn95\pm11$  &   $34.2\pm3.6$ & $(2.28\pm0.32)\times10^{13}$& \nodata & $<2.35\times10^{13}$ & $>1.0$ & \\
\medskip
1235.55 & 0.01635  & $113\pm10$     &  $48.5\pm5.1$&   $(2.45\pm0.34)\times10^{13}$ & \nodata & $<2.36\times10^{13}$   & $>1.0$ \\
\medskip
1239.05 & 0.01923 & $\phn40\pm14$ & \nodata &$(7.21\pm2.52)\times10^{12}$ & \nodata & \nodata & \nodata & 1\\
\medskip
1250.04 & 0.02827  & $219\pm07$     &  $31.4\pm1.1$  & $(6.31\pm0.32)\times10^{13}$  & \nodata & $<2.56\times10^{13}$  &$>2.5$ \\
\medskip
1254.85 & 0.03223  & $\phn93\pm09$  & $31.6\pm2.9$ &  $(2.12\pm0.27)\times10^{13}$ & \nodata & $<2.78\times10^{13}$  & $>0.8$ & 2\\
\medskip
1265.82 & 0.04125 &  $\phn81\pm17$  & $105\pm18$ &  $(1.77\pm0.40)\times10^{13}$ & $35\pm15$ &$(2.15:\pm0.80)\times10^{13}$ & $0.8\pm0.4$& 3\\
\medskip
1276.31 & 0.04996 & $\phn30\pm06$ & $16.5\pm3.2$ & $(5.25\pm1.05)\times10^{12}$ & \nodata & $<9.98\times10^{13}$ & $>0.05$ & 4\\
1287.33 & 0.05895 &  $172\pm11$     & $21, 30$  &  $(3.63\pm0.42)\times10^{13}$ & $27\pm09$& $(3.54\pm0.98)\times10^{13}$ & $1.0\pm0.3$& 5\\
\medskip
1287.69 & 0.05928 &  $15\pm05$    & $10$  &  $(2.60\pm0.87)\times10^{12}$ & \phn$5\pm05$ & $(2.45\pm0.78)\times10^{13}$ & $0.11\pm0.05$ & 6\\
\medskip
1289.49 & 0.06072 &  $\phn85\pm09$  & $55.4\pm5.8$& $(1.89\pm0.25)\times10^{13}$ & \nodata&  $<2.94\times10^{13}$ & $>0.6$& \\
\medskip
1291.58 & 0.06244 & $\phn79\pm10$   & $77.3\pm9.0$   &$(1.53\pm0.23)\times10^{13}$ & $\phn8\pm07$ & $(1.39\pm0.56)\times10^{13}$ & $1.1\pm0.5$ & 7 \\
\medskip
1303.05 & 0.07188 & $\phn36\pm06$ & $\phn9.6\pm2.0$ &$(6.17\pm1.03)\times10^{12}$ & \nodata & $<1.86\times10^{13}$ & $>0.3$ & 8 \\
\medskip
1314.09 & 0.08096 & $124\pm06$       & $24.9\pm1.0$ & $(2.79\pm0.16)\times10^{13}$  & \nodata & $<2.22\times10^{13}$ & $>1.3$ & \\
1320.06 & 0.08587 & $\phn39\pm10$    & $52\pm14$ & $(7.92\pm2.76)\times10^{12}$ & \nodata & $<2.65\times10^{13}$ & $>0.4$ & 9\\
\medskip
1320.61 & 0.08632 & $\phn20\pm08$    & $36\pm15$ & $(4.54\pm2.37)\times10^{12}$ & \nodata & $<2.20\times10^{13}$ & $>0.2$ &9\\
\medskip
1328.47 & 0.09279 & $121\pm15$  & $133\pm17$& $(2.48\pm0.47)\times10^{13}$ & \nodata & $<4.38\times10^{13}$ & $>0.6$ & 10\\
\medskip
1337.27 & 0.10003 & $\phn32\pm05$ & $23.2\pm2.2$&$(5.34\pm0.83)\times10^{12}$ & \nodata & \nodata & \nodata & 11\\
\medskip
1360.27 & 0.11895 & $138\pm09$    & $31.5\pm1.8$  & $(2.76\pm0.22)\times10^{13}$ & \nodata & $<2.78\times10^{13}$ & $>1.0$\\
\medskip
1375.54 & 0.13151 & $132\pm08$   & $28.8\pm1.3$ & $(2.55\pm0.16)\times10^{13}$ & \nodata & $<4.63\times10^{13}$ & $>0.6$\\
\medskip
 1378.21 & 0.13370 & $\phn97\pm12$ & $83.6\pm10.4$ & $(1.86\pm0.31)\times10^{13}$ & \nodata  & $<3.17\times10^{13}$& $>0.6$ & 12\\
\medskip
1384.00 & 0.13847  & $535\pm12$ & $22.4\pm0.3$ & $(1.57\pm^{0.18}_{0.14})\times10^{16}$ & $30\pm06$ & $(4.79\pm^{1.26}_{0.77})\times10^{13}$ & $328\pm^{62}_{92}$ & 13\\
1416.84 & 0.16548 & $128\pm07$ & $29.9\pm1.4$   & $(2.48\pm0.15)\times10^{13}$ & 20, 8, 8 & $(1.21\pm0.15)\times10^{14}$ & $0.20\pm0.03$& 14\\
1417.23 & 0.16580 & $\phn57\pm05$ & $33\pm6$ & $(9.00\pm1.00)\times10^{12}$ & \nodata & $<2.55\times10^{13}$ & $>0.3$\\
1417.59 & 0.16610 & $368\pm08$ & $22.0\pm0.9$ & $(4.17\pm^{0.59}_{0.48})\times10^{14}$ & \nodata & $<3.90\times10^{13}$ & $>10.7$ &15\\
\medskip
1418.44 & 0.16686 & $209\pm10$ & $38.5\pm1.2$ & $(4.75\pm0.20)\times10^{13}$ & \nodata & $<3.08\times10^{13}$  & $>1.5$ & 16\\
1426.47 & 0.17340 & $\phn66\pm03$ & $13.5\pm0.8$ & $(1.24\pm0.09)\times10^{13}$ & $43\pm09$ & $(2.72: \pm0.54)\times10^{13}$ & $0.5\pm0.1$ & 17 \\
\medskip
 1426.71 & 0.17360 & $233\pm05$ & $16.8\pm^{1.5}_{0.9}$ & $(1.28\pm^{0.26}_{0.23})\times10^{14}$ & \nodata & $<2.04\times10^{13}$ & $>6.3$ & 18
\enddata
\tablenotetext{a}{Columns include the redshift of the absorption system,
observed equivalent width of the \ion{H}{1} Ly$\alpha$ absorption,
Doppler parameter of the \ion{H}{1} absorption based on the Ly$\alpha$ line 
width unless indicated otherwise, \ion{H}{1} column density, Doppler parameter 
of the \ion{O}{6} based on a single-component fit to the observed \ion{O}{6}
lines, \ion{O}{6}
column density, \ion{H}{1} to \ion{O}{6} column density ratio, and system-specific
notes.  All errors are $1\sigma$ estimates.  Column density limits are $3\sigma$ 
estimates.}
\tablenotetext{b}{The  
\ion{H}{1} column density is based on a Voigt profile fit 
to the Ly$\alpha$ line unless indicated otherwise. }
\tablenotetext{c}{The \ion{O}{6} column density is based on the assumption of a 
linear curve of growth unless indicated otherwise. }
\tablenotetext{d}{Notes -- see next page.  Additional information can be found in the intergalactic 
absorption overview in \S5.}
\end{deluxetable}

\clearpage
\newpage
\noindent Notes -- 
(1) No known ISM feature occurs 
at the wavelength of this \ion{H}{1} line.  \ion{N}{5} $\lambda1238.821$ 
in the main Galactic absorption would be $\approx100$ \kms\ blueward of 
this position.  Tripp et al. (1998) identified a strong Ly$\alpha$ absorber at 
1239.4\,\AA\ ($z=0.01950$) with $W_{obs} = 127\pm24$\,m\AA, but that feature
is not detected in our spectrum.
(2)~\ion{H}{1} integration cut off at --60 \kms\ to 
avoid contamination from high-velocity Galactic \ion{S}{2} $\lambda1253.811$ 
absorption. 
(3)~Broad Ly$\alpha$ line, \ion{H}{1} integration range of --175 to +175 \kms.
The tentative \ion{O}{6} detection is based on the assumption of a linear COG fit to
only one line. b(\ion{O}{6}) is based on a single-component fit to this 
$\lambda1031.926$ line.
(4) No ISM or IGM lines are expected at the wavelengths of this \ion{H}{1}
feature.  
Galactic \ion{C}{1} $\lambda1276.482$ ($\log f\lambda = 0.876$) 
is excluded since it is expected to 
be a factor of 4.5 weaker than \ion{C}{1} $\lambda1280.135$ 
($\log f\lambda = 1.527$, $W_\lambda < 15$\,m\AA) and 
factors of 13 and 16 times weaker than \ion{C}{1} $\lambda1328.833$ 
($\log f\lambda = 2.003$, $18\pm6$ m\AA) 
and $\lambda1560.309$ ($\log f\lambda = 2.082$, $19\pm6$ m\AA). See 
Morton (2003) for \ion{C}{1} $f$-values.   The redshifted \ion{O}{6}
$\lambda1031.926$ line falls very close to Galactic \ion{N}{2} $\lambda1083.990$
absorption, so the \ion{O}{6} limit is based on the non-detection of the 
$\lambda1037.617$ line in the SiC channels.
(5)~Asymmetric Ly$\alpha$ line with at least two components, including a negative 
velocity wing.  Listed \ion{H}{1} b-values are
for the wing and main absorption, respectively.  The \ion{O}{6} column density
is an estimate after removal of a nearby contaminating interstellar H$_2$ line.
b(\ion{O}{6}) is based on a single-component fit to the $\lambda1031.926$ line
after removal of the interstellar H$_2$ line.  The true \ion{O}{6} b-value could 
be larger if this H$_2$ line is weaker than expected.
(6) ``Satellite'' absorber at +93 \kms\ relative to the $z=0.05895$ absorber
rest frame.  The \ion{O}{6} column 
density is based on a linear COG fit to both lines of doublet.  The \ion{O}{6} 
is offset $\sim10$ \kms\ blueward of the corresponding Ly$\alpha$ absorption.
The value of b(\ion{O}{6}) is highly uncertain because the line is narrower than 
the FUSE instrumental width.  
(7)~Broad Ly$\alpha$ line, \ion{H}{1} integration range of --150 to +150 \kms.
The tentative \ion{O}{6} detection is based on the assumption of a linear COG fit to
only one line.
(8) \ion{H}{1} blends with high-velocity Galactic \ion{O}{1} 
$\lambda1302.168$ absorption near v$_{LSR} \approx 205$ \kms.  
The Galactic \ion{O}{1} line shape at high velocities is 
inconsistent with other strong metal-line (e.g., \ion{Si}{2} and perhaps 
\ion{C}{2}) absorption at similar velocities.  See Ganguly et al. (2004) 
for a discussion of the Galactic \ion{O}{1} profile.
(9)~Weak, broad Ly$\alpha$ line, which may have two components separated by
about 125 \kms.  The listed values for these two components assume that the 
recovery between components is real.  The integration ranges were --80 to +80
\kms\ ($z=0.08587$) and --50 to +50 \kms\ ($z=0.08632$).
(10)~Broad Ly$\alpha$ line, \ion{H}{1} integration range of --150 to +125 \kms.  A
narrow Galactic \ion{C}{1} absorption feature is present within the Ly$\alpha$ profile;
this Galactic feature is not included in the Ly$\alpha$ strength estimate.
(11) The \ion{O}{6} lines in this system blend with Galactic ISM lines
or other IGM lines.  
(12)~Broad Ly$\alpha$ line, \ion{H}{1} integration range of --100 to +185 \kms.
(13) N(\ion{H}{1}) and \ion{H}{1} b-value are based on COG fit to the numerous Lyman-series
absorption lines present in this system.  Ly$\theta$ was excluded from the fit since it
is contaminated. b(\ion{O}{6}) is based on a weighted average of the FUSE and 
STIS results for a single-component fit to the $\lambda1031.926$ line.
(14)~\ion{H}{1} integration range truncated at +55 to avoid main Ly$\alpha$
component at $z=0.16610$.  Ly$\alpha$ fit yields same results for N and b-value
as COG fit to the Ly$\alpha$ and Ly$\beta$ lines. The \ion{O}{6} b-values
are estimates made by decomposing the line into three components (see text).
(15)~\ion{H}{1} integration truncated at -40 \kms.  
N and b-value are based on a COG fit to the Ly$\alpha$,
Ly$\beta$, and Ly$\epsilon$ lines.
(16)~\ion{H}{1} integration does not include weak ($W_{obs} = 27$ m\AA, 
$N=4.4\times10^{12}$ cm$^{-2}$) absorption wing from +65 to +145 \kms. 
(17)~\ion{O}{6} column density is based on the $\lambda1031.926$ line, which 
falls in the damping wing of Galactic \ion{H}{1} Ly$\beta$. The \ion{O}{6} b-value
is based on a single-component fit to the $\lambda1031.926$ line.  
(18)~\ion{H}{1} column density is based on a COG fit to the Ly$\alpha$ and 
Ly$\beta$ lines observed by STIS.  The \ion{O}{6} column density limit is 
based on the $\lambda1031.926$ line, which 
falls in the damping wing of Galactic \ion{H}{1} Ly$\beta$.

\clearpage
\newpage
\begin{deluxetable}{lcccccl}
\tabletypesize{\footnotesize}
\tablewidth{471pt} 
\tablecaption{Equivalent Widths of Intergalactic Absorption Lines at $z=0.13847$ \label{tab_ew13847}}
\tablehead{Species & $\lambda$ & $\lambda_{obs}$ & $\log f\lambda$\tablenotemark{b} & \multicolumn{2}{c}{W$_{obs}$\tablenotemark{c}}
& \multicolumn{1}{c}{Comments} \\
& (\AA) & (\AA) & & (m\AA) & (m\AA)}
\startdata
\ion{H}{1} Ly$\alpha$  & 1215.670  & 1384.004 & \phn2.704& \multicolumn{2}{c}{535$\pm$12} \\
\ion{H}{1} Ly$\beta$   & 1025.722  & 1167.754 & \phn1.909 & 350$\pm$14& 344$\pm$17 \\
\ion{H}{1} Ly$\gamma$  & \phn972.537 & 1107.204 &  \phn1.450 & 342$\pm$12 & 337$\pm$12 & Positive velocity wing? \\
\ion{H}{1} Ly$\delta$  & \phn949.743 & 1081.254 & \phn1.122 & 263$\pm$18 & 250$\pm$26 & $W_\lambda$(SiC1)=277$\pm$25\,m\AA\\
\ion{H}{1} Ly$\epsilon$  & \phn937.804 & 1067.662 & \phn0.864 & 234$\pm$12 & 213$\pm$15\\
\ion{H}{1} Ly$\zeta$   & \phn930.748 & 1059.629 & \phn0.652 & 180$\pm$10 & 190$\pm15$\\
\ion{H}{1} Ly$\eta$  & \phn926.226 & 1054.481 & \phn0.470 &  175$\pm$10 & 186$\pm$15\\
\ion{H}{1} Ly$\theta$  & \phn923.150 & 1050.979 & \phn0.311 & 184$\pm$10 & 210$\pm$18 & Blend with H$_2$ L4P1\\
\ion{H}{1} Ly$\iota$ &   \phn920.963 & 1048.489 & \phn0.170 & 140$\pm$10 & 135$\pm$16\\
\ion{H}{1} Ly$\kappa$ &  \phn919.351 & 1046.654 & \phn0.043 & \phn97$\pm$10 & 127$\pm$16\\
\ion{H}{1} Ly$\lambda$  &  \phn918.129 & 1045.262 & --0.073 & \phn97$\pm$12 & \phn80$\pm$15\\
\ion{H}{1} Ly$\mu$     &  \phn917.181 & 1044.183& --0.179 & \phn69$\pm$10 & \phn75$\pm$14\\
\ion{H}{1} Ly$\nu$     &  \phn916.428 & 1043.326& --0.283 & \nodata & \nodata & Blend with H$_2$ L5P3 \\
\\
\ion{C}{2} & \phn903.624 & 1028.749 & \phn2.181 & \phn96$\pm10$ & \phn64$\pm$14 & Possible FPN in LiF1\\
\ion{C}{2} & \phn903.962 & 1029.134 & \phn2.482 & \phn72$\pm11$ & \phn66$\pm$10 \\
\ion{C}{2} & 1036.337 & 1179.839 & \phn2.106 & \phn52$\pm10$ & \phn53$\pm$10 \\
\ion{C}{2} & 1334.532 & 1519.325 & \phn2.232 & \multicolumn{2}{c}{87$\pm$8} \\
\ion{C}{3} & \phn977.020 & 1112.388 & \phn2.872 & 118$\pm09$ & 129$\pm$10\\
\\
\ion{N}{2} & 1083.990 & 1234.090 & \phn2.048 & \multicolumn{2}{c}{48$\pm$6} \\
\ion{N}{3} & \phn989.799 & 1126.857 & \phn2.023 & \phn67$\pm$11 & \phn72$\pm$8 &
$W_\lambda$(\ion{Si}{2} $\lambda989.873$) $\lesssim10$\,m\AA\\
\ion{N}{5} & 1242.804 & 1414.895 & \phn1.988 & \multicolumn{2}{c}{5$\pm$5} \\
\ion{N}{5} & 1238.821 & 1410.361 &  \phn2.289 & \multicolumn{2}{c}{15$\pm$5} \\
\\
\ion{O}{1}  & 1302.168    & 1482.479 & \phn1.804 &  \multicolumn{2}{c}{$<21$} \\
\ion{O}{2}  & \phn832.757 & \phn948.069 & \phn1.568 &  \phn26$\pm$17 &  \phn45$\pm$12 & Line blended in SiC1 data\\
\ion{O}{6}  & 1031.926    & 1174.817 & \phn2.137 &  \phn54$\pm$13 & \phn78$\pm$15 \\
            &             &          && \multicolumn{2}{c}{84$\pm$21} & STIS E140M value \\
\ion{O}{6}  & 1037.617    & 1181.296 & \phn1.836& $<48$  & \phn70$\pm$30 & At edge of detector 2\\
            &             &          && \multicolumn{2}{c}{$<60$} & STIS E140M value \\
\\
\ion{Si}{2} & 1260.422 & 1434.953 & \phn3.148 & \multicolumn{2}{c}{63$\pm$6}\\
\ion{Si}{2} & 1193.290 & 1358.525 & \phn2.775 & \multicolumn{2}{c}{40$\pm$8}\\
\ion{Si}{2} & 1190.416 & 1355.253 & \phn2.474 & \multicolumn{2}{c}{28$\pm$10} & Excludes broad absorption nearby \\
\ion{Si}{2} & 1304.370 & 1484.872 & \phn2.086 & \multicolumn{2}{c}{18$\pm$9} & Marginal detection \\
\ion{Si}{3} & 1206.500 & 1373.564 & \phn3.304 & \multicolumn{2}{c}{98$\pm$6}\\
\ion{Si}{4} & 1393.755 & 1586.748 & \phn2.855 & \multicolumn{2}{c}{41$\pm$12}\\
\ion{Si}{4} & 1402.770 & 1597.012 & \phn2.554 & \multicolumn{2}{c}{$<30$} \\
\\
\ion{S}{2} & 1259.519 & 1433.925 & \phn1.311 & \multicolumn{2}{c}{$<12$}\\
\ion{S}{3} & 1012.502 & 1152.703 & \phn1.556 & \nodata & \nodata & Blends with Galactic \ion{P}{2} $\lambda1152.818$ \\
\\
\ion{Fe}{2} & 2382.765 & 2712.707 & \phn2.855 & \multicolumn{2}{c}{38$\pm$19} & STIS E230M measurement
\enddata
\tablenotetext{a}{Errors are $1\sigma$ estimates and limits are $3\sigma$ estimates. }
\tablenotetext{b}{$f$-values are from Morton (1991) for $\lambda_{\rm rest} > 912$\,\AA.
$f$-values are from Verner et al. (1994) for $\lambda_{\rm rest} < 912$\,\AA.}
\tablenotetext{c}{Equivalent width of feature in m\AA.  Below 1200\,\AA, two
values are listed for the LiF1 and LiF2 spectra recorded by the two FUSE detectors.   
Above 1200\,\AA,
the single value listed is the HST/STIS E140M or E230M measurement.}
\end{deluxetable}

\clearpage
\newpage
\begin{deluxetable}{lcccc}
\tabletypesize{\small}
\tablewidth{460pt} 
\tablecaption{Column Densities for the $z=0.13847$ Absorber \label{tab_col13847}}
\tablehead{Species  & b\tablenotemark{a} & N & Method\tablenotemark{b} & Note\tablenotemark{c}\\
& (\kms) & (cm$^{-2}$)  }
\startdata
\ion{H}{1} & ~~~~~~$22.4\pm0.3$~~~~~~&~~~~~~$(1.57\pm^{0.18}_{0.14})\times10^{16}$~~~~~~
& ~~~~~~COG, LL~~~~~~ & ~~~~~~1~~~~~~\\
\\
\ion{C}{2} & $9.6\pm0.7$& $>6.24\times10^{13}$ & AOD limit & 2\\
\\
\ion{C}{3}& $<13.8$ & $>3.05\times10^{13}$ & AOD limit & 3\\
\\
\ion{N}{2} & $11.1\pm1.4$ & $>5.64\times10^{13}$ & AOD limit & 4\\
\\
\ion{N}{3} & $<17.3$ & $>9.16\times10^{13}$ & AOD limit & 5\\
\\
\ion{N}{5} & \nodata & $(6.04\pm2.01)\times10^{12}$ & AOD, LC\\
\\
\ion{O}{1} & \nodata & $<2.52\times10^{13}$ & LC \\
\\
\ion{O}{2} & \nodata & $(2.16\pm0.44)\times10^{14}$ & LC & 6\\
\\
\ion{O}{6} & $25\pm5$& $(4.79\pm^{1.26}_{0.77})\times10^{13}$ & AOD & 7\\
\\
\ion{Si}{2} & $5.4\pm^{1.8}_{1.2}$ & $(7.94\pm^{4.36}_{1.92})\times10^{12}$ & COG & 8 \\
	    & 			   & $(8.6\pm2.3)\times10^{12}$ & PF & 9\\
	    & 			   & $(1.11\pm^{0.14}_{0.27})\times10^{13}$ & AOD & 10\\
\\
\ion{Si}{3} & $8.6\pm0.4$ & $>8.38\times10^{12}$ & AOD limit & 11\\
\\
\ion{Si}{4} & $7.8\pm2.0$& $(6.73\pm1.88)\times10^{12}$ & AOD & 12\\
\\
\ion{S}{2} & \nodata & $<4.62\times10^{13}$ & LC \\
\\
\ion{Fe}{2} & \nodata & $<3.32\times10^{12}$ & LC \\
\enddata
\tablenotetext{a}{Effective b-value assuming a single-component absorption.  Some of the 
absorption features may contain more than one component, but most of the column density is
confined to one dominant component or one set of blended components.}
\tablenotetext{b}{Method for calculating the column densities: apparent optical depth (AOD),
single-component Doppler-broadened curve of growth (COG), linear curve of growth (LC), 
Lyman-limit optical depth
(LL), or profile fitting (PF).  Errors are $1\sigma$ estimates.  Upper limits are 
$3\sigma$ estimates.  Lower limits derived from the AOD method are the nominal
AOD column density, which is a lower limit if unresolved saturated structures are 
present in the profiles.}
\tablenotetext{c}{Notes -- see next page.}
\end{deluxetable}

\clearpage
\newpage
\noindent 
$^{\rm c}$Notes:
(1)~Lyman-limit optical depth and COG yield consistent values of N(\ion{H}{1}).  
(2)~Limit on N(\ion{C}{2}) derived from AOD method applied to $\lambda1334.532$ line 
observed by STIS. b-value derived from a single-component fit to the $\lambda1334.532$ line.
(3) b-value limit set by single-component fit to $\lambda977.020$ line in FUSE band.
(4) b-value derived from single-component fit to $\lambda1083.990$ line in STIS band.
(5) b-value limit set by single-component fit to $\lambda989.799$ line in FUSE band.
(6) Value for N(\ion{O}{2}) is somewhat uncertain due to complexity of
spectrum near line (see Figure~\ref{o3fig}).  Value quoted is based on 
SiC2 data.
(7)~N(\ion{O}{6}) is a weighted average of the apparent column densities derived for 
the $\lambda1031.926$ line in the STIS E140M data and the FUSE LiF1B and LiF2A data.
The \ion{O}{6} b-value is from the fit to the $\lambda1031.926$ line in the FUSE 
LiF1 channel.  The b-value derived from a fit to the STIS data for the $\lambda1031.926$ 
line is $37\pm7$ \kms.
(8)~Column density and b-value are from 
\ion{Si}{2} COG fit to the $\lambda\lambda$1260.422, 1193.290, 1190.416,
and 1304.370 equivalent widths listed in Table~\ref{tab_ew13847}.
(9) PF value based on simultaneous two- or three-component fits to the $\lambda\lambda1260.422, 
1192.290, 1190.416, 1304.370$ lines.  The fit parameters could be varied widely and are 
poorly constrained.  The uncertainty in N(\ion{Si}{2}) reflects a wide range of possible
fits.
(10) AOD value based on direct integration of the $\lambda1190.416$ and $\lambda1193.290$ lines.  This value is somewhat larger than the COG value,
which is more heavily weighted by the stronger $\lambda1260.422$ line.
(11) b-value derived from a single-component fit to the $\lambda1206.500$ line.
(12)~AOD column density and b-value provided by the stronger \ion{Si}{4} $\lambda1393.755$ line.

\clearpage
\newpage
\begin{deluxetable}{lcccc}
\tabletypesize{\footnotesize}
\tablewidth{0pt} 
\tablecaption{Photoionization Constraints for the $z=0.13847$ Absorber \label{tab_mod13847}}
\tablehead{Species & log~[N(cm$^{-2}$)]\tablenotemark{a} & \multicolumn{3}{c}{$\log U$\tablenotemark{b}} \\
& & $\log Z = -1.0$ & $\log Z = -0.5$ & $\log Z = 0.0$}
\startdata

\ion{C}{2} & $>13.79$ & None & $[-2.8, -2.4]$ & $[-3.5, -1.6]$\\
\\
\ion{C}{3} & $>13.48$ & $>-3.3$ & $>-3.5$ & $>-3.5$\\
\\
\ion{N}{2} & $>13.75$ & None & None & None\\
\\
\ion{N}{3} & $>13.96$ & $>-2.5$ & $>-2.9$ & $>-3.2$\\
\\
\ion{N}{5} & $12.60-12.91$ & $[-2.3, -2.2]$ & $[-2.5, -2.4]$ & $[-2.6, -2.5]$\\
\\
\ion{O}{1} & $<13.40$ & All & All &  All\\
\\
\ion{O}{2} & $14.24-14.42$ & None & None  &  $[-2.5,-2.2]$\\
\\
\ion{O}{6} & $13.60-13.78$ &  $[-2.0, -1.9]$ & $-2.0$ & $[-2.2, -2.1]$\\
\\
\ion{Si}{2} & $12.92-13.10$ & None & None & $[-2.4, -2.2]$\\
\\
\ion{Si}{3} & $>12.92$ & $[-3.1, -1.8]$ & $[-3.4, -1.5]$ & $[-3.4, -1.3]$\\
\\
\ion{Si}{4} & $12.69-12.94$ & $[-2.6, -1.6]$ & $[-3.0, -2.8]$ & $[-3.1, -3.0]$\\
\\
\ion{S}{2} & $<13.66$ & All & All & All\\
\\
\ion{Fe}{2} & $<12.52$ & All & All & All\\
\enddata
\tablenotetext{a}{Column density range ($\pm1\sigma$) or limit ($3\sigma$)
from Table~\ref{tab_col13847}.}
\tablenotetext{b}{Ionization parameter range satisfying the observed 
column densities for the photoionization model of the $z=0.13847$ absorber
with log~N(\ion{H}{1}) = 16.20 
described in \S6.3.  Values of $\log U$ are given for three different 
metallicities $Z$.  ``None'' indicates that the constraints cannot be 
satisfied for any value of $-3.5 < \log U \le -0.5$.  ``All'' 
indicates that the constraints are satisfied for all values of 
$-3.5 < \log U \le -0.5$. }
\end{deluxetable}

\clearpage
\newpage
\begin{deluxetable}{lccccccl}
\tabletypesize{\footnotesize}
\tablewidth{0pt} 
\tablecaption{Equivalent Widths of Intergalactic Absorption Lines at $z\approx0.166$\label{tab_ew166}}
\tablehead{Species & $\lambda$\tablenotemark{a} & $\lambda_{obs}$\tablenotemark{b} & $\log f\lambda$\tablenotemark{a} & \multicolumn{3}{c}{W$_{obs}$\tablenotemark{c}}
& \multicolumn{1}{c}{Comments} \\
&       & [0.16610] & & [0.166548] & [0.16610] & [0.16686]\\
& (\AA) & (\AA) & & (m\AA) & (m\AA) & (m\AA)}
\startdata
\ion{H}{1} Ly$\alpha$  & 1215.670  & 1417.593 & 2.704 & 128$\pm$07 & 428$\pm$08 & 239$\pm$10\\
\ion{H}{1} Ly$\beta$   & 1025.722  & 1196.094 & 1.909 & \phn16$\pm$12 & 185$\pm09$ & $<68$ & \\
\ion{H}{1} Ly$\gamma$  & \phn972.537 & 1134.075 &1.450 & \nodata & \nodata & \nodata & Blended with Galactic \ion{N}{1}\\
\ion{H}{1} Ly$\delta$  & \phn949.743 & 1107.495 &1.122 & \nodata & \nodata & \nodata & Blended with other lines \\
\ion{H}{1} Ly$\epsilon$  & \phn937.804 & 1093.573 & 0.864 & \nodata & \phn29$\pm$12 & \nodata & Blended with Galactic H$_2$\\
\\
\ion{C}{2} & 1334.532 & 1556.198 & 2.232 & $<42$ & \phn42$\pm$19 & $<52$ \\ 
\ion{C}{3} & \phn977.020 & 1139.303 & 2.872 & $<36$ & \phn68$\pm$14 & $<48$ & FUSE LiF1B values \\
           &             &          &       & $<28$ & \phn72$\pm$12 & $<33$ & FUSE LiF2A values \\
\\
\ion{N}{5} & 1242.804 & 1449.234 &  1.988 & $<20$ & $<27$ & $<24$\\
\ion{N}{5} & 1238.821 & 1444.589 &  2.289 &\phn15$\pm$07 & $<36$ & $<30$ \\
\\
\ion{O}{6}  & 1031.926 &  1203.329 & 2.137 & 124$\pm$12 & $<57$ & $<45$ \\
\ion{O}{6}  & 1037.617 &  1209.965 & 1.836 & \phn72$\pm10$ & $<52$ & $<45$ \\
\\
\ion{Si}{2} & 1260.422 & 1469.778 & 3.148 & $<24$ & $<33$ & $<30$ \\
\ion{Si}{3} & 1206.500 & 1406.900 & 3.304 & $<18$ & \phn29$\pm$09 & $<27$ \\
\ion{Si}{4} & 1393.755 & 1625.258 & 2.855 & $<48$ & $<66$ & $<57$ \\
\ion{Si}{4} & 1402.770 & 1635.770 & 2.554 & $<48$ & $<72$ & $<63$ \\

\enddata
\tablenotetext{a}{Rest wavelengths and $f$-values are from Morton (1991).}
\tablenotetext{b}{Observed wavelength at $z=0.16610$. }
\tablenotetext{c}{Observed equivalent width for the three absorbers at $z=0.16548$, 0.16610, and 0.16686.
Errors are $1\sigma$ estimates.  Limits are $3\sigma$ estimates.}
\end{deluxetable}

\clearpage
\newpage
\begin{deluxetable}{lcccl}
\tablewidth{0pt} 
\tablecaption{Column Densities for the $z\approx0.166$ Absorbers \label{tab_col166}}
\tablehead{Species  & \multicolumn{3}{c}{Column Density\tablenotemark{a}} & \multicolumn{1}{c}{Method\tablenotemark{b}} \\
&  [$z=0.16548$] & [$z=0.16610$] & [$z=0.16686$]\\
& (cm$^{-2}$)  & (cm$^{-2}$) & (cm$^{-2}$)}
\startdata
\ion{H}{1} & $(2.48\pm0.15)\times10^{13}$ & $(4.17\pm^{0.59}_{0.48})\times10^{14}$ & $(4.75\pm0.20)\times10^{13}$ & PF, COG, PF\\
\\
\ion{C}{2} & $<1.79\times10^{13}$ & $(1.79\pm0.81)\times10^{13}$ & $<2.21\times10^{13}$ & LC, LC, LC\\
\\
\ion{C}{3} & $<3.73\times10^{12}$ & $(1.61\pm0.22)\times10^{13}$& $<4.40\times10^{12}$ & LC, AOD, LC\\
\\
\ion{N}{5} & $(6.03\pm2.82)\times10^{12}$  & $<1.45\times10^{13}$ & $<1.21\times10^{13}$ & LC, LC, LC\\
\\
\ion{O}{6} & $(1.21\pm0.15)\times10^{14}$  & $<3.90\times10^{13}$ & $<3.08\times10^{13}$ & AOD, LC, LC\\
\\
\ion{Si}{2} & $<9.22\times10^{11}$ & $<1.27\times10^{12}$ & $<1.15\times10^{12}$ & LC, LC, LC \\
\\
\ion{Si}{3} & $<5.04\times10^{11}$ & $(1.29\pm0.21)\times10^{12}$ & $<7.56\times10^{11}$ & LC, AOD, LC \\
\\
\ion{Si}{4} & $<3.27\times10^{12}$ & $<4.50\times10^{12}$ & $<3.88\times10^{12}$ & LC, LC, LC\\
\enddata
\tablenotetext{a}{Column densities for the three absorbers at $z=0.16548$, 0.16610, and 0.16686.
Errors are $1\sigma$ estimates.  Limits are $3\sigma$ estimates.}
\tablenotetext{b}{Method for calculating the column densities in the three absorbers: 
apparent optical depth (AOD),
single-component Doppler-broadened curve of growth (COG), linear curve of growth (LC), or
profile fitting (PF).}

\end{deluxetable}

\clearpage
\newpage
\begin{deluxetable}{ccccccccc}
\tablewidth{0pt} 
\tablecaption{Broad Ly$\alpha$ Absorbers Toward PG\,1116+215 \label{tab_broadlya}}
\tablehead{$z$ & $\lambda$ & log N(\ion{H}{1}) & b(\ion{H}{1}) & $T$\tablenotemark{a} & log $f_{\rm H}$\tablenotemark{b} & 
log N(\ion{H}{1}+\ion{H}{2})\tablenotemark{c} & \ion{O}{6}\\
& (\AA) & & (\kms) & (K) & & & Detect?}
\startdata
0.01635 & 1235.55 & $13.39\pm^{0.06}_{0.06}$& $48.5\pm5.1$ & $\le1.4\times10^5$ & $\le5.16$ & $\le18.55$ & No\\
\\
0.04125	& 1265.82 & $13.24\pm^{0.88}_{0.11}$ & $105\pm18$   & $\le6.6\times10^5$ & $\le6.35$ & $\le19.60$ & Yes\\
\\
0.06072 & 1289.49 & $13.28\pm^{0.05}_{0.06}$ & $55.4\pm5.8$ & $\le1.8\times10^5$ & $\le5.38$ & $\le18.66$ & No\\
\\
0.06244 & 1291.58 & $13.18\pm^{0.06}_{0.07}$ & $77.3\pm9.0$ & $\le3.6\times10^5$ & $\le5.91$ & $\le19.10$ & Yes\\
\\
0.08587 & 1320.06 & $12.90\pm^{0.13}_{0.19}$ & $52\pm14$ & $\le1.6\times10^5$ & $\le5.28$ & $\le18.18$ & No\\
\\
0.09279 & 1328.47 & $13.39\pm^{0.08}_{0.09}$ & $133\pm17$   & $\le1.1\times10^6$ & $\le6.66$ & $\le20.05$ &No\\ 
\\
0.13370 & 1378.21 & $13.27\pm^{0.07}_{0.08}$ & $83.6\pm10.4$& $\le4.2\times10^5$ & $\le6.03$ & $\le19.30$ & No\\
\\
0.16686 & 1418.44 & $13.67\pm^{0.02}_{0.02}$ & $38.5\pm1.2$ & $\le8.9\times10^4$ & $\le4.74$ & $\le18.42$ & No\\ 
\enddata
\tablenotetext{a}{Upper limit to the temperature 
 assuming a single component line broadened solely by thermal Doppler broadening.
The temperature limit does not account for errors in the b-value.}
\tablenotetext{b}{$\log f_{\rm H} = $log[N(\ion{H}{2})/N(\ion{H}{1}) under the assumption of collisional
ionization equilibrium at the listed temperature.}
\tablenotetext{c}{Total hydrogen column density implied by the listed values of N(\ion{H}{1}) and $f_{\rm H}$.}
\end{deluxetable}

\clearpage
\newpage

\begin{deluxetable}{lccccccc}
\tablewidth{0pc}
\tablecaption{Nearest Galaxies to \ion{O}{6} Absorption-Line Systems\label{nearestgal}}
\tablehead{\multicolumn{3}{c}{\underline{\ \ \ \ \ \ \ \ \ \ \ \ \ \ion{O}{6} Absorber\tablenotemark{a}$_{ \ }$ \ \ \ \ \ \ \ \ \ \ \ \ }} & \multicolumn{5}{c}{\underline{\ \ \ \ \ \ \ \ \ \ \ \ \ \ \ \ \ \ \ \ \ \ \ \ \ \ \ \ Nearest Galaxy\tablenotemark{b} \ \ \ \ \ \ \ \ \ \ \ \ \ \ \ \ \ \ \ \ \ \ \ \ \ \ \ }} \\
 \ $z_{\rm abs}$  & log N(\ion{O}{6}) & log N(\ion{H}{1}) & $\alpha_{2000}$ & $\delta_{2000}$ & $z_{\rm galaxy}$\tablenotemark{c} & $\mid \Delta v \mid$\tablenotemark{d} & $\rho$\tablenotemark{e} \\
 \ &  &  & (h\ \ m\ \ \ s) & ($\degr\ \ \ \arcmin\ \ \ \arcsec$) & \ & (km s$^{-1}$) & ($h_{75}^{-1}$ kpc) }
\startdata
0.04125 & $13.33\pm^{0.14}_{0.20}$ & $13.25\pm^{0.09}_{0.11}$ & 11 19 09.67 & 21 02 43.2 & 0.04108 & \phn49 & 746  \\
\\
0.05895 & $13.55\pm^{0.11}_{0.14}$ & $13.56\pm^{0.05}_{0.05}$ & 11 19 24.29 & 21 10 30.3 & 0.05916 & \phn59 & 601  \\
\\
0.05928 & $13.39\pm^{0.12}_{0.17}$ & $12.42\pm^{0.13}_{0.18}$ & 11 19 24.29 & 21 10 30.3 & 0.05916 & \phn59 & 601  \\
\\
0.06244 & $13.14\pm^{0.15}_{0.22}$ & $13.18\pm^{0.06}_{0.07}$ & 11 19 42.06 & 21 26 10.7 & 0.06134 & 311 & 677 \\
\\
0.13847 & $13.68\pm^{0.10}_{0.08}$ & $16.20\pm^{0.05}_{0.04}$ & 11 19 06.67 & 21 18 28.3 & 0.13814 & \phn87 & 127  \\
\\
0.16548\tablenotemark{f} & $14.08\pm^{0.05}_{0.06}$ & $13.39\pm^{0.03}_{0.03}$ & 11 19 18.07 & 21 15 03.5 & 0.16581 & \phn85 & 733 \\
\\
0.17340\tablenotemark{f} & $13.43\pm^{0.08}_{0.10}$ & $13.09\pm^{0.03}_{0.03}$ & 11 19 06.5\phn & 21 19 08\phantom{.0} & 0.1757\phn & 587 & \phn88 
\enddata
\tablenotetext{a}{Values for $z$, N(\ion{O}{6}), and N(\ion{H}{1}) are from Table~\ref{tab_summarylya}.}
\tablenotetext{b}{Galaxy with smallest velocity difference from the absorber ($\Delta v$) and smallest projected distance from the line of sight ($\rho$). For the calculation of projected distances, we assume H$_{0}$ = 75 km s$^{-1}$ Mpc$^{-1}$ and $q_{0}$ = 0.}
\tablenotetext{c}{Galaxy redshifts are from  Tripp et al.\ (1998), except for the galaxy at $z=0.1757$, which is 
from Ellingson et al.\ (1991).}
\tablenotetext{d}{$\Delta v = c\Delta z/(1+z)$. Galaxy redshift uncertainties are estimated to range from 40 to 150 km s$^{-1}$ (see Ellingson et al. 1991 and Tripp et al. 1998).}
\tablenotetext{e}{Projected distance from the line of sight at the distance of the galaxy, neglecting corrections for peculiar velocities.}
\tablenotetext{f}{Absorber is within 5000 km s$^{-1}$ of the QSO redshift.}
\end{deluxetable}

\clearpage
\newpage
\begin{deluxetable}{lcccccccccc}
\tablewidth{0pt} 
\tablecaption{Warm-Hot IGM Baryon Content Summary \label{tab_baryonsummary}}
\tablehead{& & \multicolumn{4}{c}{\underline{\ \ \ \ \ \ \ \ Available Path  \ \ \ \ \ \ \ }} & \multicolumn{4}{c}{\underline{\ \ \ \ \ \ \ \ \ Reduced Path\tablenotemark{a} \ \ \ \ \ \  \ \ \ }} \\
System Type& $W_r^{min}$ & $N$  & $\Delta X$ & $(\frac{{\rm d}N}{{\rm d}z})$ & $\Omega_b$~\tablenotemark{b} & $N^\prime$ & $\Delta X^\prime$ & $(\frac{{\rm d}N}{{\rm d}z})^\prime$ & $\Omega_b^\prime$~\tablenotemark{b} \\
 & (m\AA) & & & & ($h_{75}^{-1}$)& & & & ($h_{75}^{-1}$)}
\startdata
Broad \ion{H}{1} & 30 & 8 & 0.161 & 50 & $\lesssim0.020$ & 7 & 0.143 & 49 & $\lesssim0.023$\\
(${\rm b} \ge 40$ \kms) \\
\\
    Broad \ion{H}{1} & 30 & 6 & 0.161 & 37 & $\lesssim0.0046$ & 5 & 0.143 & 35 & $\lesssim0.0049$\\
($40 \le {\rm b} \le 100$ \kms) \\
\\
\tableline
\\
\ion{O}{6} 	   &  30  & 7 & 0.117 & 60 & $\gtrsim0.0045$ & 5 & 0.098 & 51 & $\gtrsim0.0026$\\
\\
\ion{O}{6} 	   &  50  & 2 & 0.117 & 17 & $\gtrsim0.0026$ & 1 & 0.098 & 10 & $\gtrsim0.0009$\\
\\
\ion{O}{6} (Survey of 6 l.o.s.)\tablenotemark{c}   &  50 &  13 & 0.905 & 14 & $\gtrsim0.0027$ & \nodata & \nodata & \nodata & \nodata\\
\\
\enddata
\tablenotetext{a}{Notes -- Primed quantities indicate values for the total PG\,1116+215 redshift path
less the 5000 \kms\ interval nearest the QSO. }
\tablenotetext{b}{Values of $\Omega_b$ and $\Omega_b^\prime$ were calculated from Equations 5 and 7.
Values for \ion{O}{6} assume an \ion{O}{6} ionization fraction $f_{\rm O\,VI} <0.2$
and a metallicity of 0.1 solar.}
\tablenotetext{c}{Values are appropriate for the redshift path probed by 
2 \ion{O}{6} absorbers toward PG\,1116+215 (this paper),
5 absorbers toward PG\,1259+593 (Richter et al.\ 2004), 2 absorbers toward PG\,0953+415
(Savage et al.\ 2002), 4 absorbers toward H\,1821+643 (Tripp et al.\ 2000; Oegerle
et al.\ 2000), and 0 absorbers toward PG\,0804+761 (Richter et al.\ 2001).}
\end{deluxetable}

\end{document}